\documentclass[twocolumn]{aastex631}

\usepackage[utf8]{inputenc}
\usepackage{graphicx}
\graphicspath{{./}{figures/}}
\usepackage{amsmath}
\usepackage{array}  
\usepackage{enumitem}
\usepackage{CJK}

\usepackage{booktabs}
\usepackage{hyperref}

\received{}
\revised{}
\accepted{}

\submitjournal{\apj}

\shorttitle{Arp 240 Merger: mK-S Law on Sub-kpc Scales}
\shortauthors{Saravia et al.}

\begin{document}

\title{The Arp 240 Galaxy Merger: A Detailed Look at the Molecular Kennicutt-Schmidt Star Formation Law on Sub-kpc Scales.}

\correspondingauthor{Alejandro Saravia}
\email{mas6um@virginia.edu}

\author[0000-0003-4546-3810]{A. Saravia}
\altaffiliation{Grote Reber Fellow of the National Radio Astronomy Observatory}
\affiliation{Department of Astronomy, University of Virginia, 530 McCormick Road, Charlottesville, VA 22903, USA}

\author[0000-0001-6956-0987]{E. Rodas-Quito}
\affiliation{Universidad Nacional Autónoma de Honduras, Ciudad Universitaria, Tegucigalpa, Honduras}

\author[0000-0003-0057-8892]{L. Barcos-Muñoz}
\affiliation{National Radio Astronomy Observatory, 520 Edgemont Road, Charlottesville, VA 22903, USA}
\affiliation{Department of Astronomy, University of Virginia, 530 McCormick Road, Charlottesville, VA 22903, USA}

\author[0000-0003-2638-1334]{A. S. Evans}
\affiliation{Department of Astronomy, University of Virginia, 530 McCormick Road, Charlottesville, VA 22903, USA}
\affiliation{National Radio Astronomy Observatory, 520 Edgemont Road, Charlottesville, VA 22903, USA}

\author[0000-0002-1568-579X]{D. Kunneriath }
\affiliation{National Radio Astronomy Observatory, 520 Edgemont Road, Charlottesville, VA 22903, USA}

\author[0000-0003-3474-1125]{G. Privon}
\affiliation{Department of Astronomy, University of Virginia, 530 McCormick Road, Charlottesville, VA 22903, USA}
\affiliation{National Radio Astronomy Observatory, 520 Edgemont Road, Charlottesville, VA 22903, USA}
\affiliation{Department of Astronomy, University of Florida, 1772 Stadium Road, Gainesville, FL 32611, USA}
\author[0000-0002-3139-3041]{Y. Song}
\affiliation{European Southern Observatory, Alonso de Córdova, 3107, Vitacura, Santiago 763-0355, Chile}
\affiliation{Joint ALMA Observatory, Alonso de Córdova, 3107, Vitacura, Santiago 763-0355, Chile}

\author[0000-0001-9163-0064]{I. Yoon}
\affiliation{National Radio Astronomy Observatory, 520 Edgemont Road, Charlottesville, VA 22903, USA}

\author[0000-0001-6527-6954]{K. L.~Emig}

\affiliation{National Radio Astronomy Observatory, 520 Edgemont Road, Charlottesville, VA 22903, USA}

\author[0000-0003-4286-4475]{M. S\'{a}nchez-Garc\'{i}a}
\affiliation{Institute of Astrophysics, Foundation for Research and Technology-Hellas (FORTH), Heraklion, 70013, Greece}

\author[0000-0002-1000-6081]{S. T. Linden}
\affiliation{Steward Observatory, University of Arizona, 933 N Cherry Avenue, Tucson, AZ 85721, USA}

\author[0009-0002-2049-9470]{K. Green}
\affiliation{Department of Astronomy, University of Virginia, 530 McCormick Road, Charlottesville, VA 22903, USA}

\author[0000-0001-7690-3976]{M. Johnstone}
\affiliation{Department of Astronomy, University of Virginia, 530 McCormick Road, Charlottesville, VA 22903, USA}

\author[0009-0002-6248-3688]{J. Nagarajan-Swenson }
\affiliation{Department of Astronomy, University of Virginia, 530 McCormick Road, Charlottesville, VA 22903, USA}

\author[0009-0006-6594-1516]{G. A. Meza }
\affiliation{Universidad Nacional Autónoma de Honduras, Ciudad Universitaria, Tegucigalpa, Honduras}

\author[0000-0003-3168-5922]{E. Momjian}
\affiliation{National Radio Astronomy Observatory, P.O. Box O, Socorro, NM 87801, USA}

\author[0000-0003-3498-2973]{L. Armus}
\affiliation{IPAC, California Institute of Technology, 1200 E. California Boulevard, Pasadena, CA 91125, USA}

\author[0000-0002-2688-1956]{V. Charmandaris}
\affiliation{Institute of Astrophysics, Foundation for Research and Technology-Hellas (FORTH), Heraklion, 70013, Greece}
\affiliation{School of Sciences, European University Cyprus, Diogenes Street, Engomi, 1516 Nicosia, Cyprus}
\affiliation{Department of Physics, University of Crete, Heraklion, 71003, Greece}

\author[0000-0003-0699-6083]{T. Diaz-Santos}
\affiliation{Institute of Astrophysics, Foundation for Research and Technology-Hellas (FORTH), Heraklion, 70013, Greece}

\author[0000-0002-1185-2810]{C. Eibensteiner}

\affiliation{National Radio Astronomy Observatory, 520 Edgemont Road, Charlottesville, VA 22903, USA}

\author[0000-0001-6028-8059]{J. Howell}
\affiliation{IPAC, California Institute of Technology, 1200 E. California Boulevard, Pasadena, CA 91125, USA}

\author[0000-0003-4268-0393]{H. Inami}
\affiliation{Hiroshima Astrophysical Science Center, Hiroshima University, 1-3-1 Kagamiyama, Higashi-Hiroshima, Hiroshima 739-8526, Japan}

\author[0000-0002-6650-3757]{J. Kader}
\affiliation{4129 Frederick Reines Hall, Department of Physics and Astronomy, University of California, Irvine, CA 92697, USA}

\author[0000-0001-5231-2645]{C. Ricci}
\affiliation{Núcleo de Astronomía de la Facultad de Ingeniería y Ciencias, Universidad Diego Portales, Santiago, 8320000, Chile}

\author[0000-0001-7568-6412]{E. Treister}
\affiliation{Instituto de Astrofísica, Facultad de Física, Pontificia Universidad Católica de Chile, Campus San Joaquín, 7820436,Chile}

\author[0000-0002-1912-0024]{V. U}
\affiliation{4129 Frederick Reines Hall, Department of Physics and Astronomy, University of California, Irvine, CA 92697, USA}

\author[0000-0002-4375-254X]{T. Bohn}
\affiliation{Hiroshima Astrophysical Science Center, Hiroshima University, 1-3-1 Kagamiyama, Higashi-Hiroshima, Hiroshima 739-8526, Japan}

\author[0000-0002-1233-9998]{D. B. Sanders}
\affiliation{Institute for Astronomy, University of Hawaii, 2680 Woodlawn Drive, Honolulu, HI 96822, USA}


\begin{abstract}
The molecular Kennicutt-Schmidt (mK-S) Law has been key for understanding star formation (SF) in galaxies across all redshifts. However, recent sub-kpc observations of nearby galaxies reveal deviations from the nearly unity slope (N) obtained with disk-averaged measurements. We study SF and molecular gas (MG) distribution in the early-stage luminous infrared galaxy merger Arp240 (NGC5257-8). Using VLA radio continuum (RC) and ALMA CO(2-1) observations with a uniform grid analysis, we estimate SF rates and MG surface densities ($\Sigma_{\mathrm{SFR}}$ and $\Sigma_{\mathrm{H_2}}$, respectively). In Arp 240, N is sub-linear at 0.52 $\pm$ 0.17. For NGC 5257 and NGC 5258, N is 0.52 $\pm$ 0.16 and 0.75 $\pm$ 0.15, respectively. We identify two SF regimes: high surface brightness (HSB) regions in RC with N $\sim$1, and low surface brightness (LSB) regions with shallow N (ranging 0.15 $\pm$ 0.09 to 0.48 $\pm$ 0.04). Median CO(2-1) linewidth and MG turbulent pressure (P$_{\mathrm{turb}}$) are 25 km~s$^{-1}$ and 9 $\times$10$^{5}$ K~cm$^{-3}$. No significant correlation was found between $\Sigma_{\mathrm{SFR}}$ and CO(2-1) linewidth. However, $\Sigma_{\mathrm{SFR}}$ correlates with P$_{\mathrm{turb}}$, particularly in HSB regions ($\rho >$0.60). In contrast, SF efficiency moderately anti-correlates with P$_{\mathrm{turb}}$ in LSB regions but shows no correlation in HSB regions. Additionally, we identify regions where peaks in SF and MG are decoupled, yielding a shallow N ($\leq$ 0.28 $\pm$ 0.18). Overall, the range of N reflects distinct physical properties and distribution of both the SF and MG, which can be masked by disk-averaged measurements.

\end{abstract}


\section{Introduction}\label{Intro}

Star formation (SF), a process that fundamentally shapes the evolution of galaxies, occurs within molecular gas clouds. As stars form, their jets, radiation, winds, and explosions inject energy and momentum into the interstellar medium (ISM) over a wide range (sub-pc to 10s kpc) of scales, a process referred to as stellar feedback \citep[e.g.,][]{Kennicutt_n_Evans_2012}.

The rate and efficiency at which stars form is thereby governed by a complex interplay of physical processes. Stellar feedback, the presence of an active galactic nucleus (AGN), the accretion of gas from the larger environment, the mass and weight of a galaxy, and the presence of other galaxies can all influence the spatial distribution and physical conditions of molecular gas within galaxies, and thus SF \citep[e.g.,][]{Leroy_2008, Dekel_2009, Krumholz_2009, Murray_2010, Kormendy_2013}. Both large samples and individual dedicated studies are required to understand SF across a wide variety of environments.

SF, while complex, is often assessed via a simple relationship between the SF rate (SFR) and total cold (neutral and molecular) gas surface densities, $\Sigma_{\mathrm{SFR}}$ and $\Sigma_{\mathrm{gas}}$ respectively. The Kennicutt-Schmidt (K-S) Law, is formulated as a power law $\Sigma_{\mathrm{SFR}}\propto (\Sigma_{\mathrm{gas}})^N$, and it is correlated across seven orders of magnitude in both quantities for integrated measurements \citep[e.g.,][]{Schmidt_1959, Kennicutt98, Kennicutt2021}. Additionally, a variant of this relation, the Molecular Kennicutt-Schmidt (mK-S) Law, $\Sigma_{\mathrm{SFR}} \propto (\Sigma_{\mathrm{H_2}})^N$, specifically links the $\Sigma_{\mathrm{SFR}}$ to the molecular gas surface density $\Sigma_{\mathrm{H_2}}$, exhibiting a tighter correlation \citep[e.g.,][]{Wong_2002}. In both cases, the power law index (N), or the slope in logarithmic space, indicates how the efficiency of SF varies with $\Sigma_{\mathrm{H_2}}$ \citep[e.g.,][]{Semenov_2019}. Typically, when averaged over an entire galaxy, the slope has been found to be around 1.4-1.5 for the standard K-S law and closer to unity in the case of the mK-S law \citep[e.g.,][]{Kennicutt2021}. 

To further investigate the robustness of these laws and understand what drives the correlations seen at global scales, it has become essential to study their applicability at smaller scales (e.g., sub-kpc) in a wide range of environments \citep[e.g.,][]{Bigiel_2008, Leroy_2013, Shi_2018, Sanchez-Garcia_2022,Sun_2022}. In recent years, enabled by the Atacama Large Millimeter/Sub-millimeter array (ALMA), the Physics at High Angular Resolution in Nearby Galaxies -- ALMA (PHANGS-ALMA) collaboration \citep{Phangs2021} has made  significant strides in addressing this issue, studying SF and gas conditions within normal nearby star-forming galaxies. In their sample, they found that the scatter of the mK-S law increases with improving resolution \citep[e.g.,][]{Kreckel_2018, Pessa_2021,Sun_2023}. Similar studies in local luminous infrared galaxies (LIRGs; L$_{\mathrm{IR}}$~$>$ 10$^{11}$ L$_{\odot}$) have found several examples of significant deviations in the slope of the mK-S law at sub-kpc resolution when contrasting the central regions of a galaxy and its outskirts \citep[e.g.,][]{Sanchez-Garcia_2022}.

\begin{figure*}[ht]
    \centering
    \includegraphics[width=1\textwidth]{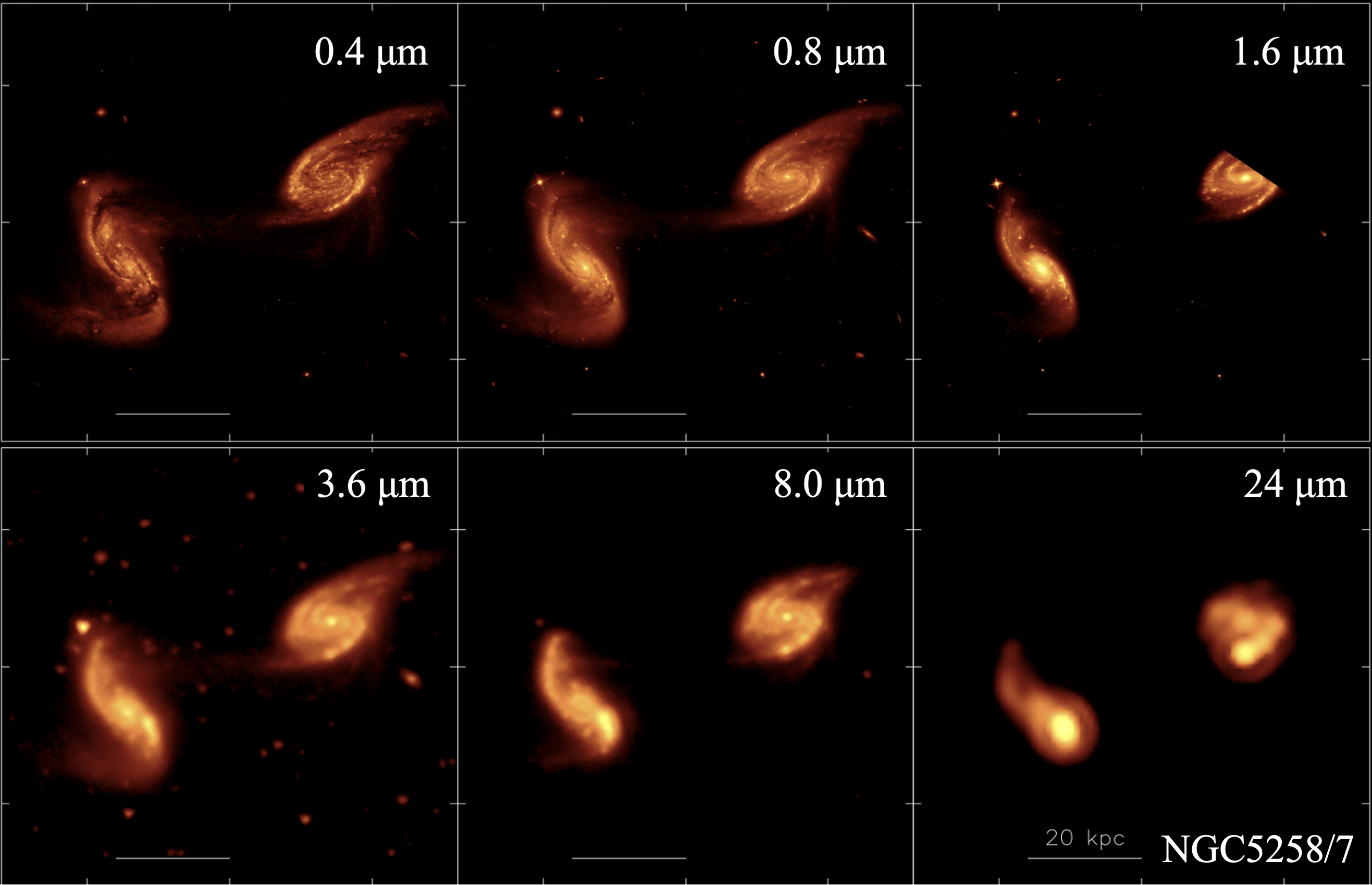}
    \caption{\small The Arp 240 system, seen from the optical with HST (0.4 \(\mu\)m) to the mid infrared (24 \(\mu\)m) with Spitzer. In this Figure, North is up and east is to the left. In each panel, NGC 5258 is the eastern galaxy, while NGC 5257 is the western galaxy. The sequence shows how the extra nuclear regions in both galaxies become the dominant components at longer wavelengths.}
    \label{fig:composite}
\end{figure*}

In this study, we explore the mK-S law for sub-kpc regions within a system of two massive galaxies that are in the early stage of their merger. This phase provides a unique window into the initial interactions and their impact on molecular gas distribution and SFR in colliding galaxies \citep[e.g.,][]{Sanders1996, Hopkins_2008}. 
We focus on Arp 240 (NGC5257/8; L$_{\mathrm{IR}} = 10^{11.4} \, \text{L}_{\odot}$), which is part of the Great Observatories All-sky LIRG Survey (GOALS; \citealt{GOALS_2009}). Multiwavelength images of Arp 240 are presented in Figure \ref{fig:composite}. This system provides a striking example of enhanced SF triggered by the tidal interaction of two massive spiral galaxies. N-body simulations have been used to determine that we are observing this system only $\sim$ 240 Myr after the  first peri-centric passage \citep{Privon_2013}, making it an excellent target for studying SF and ISM properties during the initial phases of a major merger event. 

\begin{figure*}
    \centering
    \includegraphics[width=\textwidth]{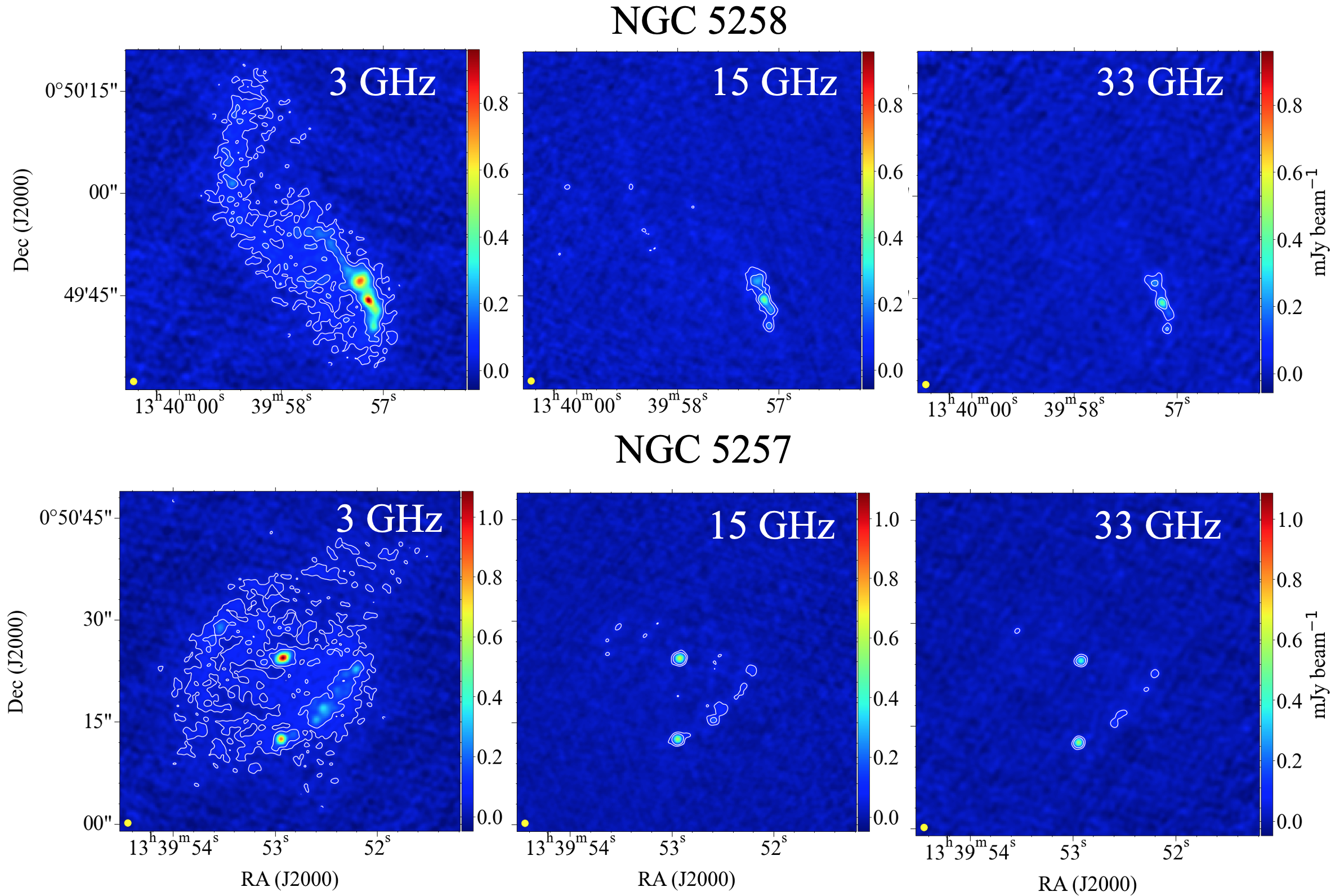}
    \caption{\small Radio continuum images of NGC 5258 (top) and NGC 5257 (bottom) at resolution 1$\arcsec$ $\sim$ 500 pc. The yellow filled circle in the lower left corner represents the synthesized beam size. Images at 3 GHz (S-Band) and 15 GHz (Ku-Band) are a combination of configuration A and C of the VLA; and  for 33 GHz (Ka-Band) only configuration C is presented. White contours enclose emission at 3-$\sigma$ and 7-$\sigma$. The rms values are 16 $\mu Jy$ at S-Band, 18.4 $\mu Jy$ at Ku-Band and 22.7 $\mu Jy$ at Ka-Band. A full description of the data is presented in \cite{Linden_2019} and \cite{Song_2022}. The sequence showcases Arp 240’s radio SED and the sensitivity of the S-Band to low surface brightness emission.}  
\label{fig:radio_S_ku_Ka}
\end{figure*}

Notably, some of its brightest SF regions, which contribute significantly to its infrared and radio luminosity, are predominantly located outside the nuclei, as shown in Figures \ref{fig:composite} and \ref{fig:radio_S_ku_Ka}. This is in stark contrast to the nuclear concentrations of SF occurring in late-stage mergers \citep[e.g.,][]{Haan_2011, Barcos-Muñoz_2017,Song_2022, Evans_2022, Rich_2023}. The substantial extranuclear star-forming complexes in Arp 240 make this system an ideal laboratory to characterize pure SF in extreme environments with no potential contamination from an AGN.

\begin{figure*}
    \centering
    \includegraphics[width=0.49\textwidth]{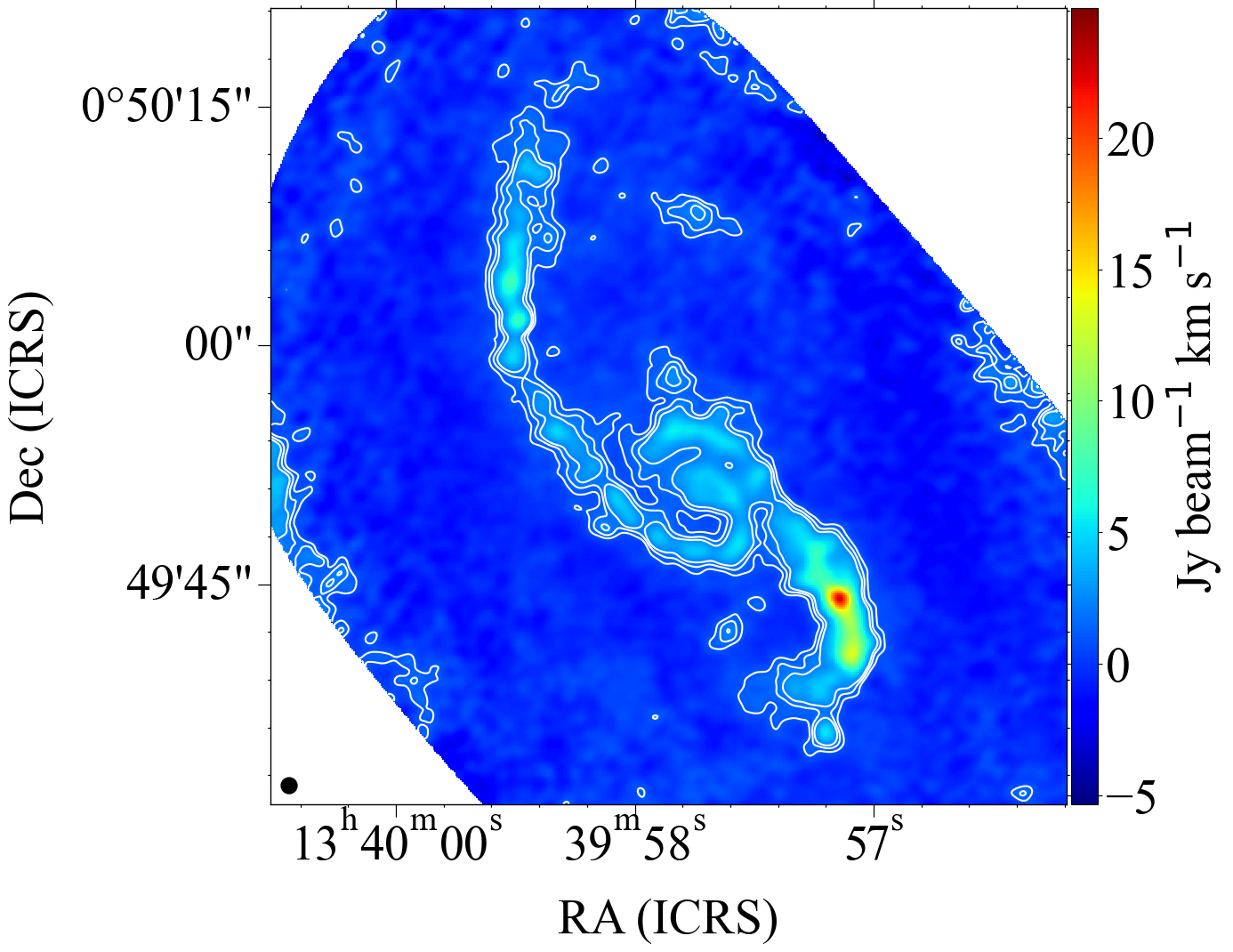}
    \includegraphics[width=0.49\textwidth]{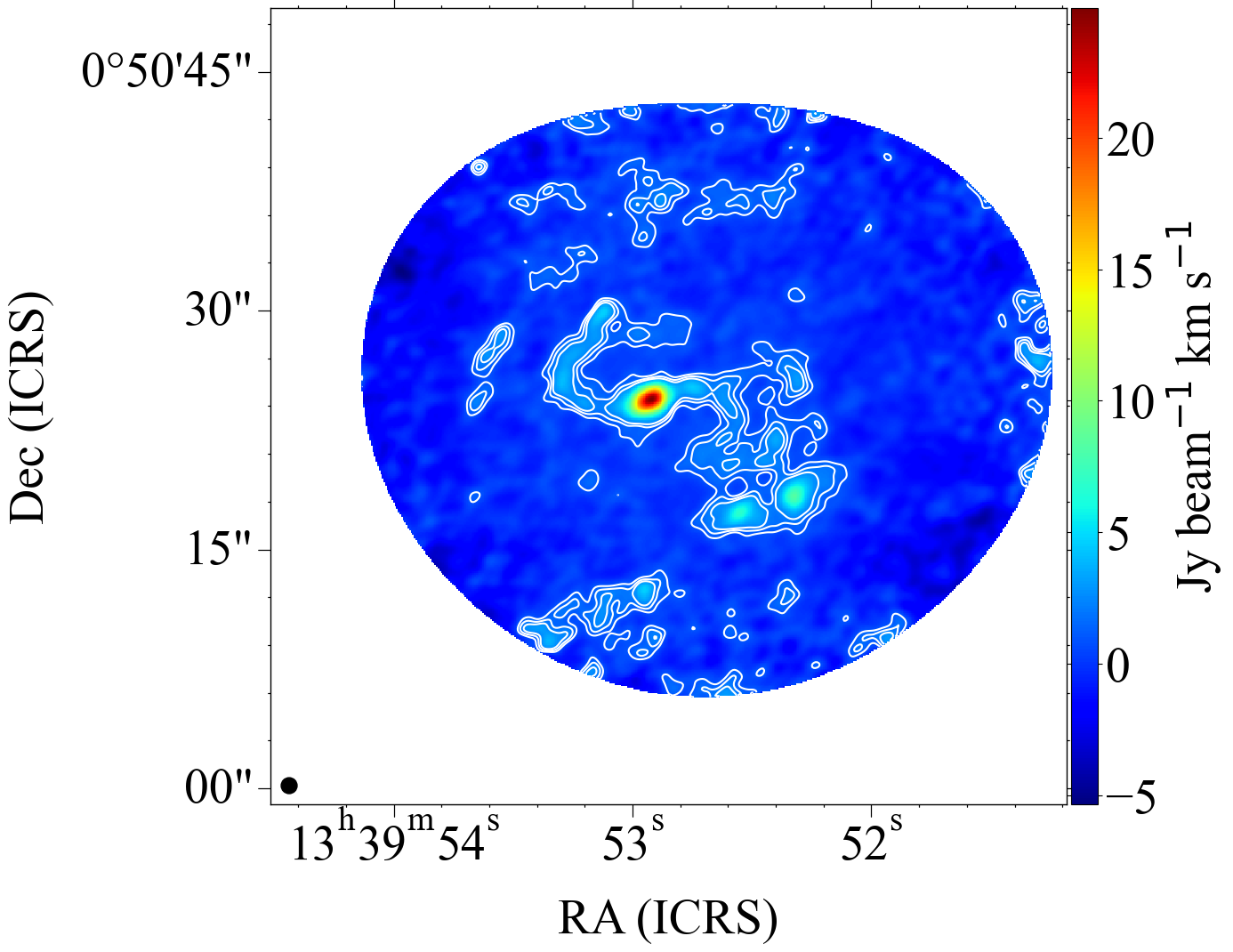}
    \caption{\small Total line intensity (moment-zero) maps of NGC 5258 (Left) and NGC 5257 (Right) at 1$\arcsec$ $\sim$ 500 pc from the ALMA CO($2\rightarrow1$) observations. White contours correspond to emission at 3-$\sigma$, 5-$\sigma$ and 7-$\sigma$. The black filled circle in the lower left corner of each panel represents the synthesized beam size.} 
    \label{fig:CO+VLA}
\end{figure*}

We will derive estimates of SFR using radio continuum observations from the Karl G. Jansky Very Large Array (VLA), and ALMA carbon monoxide (CO $2\rightarrow1$) observations to measure properties of the molecular gas.
Previous studies of this system using radio continuum observations have either estimated SFR integrated over the entire galaxies \cite[e.g.,][]{Yun_2001, Iono_2005}, or focused on the brightest compact regions \citep[e.g.,][]{Linden_2019, Song_2022}. Specifically, \cite{He_2019} conducted a detailed study of Arp 240 that combined radio continuum data with several CO transitions, yet their SFR estimates remained limited to the brightest complexes in radio continuum. In this paper, we expand upon these previous studies by employing a uniform grid analysis across the entire extension of both galaxies. This approach allows us to probe regions of both high and low surface brightness in radio continuum and molecular gas. We examine the mK-S law and the dynamical properties of the gas within regions that are 500~pc in radius, spanning the full extent of each galaxy. 

This paper is structured as follows: In \S \ref{data}, we provide a description of the datasets and the imaging process. \S \ref{calculations} is a presentation of our methodology for region selection and threshold determination, photometry results, and the derived quantities.  In \S \ref{discussion}, we discuss the results of applying the mK-S law to regions within Arp 240. Finally, \S \ref{Summary} is a summary of our findings and conclusions. 

We used a cosmology with Hubble constant of $H_{0}$ = 70 $\mathrm{km\,s^{-1}\,Mpc^{-1}}$, $\Omega_{vacuum} = 0.72$, $\Omega_{matter} = 0.28$ and adopt a redshift of $z=0.02250$ (equivalent to a luminosity distance of 98~Mpc and a scale of 1\arcsec $\sim$ 500~pc). 



\section{Observations and data reduction}\label{data}

\subsection{VLA}\label{VLA}
To study SF in Arp 240, we sample a significant range of the radio continuum spectral energy distribution (SED). We made use of three well-separated frequencies ($\nu$): 3 GHz (S-Band), 15 GHz (Ku-Band) and 33 GHz (Ka-Band) as shown in Figure \ref{fig:radio_S_ku_Ka}. A full description of the observations is presented in \cite{Linden_2019} and \cite{Song_2022}. We used observations from A- and C-configurations of the VLA for S- and Ku-Bands. Combining measurements from various configurations optimizes resolution and sensitivity. This multi-configuration method improves image quality through increased $u,v$ coverage, enhancing sensitivity to detect both compact regions and extended emission. In the case of Ka-Band, we only used measurements from C-configuration, whose angular resolution is similar to that of S-Band at A-configuration (0.63\arcsec). Radio emission at frequencies near Ka-Band is expected to be significantly fainter than the emission at lower frequencies, as the SED slope follows a decreasing power law \citep[e.g.,][]{Condon1990,Murphy_2012}. Therefore, detection at high frequencies requires observations with integration times considerably longer than those of our datasets to achieve comparable signal-to-noise ratio (SNR) of the S-Band data. Nonetheless, observing at Ka-Band with the C-configuration is suitable for our study, considering the final common resolution of 1$\arcsec$ $\sim$ 500~pc and its largest angular scale of 44$\arcsec$ $\sim $ 21 kpc. The largest angular scales at 3~GHz are 18$\arcsec$ and 490$\arcsec$ for the A- and C-configurations, respectively, while at 15~GHz the largest angular scales are 3.6$\arcsec$ and 97$\arcsec$ for the A- and C-configurations, respectively.
\par 
We visually inspected each measurement set and manually removed the defective data, bad antennae, and radio frequency interference (RFI) using the command $\tt{flagdata}$ in CASA \citep{CASA, CASA_2022}. To produce a single, full-bandwidth continuum image at each frequency band, we combined the measurement sets from each configuration using the $\tt{concat}$ task in CASA. Similarly to \cite{Song_2021}, we down-weighted shorter baselines (C-configuration) to account for their denser spatial distribution and $u,v$ coverage relative to longer baselines. After testing a range of weighting factors, we adopted a 4:1 (A:C) ratio, which produced the most centrally concentrated Point Spread Function (PSF) for all images. This choice represents the minimum ratio needed to maintain an optimal PSF shape for the combined image, effectively balancing the enhancement of the central concentration without excessively downweighting the data from the C-configuration.

 We used $\tt{tclean}$ in CASA v6.5 for imaging, using Briggs weighting and a robust parameter of 0.5. For the S-band, we used the $\textit{Hogbom}$ deconvolver \citep{Hogbom_1974}, while for the Ku- and Ka-Bands, we used $\textit{Multi-term (Multi Scale) Multi-Frequency Synthesis}$ \citep{msmf_deconvolver} with nterms = 2. Self-calibration, in amplitude and phase, was possible for the S-band due to its high signal-to-noise ratio. Additionally for all images, we configured the $\tt{tclean}$ function with the following parameters: restoringbeam = 1$\arcsec$, uvtaper = 0.75$\arcsec$, and a pixel size of (cell) = 0.1$\arcsec$, which resulted on an image with a beam full width half maximum (FWHM) of 1$\arcsec$. This enables direct measurements and pixel-by-pixel comparisons across all images. We determined the root mean square (rms) noise of the radio continuum image by using the $\tt{imstat}$ function in CASA. The rms noise values are 16 $\mu$Jy~beam$^{-1}$ at S-Band, 18.4 $\mu$Jy~beam$^{-1}$ at Ku-Band and 22.7 $\mu$Jy~beam$^{-1}$ at Ka-Band.

\subsection{ALMA}\label{ALMA}
Our analysis of molecular gas was conducted using the CO ($2\rightarrow1$) transition, employing ALMA pipeline-calibrated datasets (ID 2015.1.00804.S, PI: Kazimierz Sliwa). Observations were performed using both ALMA 7m and 12m arrays. The native angular resolution were 4$\arcsec$ ($\sim$ 2 kpc) for the 7m datasets and 0.6$\arcsec$ ($\sim$300 pc) for the 12m datasets. Their sensitivity to largest angular scale were 30$\arcsec$ ($\sim$15 kpc) and 7$\arcsec$ ($\sim$3.5 kpc), respectively. 

We combined the 7m and 12m datasets using CASA's $\tt{concat}$ task with equal weighting, and re-imaged them with $\tt{tclean}$ to match the angular resolution and pixel size of our radio continuum image, setting the restoring beam to 1$\arcsec$ and cell size to 0.1$\arcsec$. Our spectral resolution is 3 km~s$^{-1}$ per channel. We utilized automasking \citep{Kepley_2020} following the CASA documentation and adopted their recommended values: rms factor of 3, sidelobethresh~=~2.0, noisethresh~=~4.25, minbeamfrac~=~0.3, lownoisethresh~=~1.5, negativethresh~=~0.0. We computed the average rms noise per line-free channel, obtaining values of 4.7 mJy~beam$^{-1}$ for NGC 5258 and 6.6 mJy~beam$^{-1}$  for NGC 5257. Integrated intensity images (moment-zero maps) are presented in Figure \ref{fig:CO+VLA}, with white contours highlighting emissions at 3-$\sigma$, 5-$\sigma$ and 7-$\sigma$ levels.

\section{Calculations and Results}\label{calculations}
\subsection{Hexagonal grid and region selection}\label{hex_sel}
To study the emission from both radio continuum and CO ($2\rightarrow1$), we adopted a uniform grid of hexagon-shaped tiles covering the entire field of view for each image. The area of each tile corresponds to the area of the synthesized beam (1.13 arcsecond$^2$ $\sim$ 0.26~kpc$^2$), which is consistent across all images. Each tile was applied as an aperture upon the image to conduct photometric analysis. This approach allows us to associate each tile, or region, with a specific location within the galaxy, select thresholds to  include or exclude regions, and group them based on their properties. Moreover, it allows us to generate synthetic maps of all the measurable quantities from our images. This uniform hexagonal grid leverages the capabilities of ALMA and the VLA at 3~GHz to detect a wide range in surface brightness, as illustrated in Figures \ref{fig:radio_S_ku_Ka} and \ref{fig:CO+VLA}, thus enabling a more thorough spatial sampling.

For region selection, we applied a threshold of three times the rms noise value (3-$\sigma$) to each radio continuum image at each frequency band. For the CO ($2\rightarrow$1) data cubes, to ensure robust line detection and maximum flux density recovery, we chose hexagons showing emissions exceeding twice the rms noise across at least nine consecutive channels. This selection criterion was determined after testing various combinations of rms values and channel numbers, finding that our chosen parameters significantly enhanced our ability to detect fainter emission. Furthermore, we visually inspected each hexagon's line profile to ensure the reliability of each detected line, thereby maximizing the overall number of lines of sight with detection.

Utilizing the extended emission observed at 3~GHz, we mapped the spatial overlaps between the radio continuum and CO ($2\rightarrow1$) detections, extending our analysis beyond the brightest knots of SF seen at all frequency bands. We identified a total of 341 hexagons\footnote{Throughout the paper, the terms ``hexagons" and ``regions" are used interchangeably, indicating that each hexagonal aperture encloses a region measuring 500~pc across.} with overlapping detections in both the CO ($2\rightarrow1$) data and radio continuum at 3 GHz: 214 in NGC 5258 and 127 in NGC 5257. In addition to the overlapping detections, we also found hexagons with emission detected in one tracer but not the other. Specifically, we found 48 regions (36 in NGC 5258 and 12 in NGC 5257) with CO~($2\rightarrow1$) detection only, and 264 regions (48 in NGC 5258 and 216 in NGC 5257) with only radio continuum detection. 

For most of our analysis, we primarily use hexagons from the overlapping category. However, single detections are considered to place upper limits for each corresponding tracer. We will discuss the single detections in more detail in $\S$ \ref{section_KS} and $\S$ \ref{discussion}.

\begin{figure}
    \centering
    \includegraphics[width=1\columnwidth]{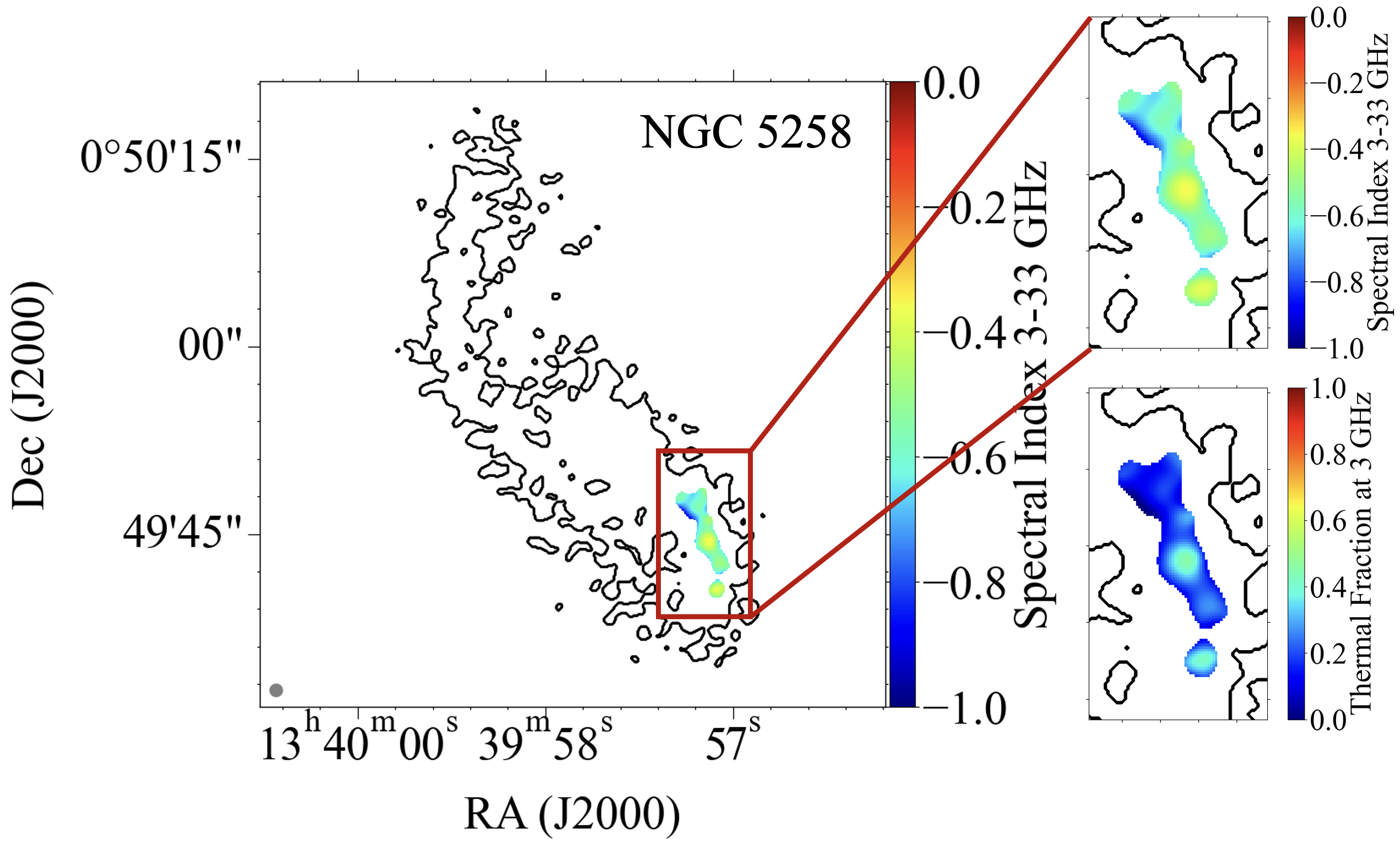}
    \includegraphics[width=1\columnwidth]{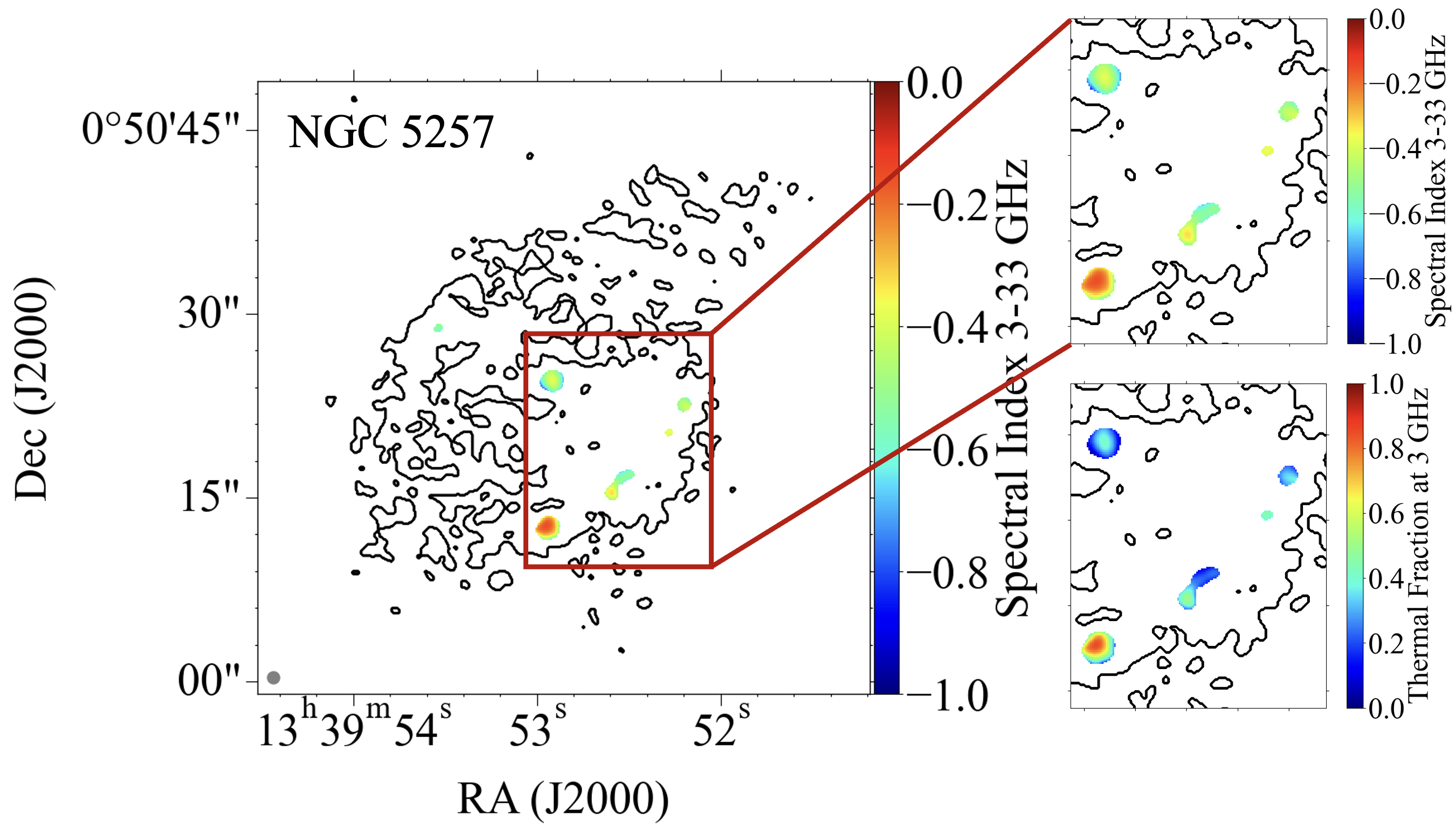}
    \caption{\small Maps of spectral indices between 3 GHz and 33 GHz for the brightest regions of NGC 5258 (top) and NGC 5257 (bottom). Black contours represent 3 GHz radio continuum emission at 3-$\sigma$. The right panels show a zoom-in of the 3-33 GHz spectral index and the corresponding thermal fraction at 3 GHz. These values are consistent with those found in star-forming galaxies in the literature. The area left blank inside the contours corresponds to regions with detection only at 3 GHz but not at 33 GHz.} 
    \label{fig:Th_F}
\end{figure}



\subsection{Radio Continuum and Star Formation Rates}\label{sfr}
The most noticeable feature in Figure \ref{fig:radio_S_ku_Ka} is the decrease in flux density with increasing frequency, which is expected for SF dominated galaxies \cite[e.g.,][]{Condon1990, Murphy_2012}. Additionally, two distinct surface brightness regimes are observed: the brightest and compact star-forming complexes and the low surface brightness extended emission. The bright and compact complexes are detected across all frequency bands in our datasets, while the low surface brightness regions are only detected with the 3~GHz. To capture these low surface brightness components at 15 GHz and 33 GHz, observation times would need to be extended to at least ten times those currently utilized at these frequencies to achieve the necessary sensitivity. We therefore leverage the higher SNR available at 3 GHz to comprehensively calculate the SFR across the full extent of both galaxies, including the low surface brightness ISM, which remains largely undetected at higher frequencies. Our choice of 3~GHz as an SFR tracer significantly expands upon previous studies \citep[e.g.,][]{He_2019}, by increasing the number of line of sights analyzed.

To estimate SFR, we adopted the prescription outlined in \cite{Murphy_2012}, for radio continuum emission:
\begin{equation}\label{eq_sfr}
\begin{split}
\left( \frac{\mathrm{SFR_\nu}}{\text{M}_{\odot} \text{ yr}^{-1}} \right) &= 10^{-27}  \left[ 2.18 \left( \frac{T_{e}}{10^{4} \text{ K}} \right)^{0.45} \left( \frac{\nu}{\text{GHz}} \right)^{-0.1} + \right. \\
& \left. 15.1 \left( \frac{\nu}{\text{GHz}} \right)^{\alpha_{\text{NT}}} \right]^{-1}  \left( \frac{L_{\nu}}{ \text{ erg s}^{-1} \text{ Hz}^{-1}} \right)
\end{split}
\end{equation}

\noindent
where $\nu$ = 3 GHz, $L_{\nu}$ is the luminosity of each region at 3 GHz and $T_{\mathrm{e}}$ is the electron temperature, which we have assumed to be 10$^{\mathrm{4}}$ K \citep[e.g.,][]{Murphy_2011}.

This approach is premised on the understanding that the measured intensity combines spectral contributions from both thermal ($\propto \nu^{\mathrm{-0.1}}$) and non-thermal ($\propto \nu^{\mathrm{\alpha_{NT}}}$) emission. The prescription is based on Starburst99 models \citep{Leitherer_1999}, which assume a Kroupa  initial mass function (IMF) \citep{Kroupa_2001} and continuous SF over a 100 Myr timescale \citep{Murphy_2012}. \citealt{Linden_2019} showed that for continuous SF, the Starburst99 models predict a 3-33~GHz spectral index of approximately -0.5 at 100 Myr, assuming a non-thermal spectral index of $\alpha_{NT}$ = -0.85. This corresponds to a fixed thermal fraction of 0.26 at 3 GHz.

To validate the use of Equation \ref{eq_sfr} for Arp 240, we calculated the spectral indices between 3 GHz and 33 GHz for the high surface brightness regions in both galaxies. This was done by fitting the data from all three frequency bands, selecting emission above a 3-$\sigma$ threshold.
 
Figure \ref{fig:Th_F} shows the resulting spectral index maps, with median values of -0.56 for NGC 5258 and -0.45 for NGC 5257. The corresponding error maps (not displayed) yield a median error of $\pm$ 0.06, calculated from the rms noise of each image. These values are consistent with the models discussed earlier. Additionally, the lower lateral panels of Figure \ref{fig:Th_F} present the thermal fraction maps for both galaxies at 3 GHz, derived using Equation 11 from \cite{Murphy_2012}. The median thermal fractions are 0.20 for NGC 5258 and 0.32 for NGC 5257, with a median error of $\pm$ 0.05. In regions detected exclusively at 3 GHz, the upper limit for the thermal fraction is 0.52, allowing for consistency with the theoretical expectation. 

Nevertheless, there is one region in NGC 5257 where the thermal fraction reaches an unusually high value of 0.85, which appears unrealistic. This region is confined to a single hexagon in the applied grid. The overestimation of the thermal fraction in this area may stem from our assumption of a fixed $\alpha_{\mathrm{NT}}$, which could vary locally. In a forthcoming paper (Saravia et al., in preparation), we will revisit this calculation using 100~pc scales observations and wider frequency coverage, allowing us to directly fit both the thermal and non-thermal components, as well as $\alpha_{\mathrm{NT}}$. For instance, allowing $\alpha_{\mathrm{NT}}$ to vary between $-1$ and $-0.6$ \citep[e.g.,][]{Murphy_2012}, results in a corresponding variation in SFR within $\pm$ 20$\%$. 

We further utilized the median spectral indices to convert the integrated 3 GHz luminosity to 1.4 GHz to verify consistency with the well-established far-infrared–radio correlation \citep[e.g.,][]{Helou} for star-forming galaxies. Following \cite{Murphy_2011}, we used the integrated total infrared luminosity and derived a value of q$_{\mathrm{IR}}$ =2.56 $\pm$ 0.04, which is consistent with the results of \cite{Murphy_2011} for a star-forming galaxy.

Overall, our results agree with theoretical expectations and with previous studies of star-forming galaxies \citep[e.g.,][]{Niklas_97, Murphy_2011}. Consequently, we applied Equation \ref{eq_sfr} to compute the SFR, assuming a non-thermal spectral index of $\alpha_{NT} = -0.85$ for all hexagons detected at 3~GHz.

\begin{figure*}
    \centering
    \begin{tabular}{cc}
        \includegraphics[width=0.45\linewidth]{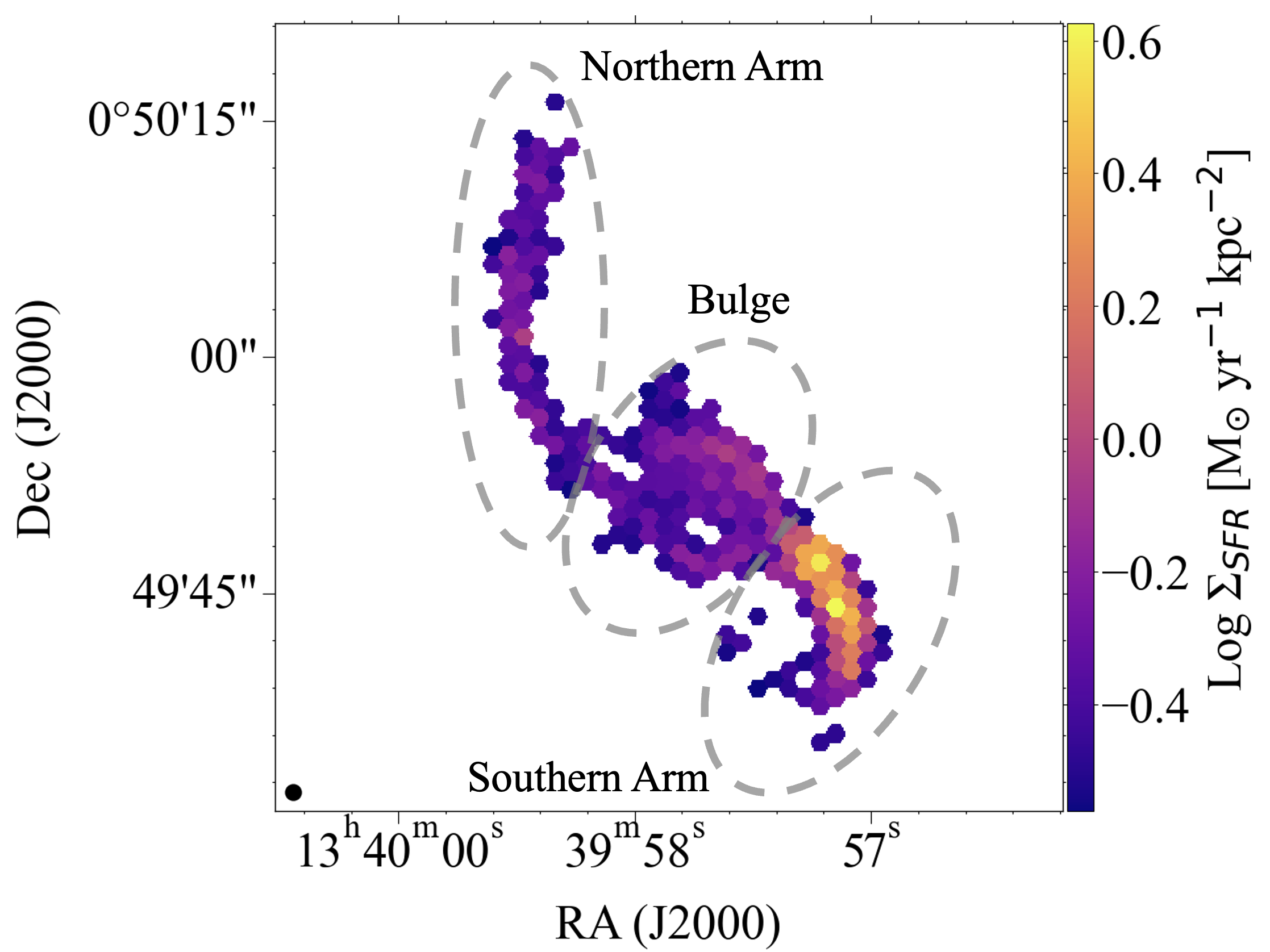} &
        \includegraphics[width=0.45\linewidth]{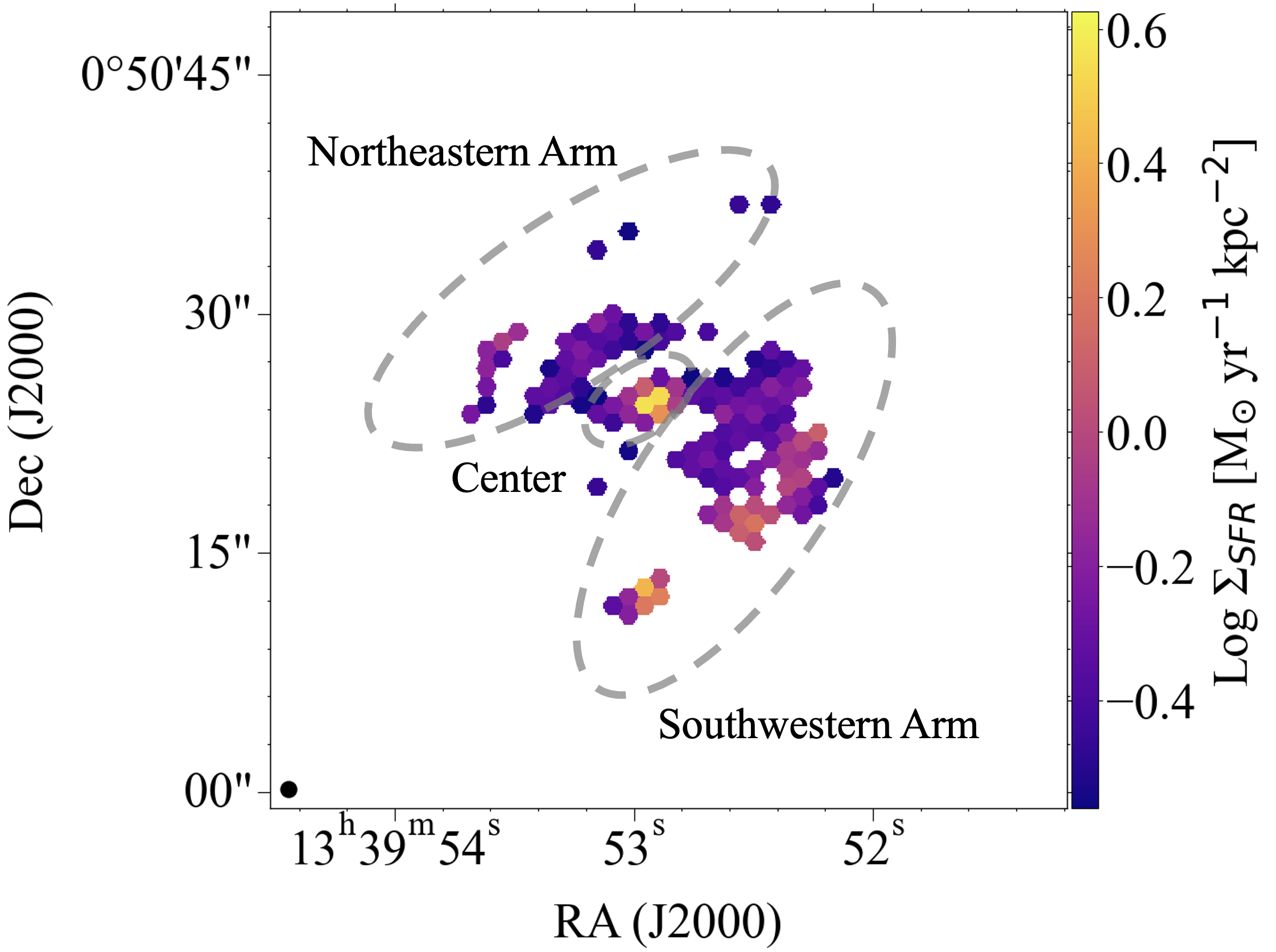} \\
        \textbf{(a)} $\Sigma_{SFR}$ NGC 5258 & \textbf{(b)} $\Sigma_{SFR}$ NGC 5257 \\
    \end{tabular}
    \\[2ex] 

    \begin{tabular}{cc}
        \includegraphics[width=0.45\linewidth]{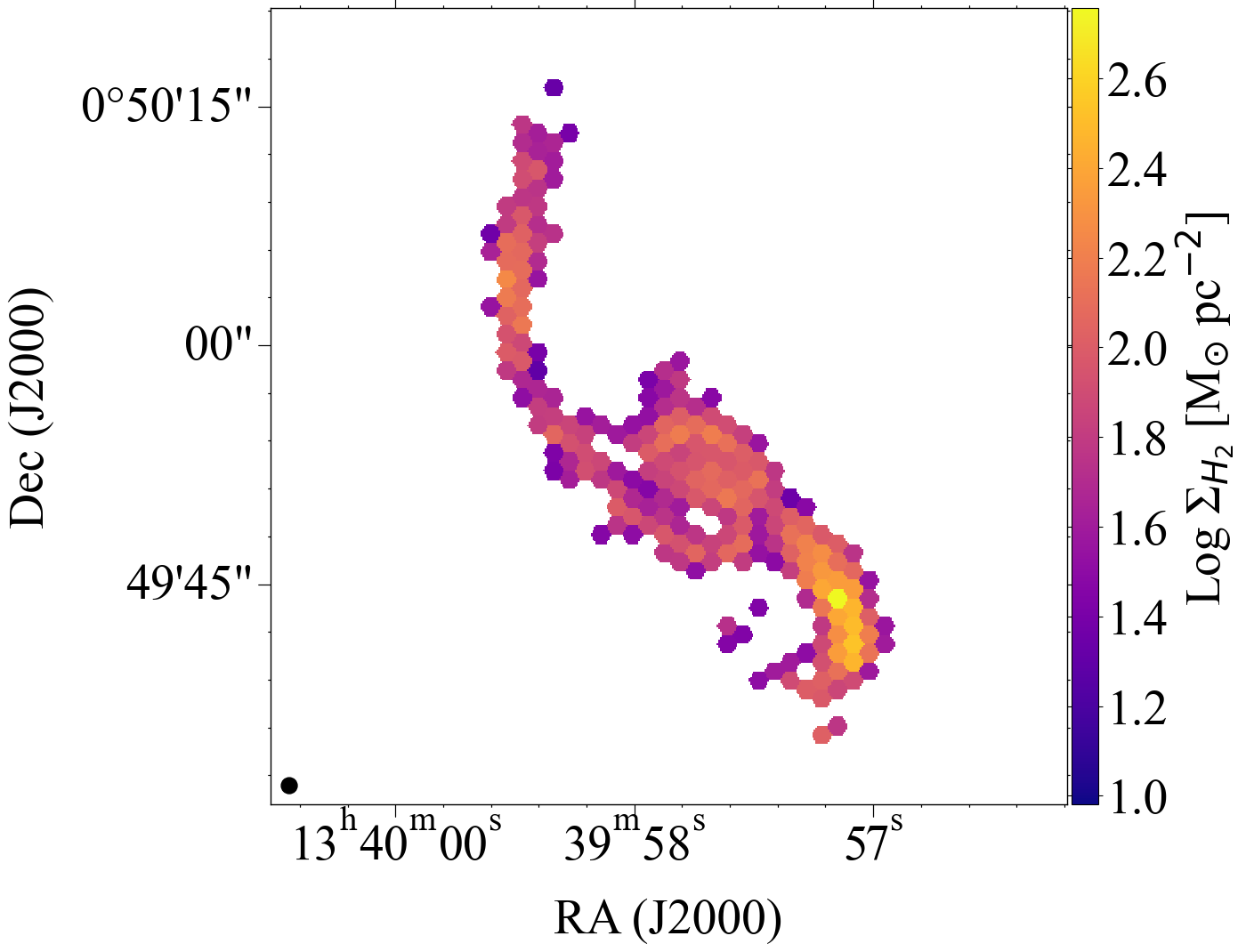} &
        \includegraphics[width=0.45\linewidth]{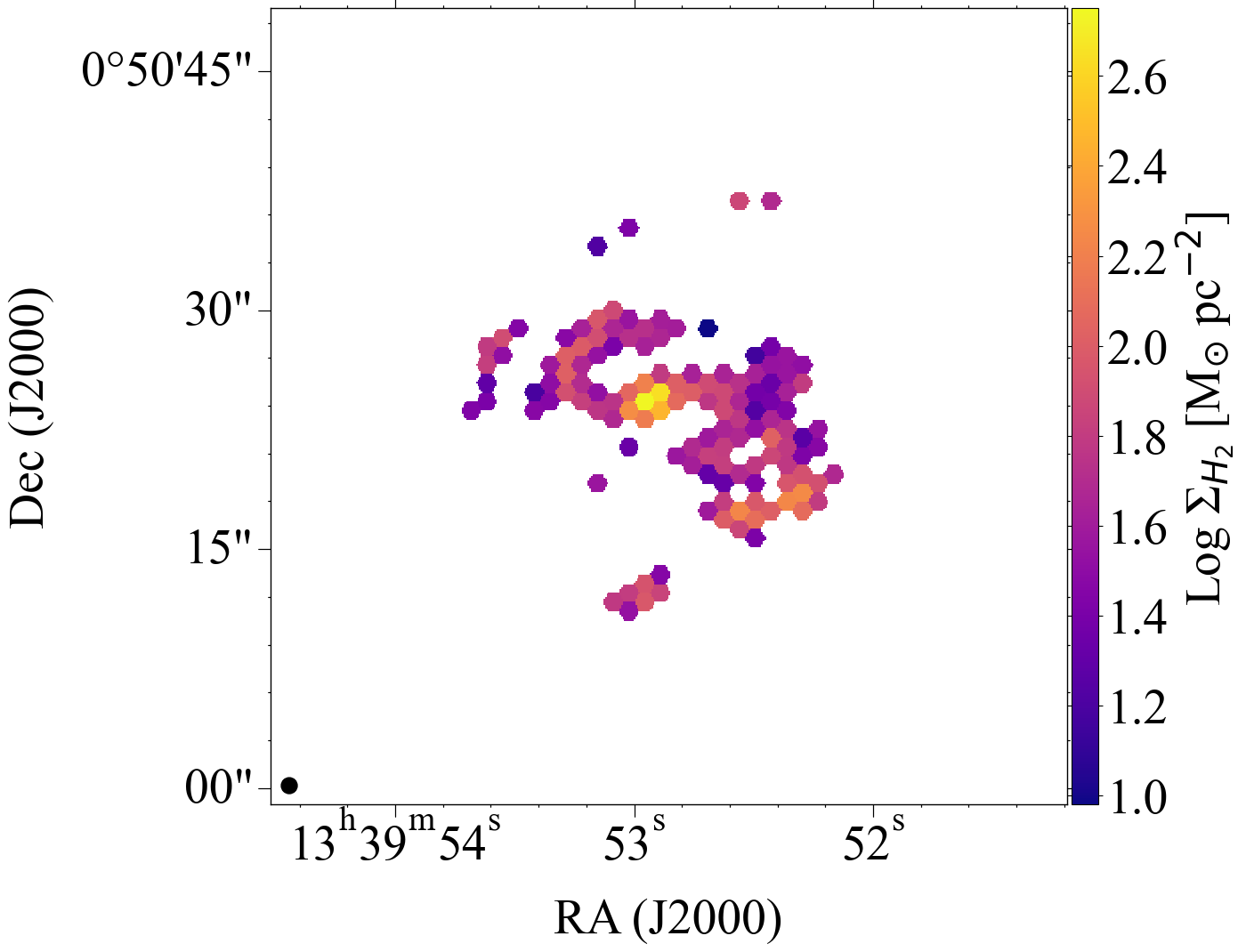} \\
        \textbf{(c)} $\Sigma_{H_2}$ NGC 5258 & \textbf{(d)} $\Sigma_{H_2}$ NGC 5257 \\
    \end{tabular}
    \\[2ex]

    \begin{tabular}{cc}
        \includegraphics[width=0.45\linewidth]{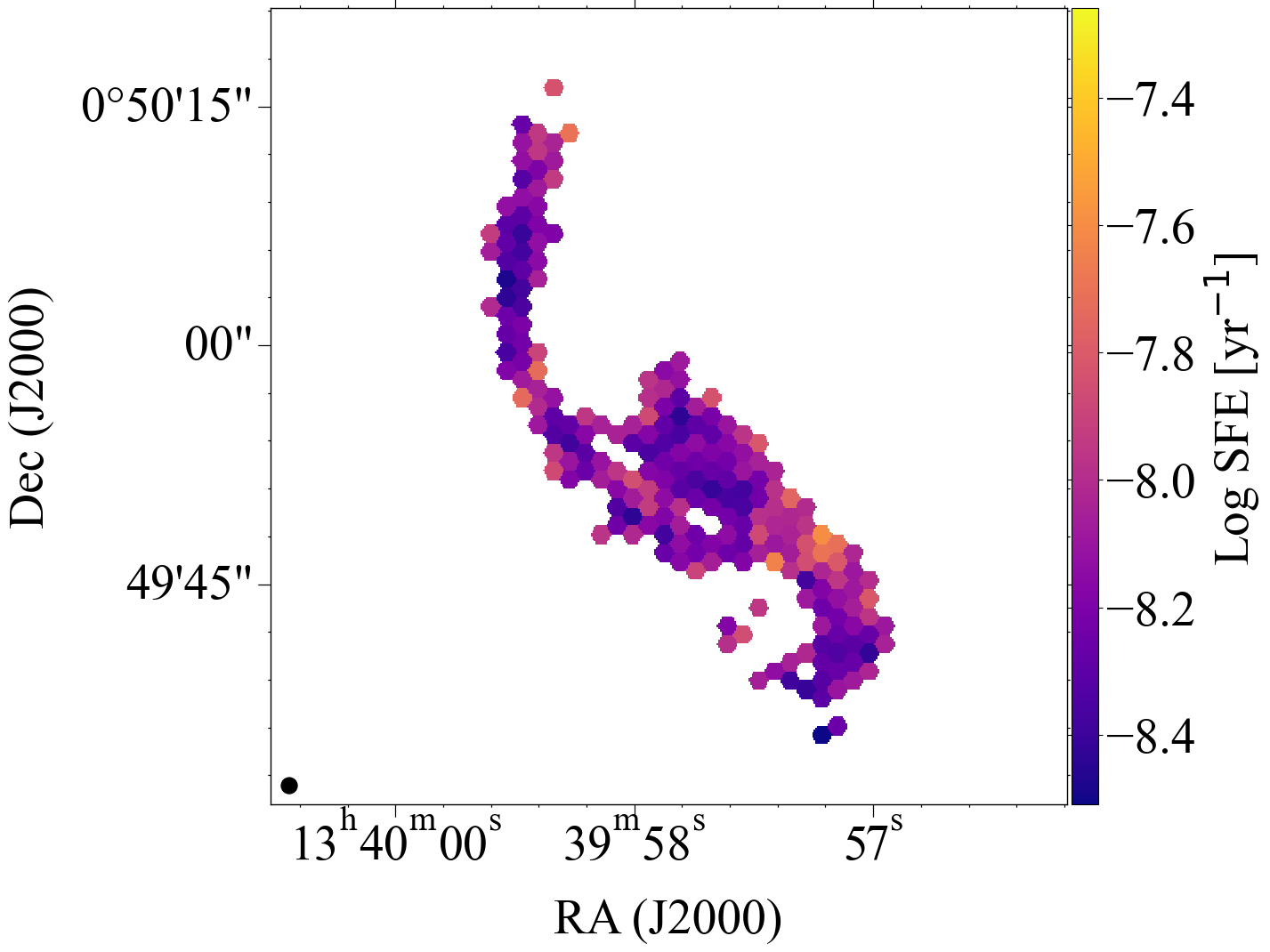} &
        \includegraphics[width=0.45\linewidth]{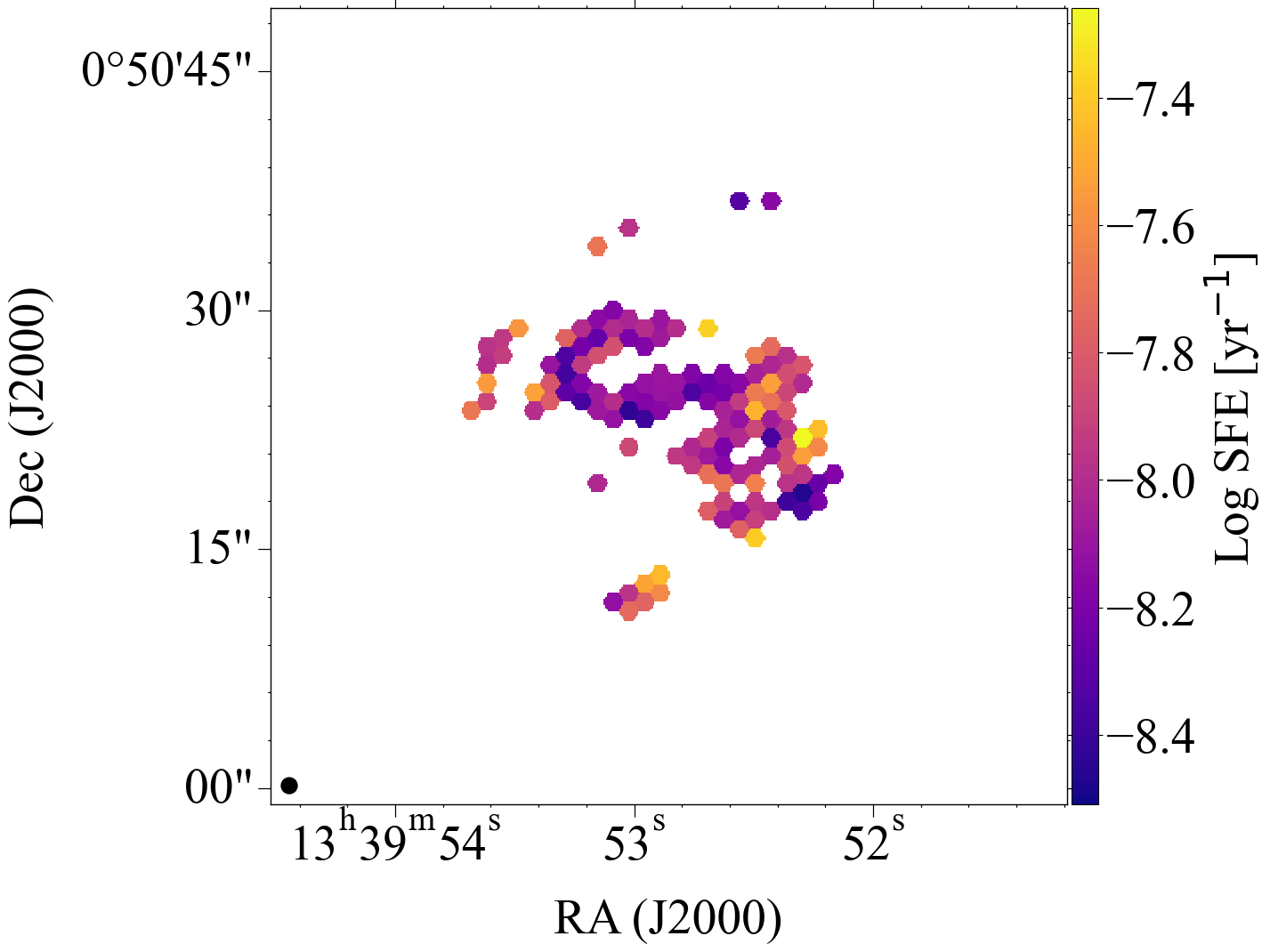} \\
        \textbf{(e)} SFE NGC 5258 & \textbf{(f)} SFE NGC 5257 \\
    \end{tabular}

    \caption{\small From top to bottom: $\Sigma_{\mathrm{SFR}}$, $\Sigma_{\mathrm{H_2}}$, and SFE. We present hexagons where both radio continuum and CO ($2\rightarrow1$) overlap. The black filled circle in the lower-left corner of each panel represents the beam size. The gray dashed ellipses in the top panels enclose hexagons from distinct morphological components of each galaxy.}
    \label{fig:maps}
\end{figure*}

To calculate $\Sigma_{\mathrm{SFR}}$, we divided the SFR of each hexagon by the area of the beam, 0.26 kpc$^2$. The top panels of Figure \ref{fig:maps} show the $\Sigma_{\mathrm{SFR}}$ for regions with detected radio continuum and CO ($2\rightarrow1$) in both galaxies.  In our analysis, we focus on examining both bright and low surface brightness regions within their specific morphological environments, such as spiral arms and nuclear regions, as delineated by the gray dashed ellipses in the top panels of Figure \ref{fig:maps}. This grouping criterion is consistently applied throughout the paper. 

The brightest hexagon in NGC 5258, located in the southern spiral arm, has a $\Sigma_{\mathrm{SFR}}$ of 4.22 $\pm$ 0.23 M$_{\odot}$~yr$^{-1}$~kpc$^{-2}$ (SFR = 1.08 M$_{\odot}$~yr$^{-1}$), while the maximum value for NGC 5257 is found in the nucleus and has a $\Sigma_{\mathrm{SFR}}$ of 3.61 $\pm$ 0.20 M$_{\odot}$~yr$^{-1}$~kpc$^{-2}$ (SFR = 0.92 M$_{\odot}$~yr$^{-1}$). It is worth noting that the nuclear emission in NGC 5257 has been classified as a nuclear starburst, with no evidence of an AGN \citep[e.g.,][]{Stierwalt_2013,Song_2022}. Outside the nucleus, the three brightest hexagons of NGC 5257 are all located in the southwestern spiral arm (see the southernmost group of hexagons in top-right panel of Figure \ref{fig:maps}), and have $\Sigma_{\mathrm{SFR}}$ of 2.66 $\pm$ 0.16, 1.72 $\pm$ 0.13 and 1.61 $\pm$ 0.12 M$_{\odot}$ yr$^{-1}$ kpc$^{-2}$. Meanwhile, the faintest region in both galaxies is determined by the sensitivity of our images and the 3-$\sigma$ threshold, which corresponds to a $\Sigma_{\mathrm{SFR}}$ of 0.27 $\pm$ 0.09 M$_{\odot}$ yr$^{-1}$ kpc$^{-2}$ (SFR = 0.02 M$_{\odot}$~yr$^{-1}$). Note that our choice of the VLA 3 GHz frequency enhances our sensitivity to detect SFR across two orders of magnitude, allowing us to discern diverse environments throughout the galaxies.

\subsection{CO ($2\rightarrow1$) Luminosity and $\Sigma_{H_2}$}\label{h2}
We derived the mass of the molecular gas by using the relation:
\begin{equation}
    M_{H_{2}} = \alpha_{CO} \times L_{CO}^{'}
\end{equation}
\noindent
where $L_{CO}^{'}$ is the luminosity of CO ($1\rightarrow0$) and $\alpha_{CO}$ is the CO-to-H$_{2}$ conversion factor.
To transform our measurements of CO ($2\rightarrow1$) to CO ($1\rightarrow0$), we used the average ratio reported by \cite{He_2019} for Arp 240, CO ($2\rightarrow1$)/($1\rightarrow0$) = 0.85 $\pm$ 0.02. We employed the prescription in \cite{Bolatto_2013} to compute $L_{CO}^{'}$:


\begin{equation}\label{eq_Lco}
\begin{split}
    L_{\scriptscriptstyle \mathrm{CO}(J \rightarrow J-1)}^{'} = & \, 2453 \,  \left( \frac{S_{\mathrm{CO}} \Delta v}{\mathrm{Jy \, km \, s^{-1}}} \right)  \left( \frac{D_L^2}{\mathrm{Mpc^2}} \right) \, J^{-2} \, (1+z)^{-1} \\
    & \times \left( \mathrm{K \, km \, s^{-1} \, pc^2} \right)    
\end{split}
\end{equation}

\noindent
where $S_{CO}\Delta v$ is the integrated flux density, $D_{L}$ is the luminosity distance at redshift $z$ and J is the CO transition.
To calculate $S_{CO}\Delta v$, we directly applied the grid to the data cubes and extracted the line profile for each hexagon. We then summed over all channels, multiplying by the channel width of 3 km~s$^{-1}$.

The conversion factor, $\alpha_{CO}$, is dependent on the properties of the gas, including temperature, density, and metallicity. However, its value is still a topic of debate and may vary for different environments, especially for extreme ones like those in LIRGs \citep{Bolatto_2013}.

To calculate the molecular gas mass, we adopted an $\alpha_{CO}$ value of $1.1$~M$_{\odot}$~(K~km~s$^{-1}$~pc$^{2}$)$^{-1}$, which is typically used for LIRGs and is close to the values reported by \cite{He_2019} for the brightest regions in CO ($2\rightarrow1$) in Arp 240. We will discuss the implications of our selection of $\alpha_{CO}$ in $\S$ \ref{vel_mkS}. Lastly, we calculate $\Sigma_{\mathrm{H_2}}$ by dividing M$_{\mathrm{H_2}}$ by the area of the synthesized beam, expressed in pc$^2$.

Figure \ref{fig:maps} shows maps of the $\Sigma_{\mathrm{H_2}}$ for both galaxies. For NGC 5258, the brightest hexagon in $\Sigma_{\mathrm{H_2}}$ corresponds to the brightest hexagon in $\Sigma_{\mathrm{SFR}}$ in the southern spiral arm, and has a value of (5.67 $\pm$ 0.29) $\times$ 10$^2$ M$_\odot$~pc$^{-2}$. In NGC 5257, the highest concentrations of H$_2$ are located in the nucleus with a peak hexagon of (5.49 $\pm$ 0.28) $\times$ 10$^2$ M$_\odot$ pc$^{-2}$. Outside the nucleus, the three brightest hexagons in $\Sigma_{\mathrm{H_2}}$ are located in the middle of the southwestern spiral arm, contrasting with the three brightest hexagons in $\Sigma_{\mathrm{SFR}}$, which are situated in the southernmost group of hexagons at the tip of the arm. Their values are (1.79 $\pm$ 0.09) $\times$ 10$^2$, (1.74 $\pm$ 0.09) $\times$ 10$^2$ and (1.72 $\pm$ 0.10) $\times$ 10$^2$ M$_\odot$ pc$^{-2}$. We will discuss this mismatch between the extra-nuclear hexagons with the highest $\Sigma_{\mathrm{H_2}}$ and $\Sigma_{\mathrm{SFR}}$ values in $\S$ \ref{discussion}. 

Meanwhile, the hexagons with the smallest values are (1.98 $\pm$ 0.75) $\times$ 10$^1$ and (0.96 $\pm$ 0.26) $\times$ 10$^1$ M$_\odot$ pc$^{-2}$, for NGC 5258 and NGC 5257 respectively.  

\subsection{Star formation Efficiency and Gas Depletion Times}\label{section_SFE}
The ratio between $\Sigma_{\mathrm{SFR}}$ and $\Sigma_{\mathrm{H_2}}$ is defined as the star formation efficiency, SFE = $\Sigma_{\mathrm{SFR}}$/$\Sigma_{\mathrm{H_2}}$, with units of inverse time \citep[e.g.,][]{Leroy_2008}. The SFE gives us a sense of how quickly the galaxy will exhaust its cold gas reservoir if it continues to produce stars at the present SFR. The characteristic timescale of this process is referred to as the depletion time, $t_{d}$ = 1/SFE. 

In the bottom panels of Figure \ref{fig:maps}, we present maps of the SFE for both galaxies. The distribution of SFE is heterogeneous across each galaxy. In NGC 5258, 67$\%$ of hexagons (43 out of 64) in the southern spiral arm exceed the galaxy's median\footnote{Throughout the paper, we report the median value for each quantity, accompanied by the median of all calculated errors.} SFE of (7.24 $\pm$ 0.22) $\times$ 10$^{-9}$ yr$^{-1}$ (t$_d$ $\sim$ 0.14 Gyr). In contrast, in the bulge and northern spiral arm, only 43$\%$ (64 out of 150) exceed this median. The hexagons connecting the southern spiral arm to the galactic bulge exhibit the highest SFE in NGC 5258, reaching (2.51 $\pm$ 0.39) $\times$ 10$^{-8}$ yr$^{-1}$ (t$_d$ $\sim$ 0.04 Gyr).

The median SFE for NGC 5257 is (1.00 $\pm$ 0.26) $\times$ 10$^{-8}$ yr$^{-1}$ (t$_d$ $\sim$ 0.10 Gyr) . Notably, no hexagon in the center exceeds this value. Roughly half of the hexagons in the northeastern spiral arm, 48$\%$ (23 out of 44), are above the median SFE; while the southwestern spiral arm has a slightly higher fraction with 56$\%$ (42 of 73). The ten highest SFE values, peaking at (5.46 $\pm$ 0.14) $\times$ 10$^{-8}$ yr$^{-1}$ (t$_d$ $\sim$ 0.02 Gyr or 20 Myr), are located in this arm, with three of these within the southernmost clump.
It is worth noting that the median t$_d$ of 0.12~Gyr in Arp~240 is significantly shorter than the 1.2~Gyr observed in central regions of spiral galaxies in the PHANGS survey \citep{Querejeta_2021}, highlighting the extreme ISM conditions in Arp 240, even beyond its nuclear regions.

\begin{figure}
    \centering
    \includegraphics[width=1\columnwidth]{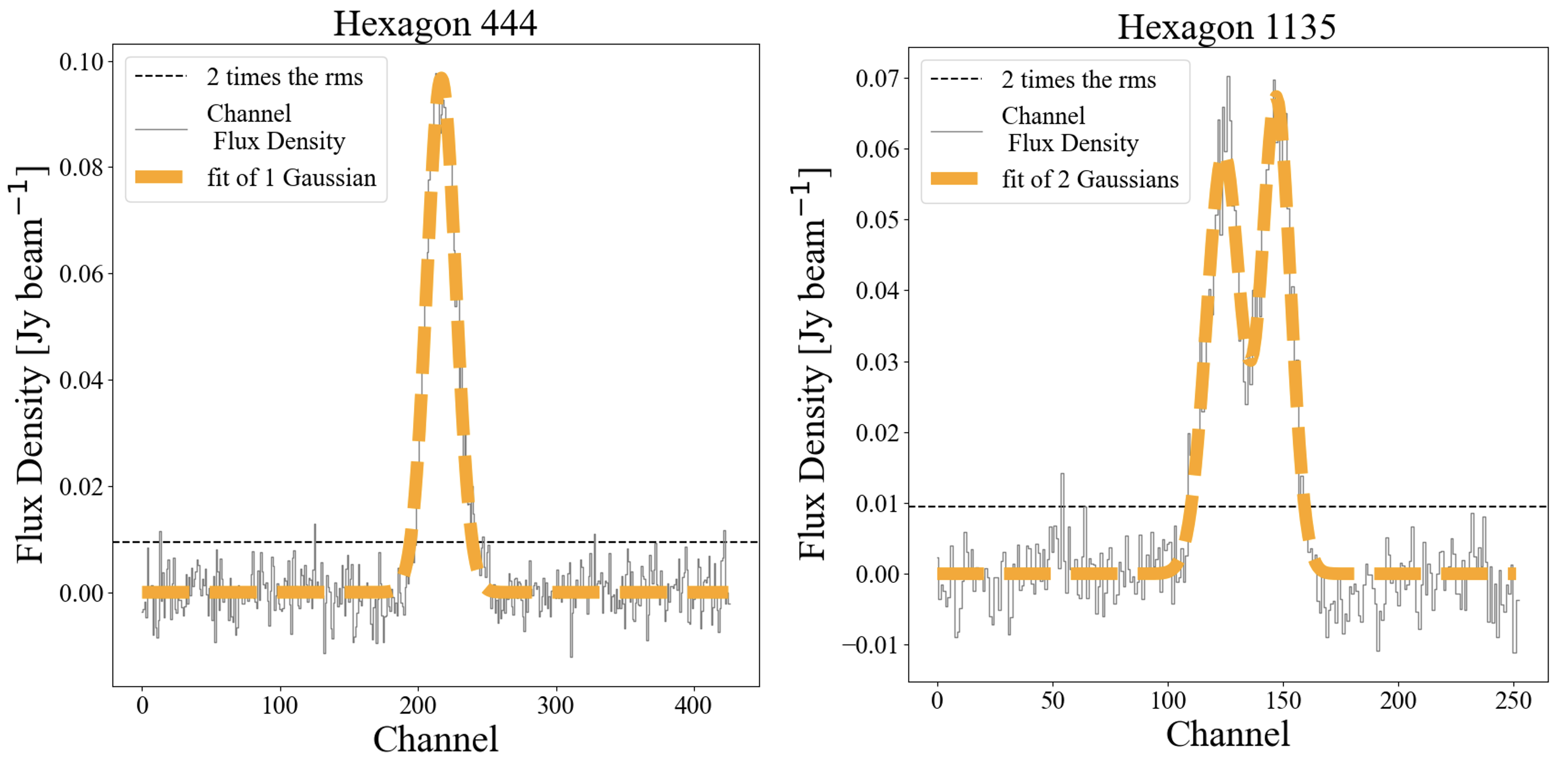}
    \caption{\small Example of linewidth fitting within individual hexagons, showing single (left) and double (right) Gaussian components. Gray lines represent the flux density per channel within hexagons. Yellow lines indicate the Gaussian fits. For hexagons with multiple components, the average sigma from the fits is used to represent the linewidth of the hexagon.} 
    \label{fig:lnwith-fit}
\end{figure}



\begin{figure*}
    \centering

    \begin{tabular}{ccc}
        \includegraphics[width=0.3\textwidth]{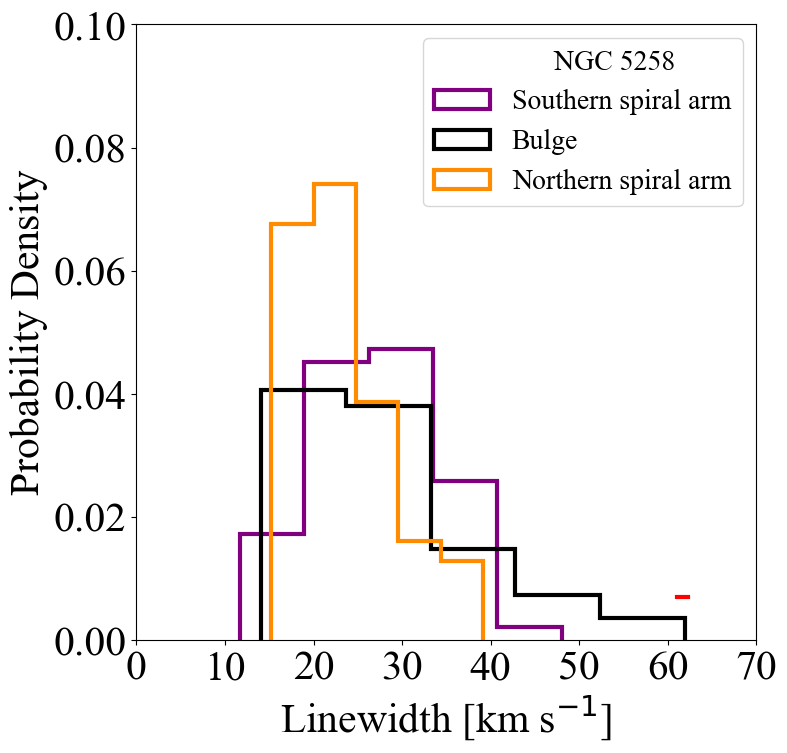} &
        \includegraphics[width=0.3\textwidth]{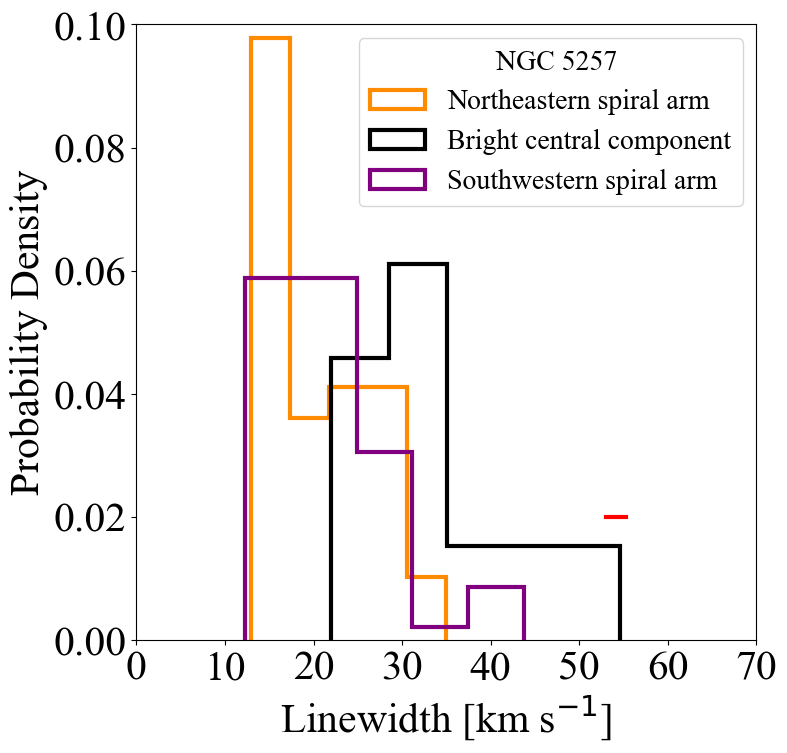} &
        \includegraphics[width=0.3\textwidth]{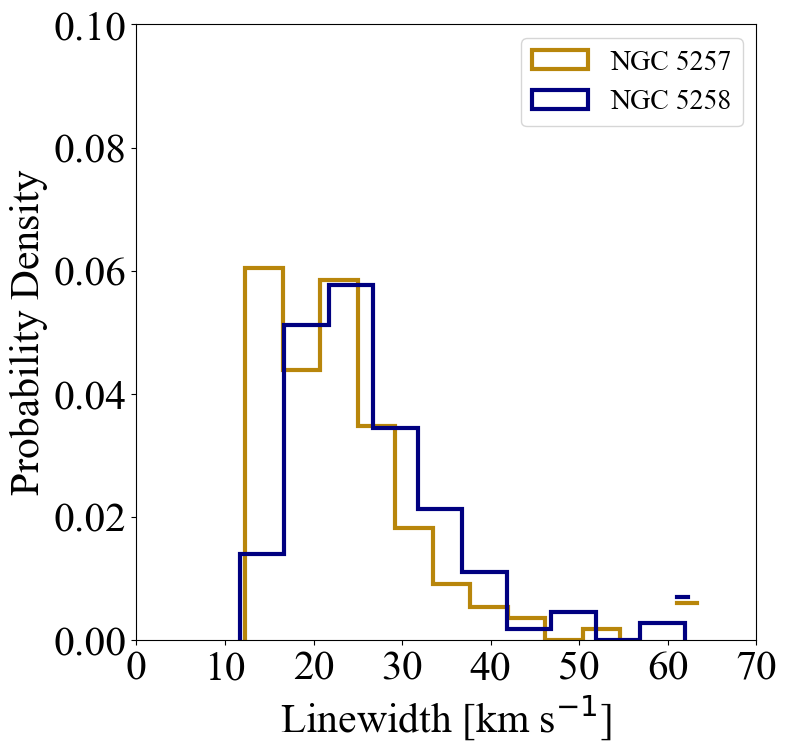} \\
    \end{tabular}
    \\[2ex]

    \begin{tabular}{cc}
        \includegraphics[width=0.45\textwidth]{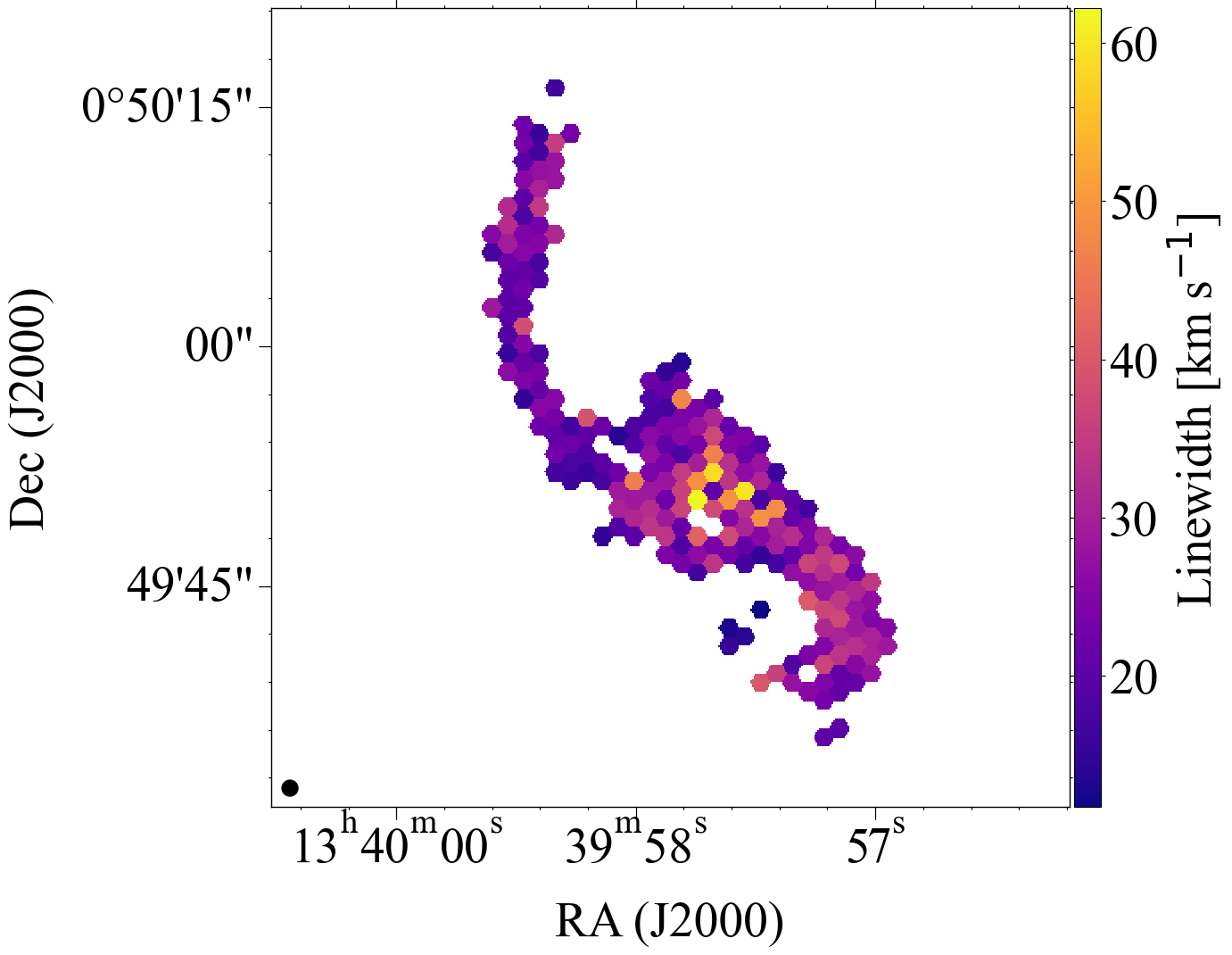} &
        \includegraphics[width=0.45\textwidth]{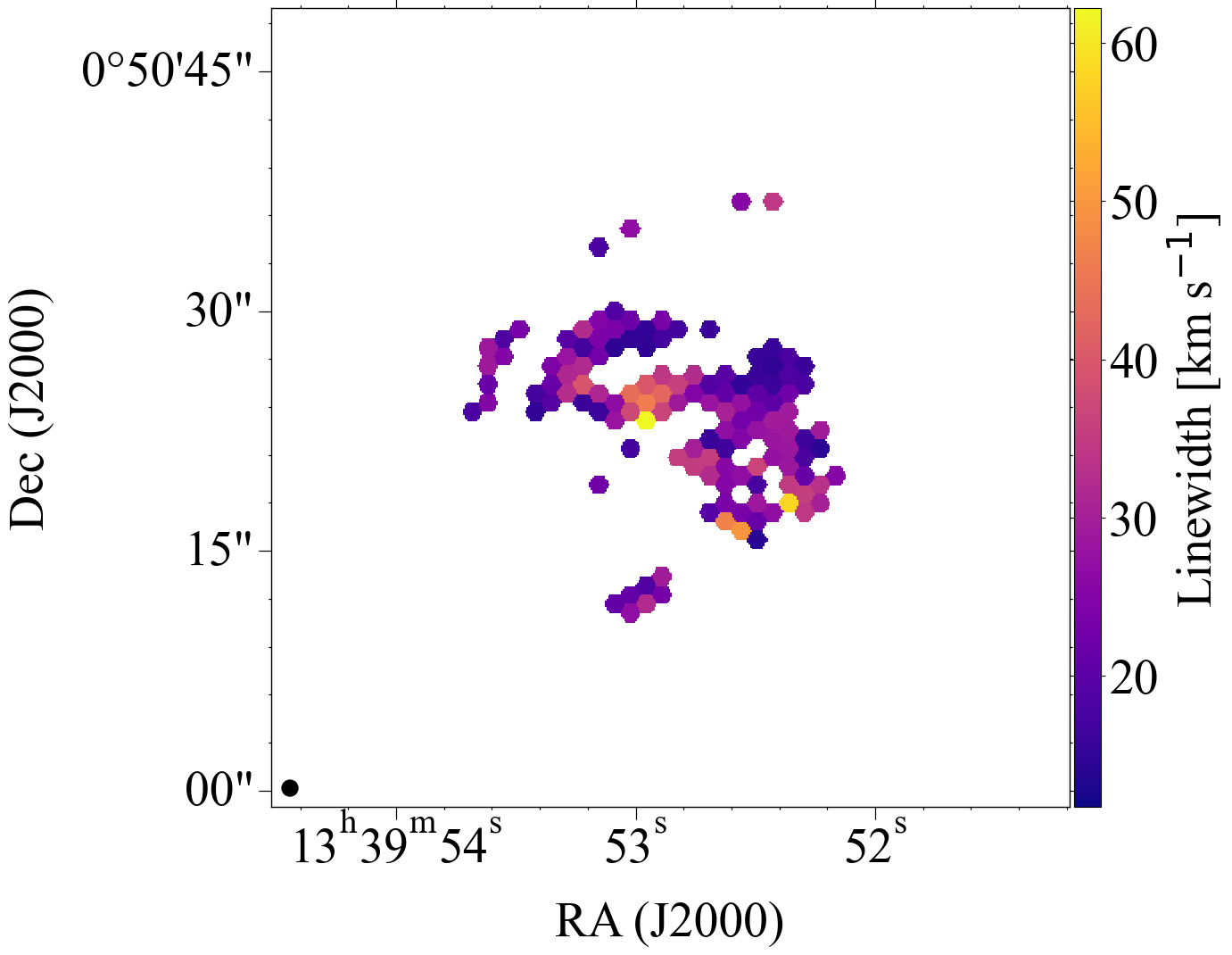} \\
    \end{tabular}

    \caption{\small
        Top panels: Probability density distribution of linewidths. We estimated the CO ($2\rightarrow1$) linewidth by fitting single or multiple Gaussians to the line profile of each hexagon. The left and middle panels show linewidths for different morphological components in NGC 5258 and NGC 5257, respectively. The right panel shows linewidths for both galaxies. The minimum value was 11.68 $\pm$ 0.40 km~s$^{\mathrm{-1}}$ for NGC 5258 and 12.30 $\pm$ 0.99 km~s$^{\mathrm{-1}}$ for NGC 5257. The median linewidths are 24.72 $\pm$ 0.64 km~s$^{\mathrm{-1}}$ for NGC 5258 and 22.06 $\pm$ 1.13 km~s$^{\mathrm{-1}}$ for NGC 5257. The red bar in the lower-right corner of the left and middle panels indicates the median error. Likewise, in the right panels, the blue and gold bars represent the median errors for each galaxy. Bottom panels: Synthetic maps displaying the linewidths within NGC 5258 (left) and NGC 5257 (right). The black solid circle in the left-lower corner of the bottom panels represents the synthesized beam.
    }
    \label{fig:linewidth}
\end{figure*}

\subsection{Molecular Gas Kinematics}\label{dyn}
\subsubsection{Linewidth}\label{sec_linewidth}
We model gas motion by measuring the linewidth of velocity components in each hexagon. Given the spatial resolution of our images, multiple components were observed within some hexagons, as illustrated in Figure \ref{fig:lnwith-fit}. At 500 pc scales, rotation and the blending of individual velocity profiles (e.g., from giant molecules clouds (GMC) or associations of GMCs) are expected to influence the observed linewidths. The fractions of multiplicity are: 35$\%$ of hexagons (74 out of 214) for NGC 5258 and 24$\%$ of hexagons (31 out of 127) for NGC 5257 showing multiple components.

We identified the number of peaks within each hexagon and modeled them by fitting multiple Gaussian profiles to the spectral-line velocity components. The linewidth of each component was determined by fitting these profiles using $\tt{scipy.optimize.leastsq}$ in Python \citep{SciPy}. We characterize the linewidth for each hexagon using the sigma value from the Gaussian fit. For hexagons with multiple components, as shown in the right panel of Figure \ref{fig:lnwith-fit}, we use the average of these sigma values to represent the linewidth.

In the top panels of Figure \ref{fig:linewidth}, we present the probability density distribution of the best-fit linewidths. We follow the same scheme as introduced in $\S$ \ref{sfr}, where we group results from areas of both bright and low surface brightness emission according to their corresponding morphology. The left and middle panels display measurements for different morphological components, while the right panel presents data from all regions of each galaxy. The minimum value is 11.68 $\pm$ 0.40 km~s$^{\mathrm{-1}}$ for NGC 5258 and 12.30 $\pm$ 0.99 km~s$^{\mathrm{-1}}$ for NGC 5257; in both cases, these values are observed in the spiral arms away from the brightest SF complexes, as shown in the bottom panels of the Figure \ref{fig:linewidth}. The central regions of both galaxies exhibit the broadest linewidths, with peaks at 61.91 $\pm$ 1.53 km~s$^{\mathrm{-1}}$ for NGC 5258 and 54.66 $\pm$ 1.45 km~s$^{\mathrm{-1}}$ for NGC 5257. These peaks are indicated by yellow hexagons in their respective maps, as shown in the bottom panel of Figure \ref{fig:linewidth}. We visually inspected the profile and the fit of each of these peaks. We found no evidence of multiple components at the spatial and spectral resolution of our observations; instead, each peak was fitted with a single Gaussian component. However, due to the resolution constraints of our data, some degree of beam smearing may still be present.

The top-right panel of Figure \ref{fig:linewidth} presents a comparison of the linewidth distributions for the two galaxies, indicating similar behavior. The median linewidths are 24.72 $\pm$ 0.64 km~s$^{\mathrm{-1}}$ for NGC 5258 and 22.06 $\pm$ 1.13 km~s$^{\mathrm{-1}}$ for NGC 5257. We further explore the implications of these linewidth results and their relationship to $\Sigma_{\mathrm{SFR}}$ and $\Sigma_{\mathrm{H_2}}$ in $\S$ \ref{SFR_n_kinematics} and \ref{vel_mkS}.

\begin{figure*}
    \centering

    \begin{tabular}{cc}
        \includegraphics[width=0.45\textwidth]{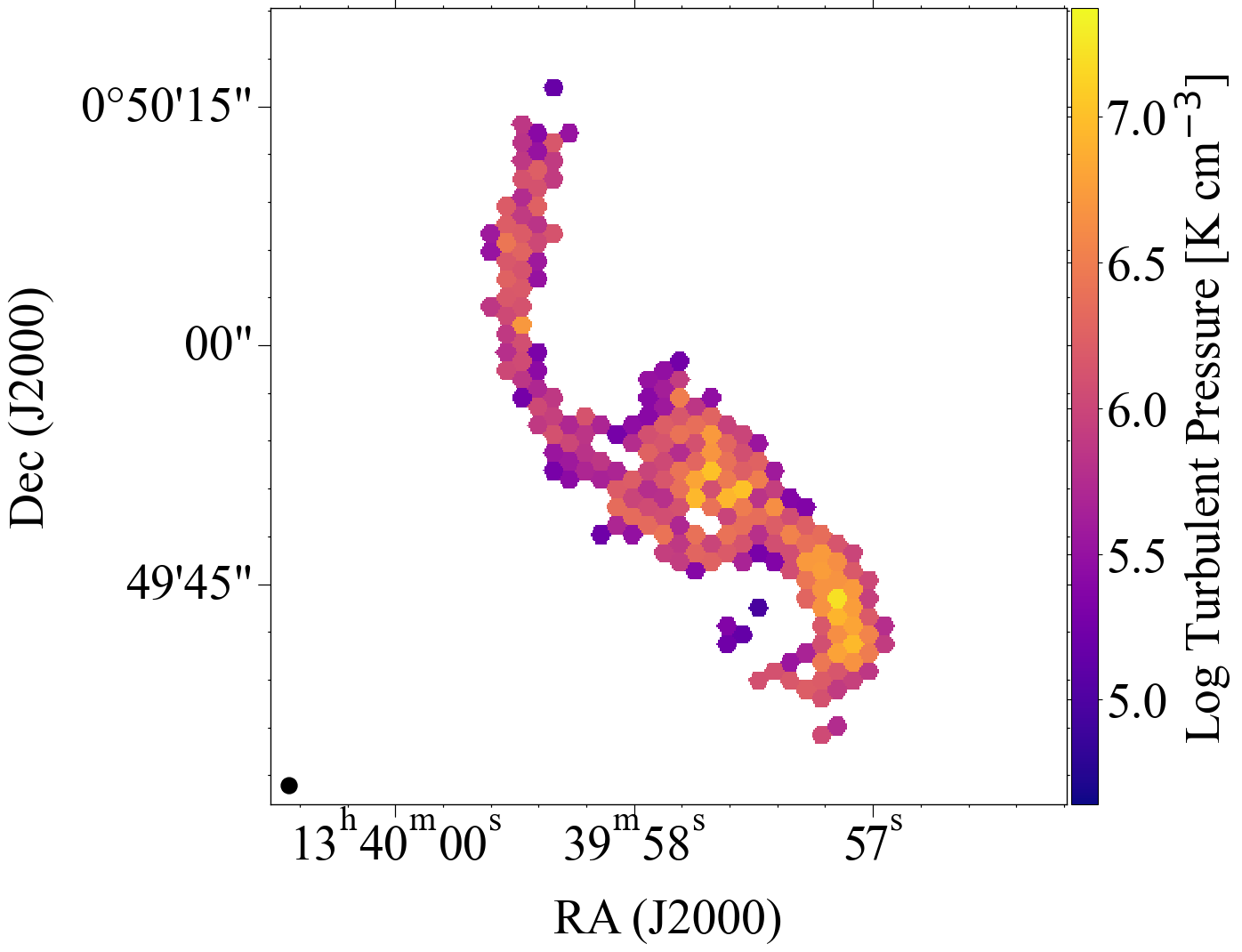} &
        \includegraphics[width=0.45\textwidth]{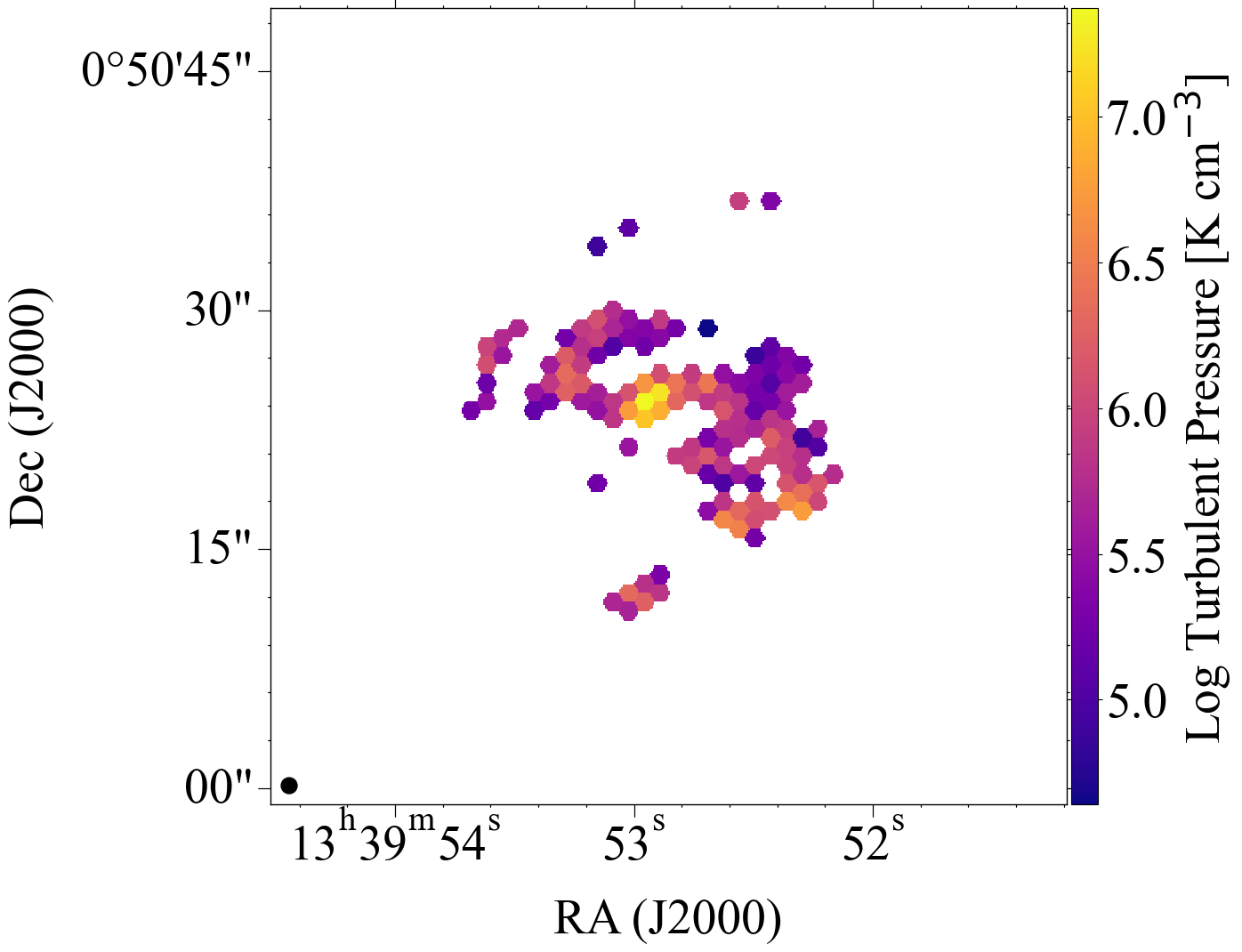} \\
        \textbf{(a)} P$_{\mathrm{turb}}$ NGC 5258 & \textbf{(b)} P$_{\mathrm{turb}}$ NGC 5257 \\
    \end{tabular}
    \\[2ex]

    \begin{tabular}{cc}
        \includegraphics[width=0.45\textwidth]{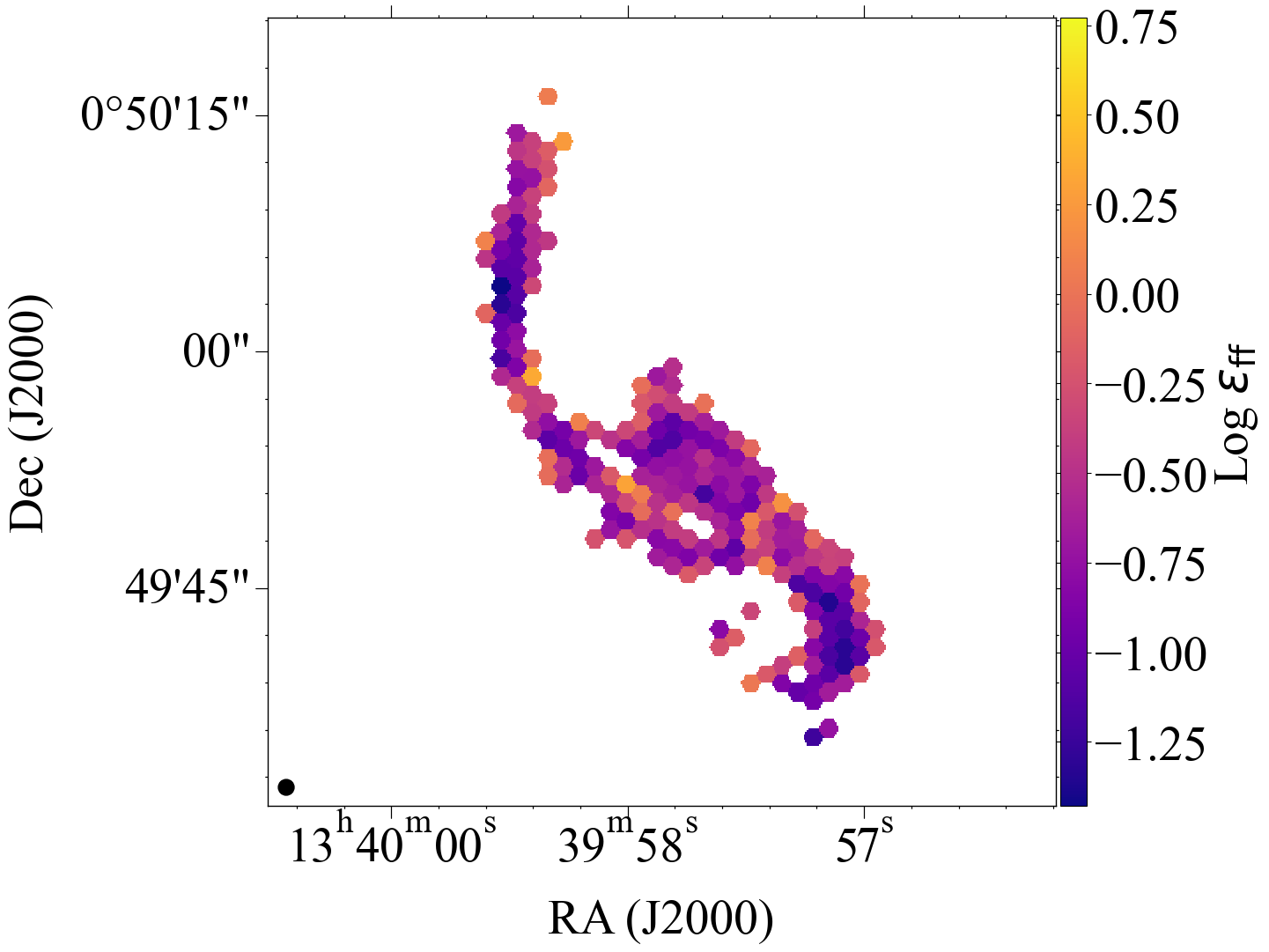} &
        \includegraphics[width=0.45\textwidth]{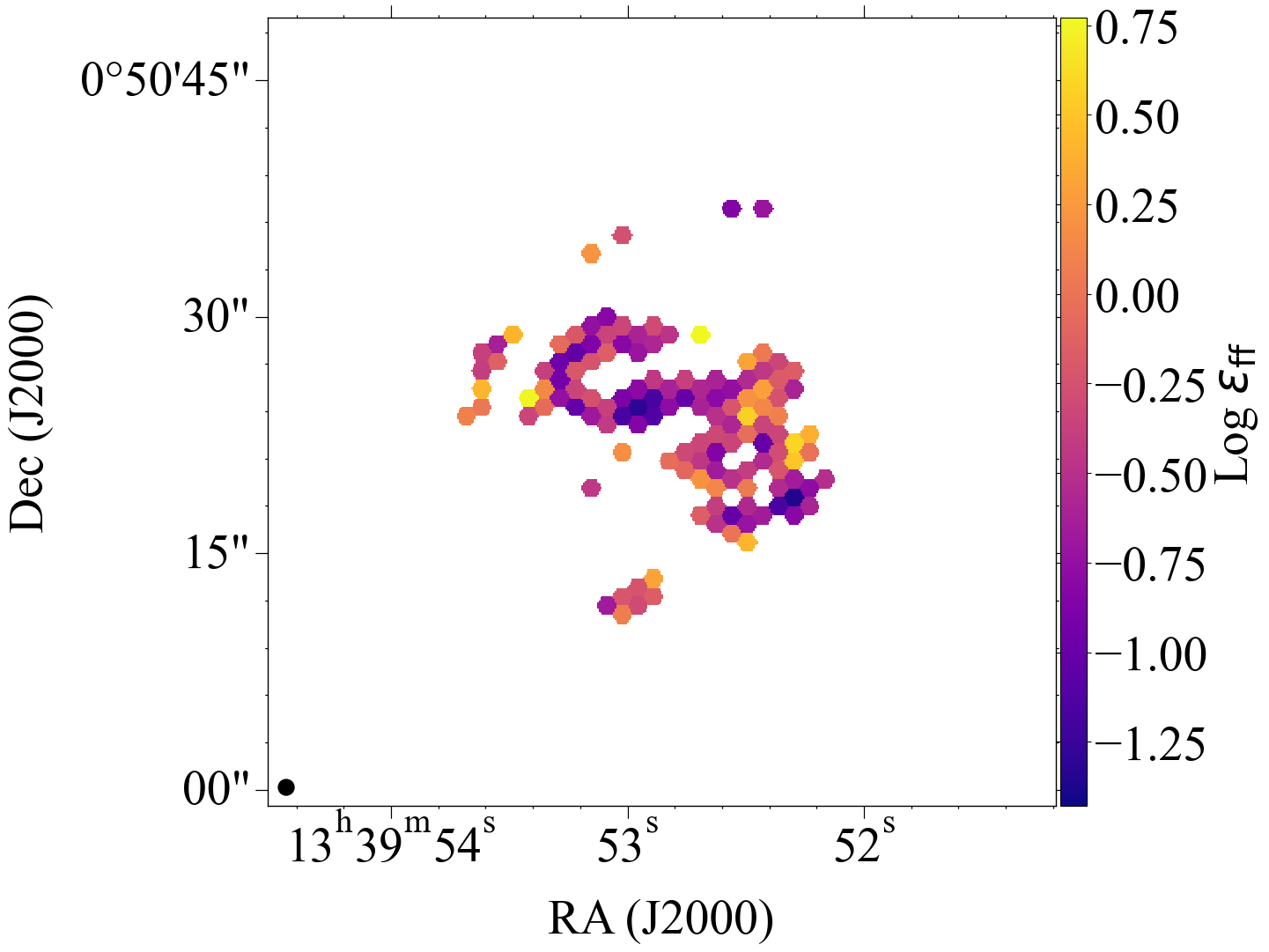} \\
        \textbf{(c)} $\epsilon_{\mathrm{ff}}$ NGC 5258 & \textbf{(d)} $\epsilon_{\mathrm{ff}}$ NGC 5257 \\
    \end{tabular}
    \\[2ex]

    \begin{tabular}{cc}
        \includegraphics[width=0.38\textwidth]{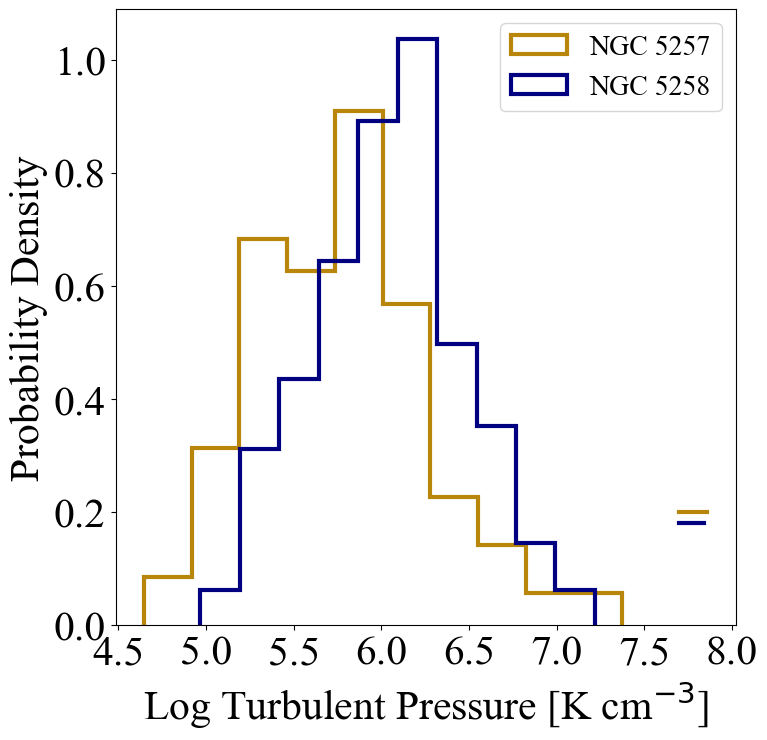} &
        \includegraphics[width=0.38\textwidth]{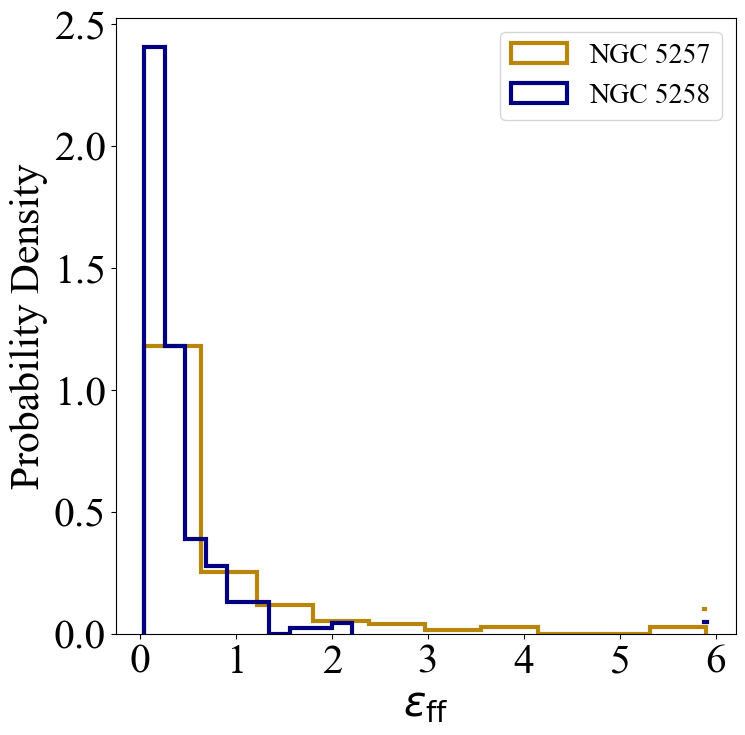} \\
        \textbf{(e)} P$_{\mathrm{turb}}$ Probability Density Distribution &
        \textbf{(f)} $\epsilon_{\mathrm{ff}}$ Probability Density Distribution \\
    \end{tabular}

    \caption{\small
        Top panels (a-b): Maps of the Log P$_{\mathrm{turb}}$. Middle panels (c-d): Maps of Log $\epsilon_{\mathrm{ff}}$. The black solid circle in the left-lower corner of each map represents the synthesized beam. Bottom panels: Probability density distribution of P$_{\mathrm{turb}}$ (e) and $\epsilon_{\mathrm{ff}}$ (f) for both galaxies. The blue and golden bars in the lower-right corner represent the median error.
    }
    \label{fig:figure_P_E}
\end{figure*}

\subsubsection{Turbulent Pressure}\label{sec_P_turb}
Assuming the linewidth is turbulently broadened, we utilized the methodology outlined by \cite{Sun2018} to calculate\footnote{It is important to note that this calculation may be affected by residual beam smearing effects, and caution should be exercised when interpreting the results.} the mean turbulent pressure of the ISM within our hexagonal apertures, represented as P$_{\mathrm{turb}}$: 

\begin{equation}
    P_{\mathrm{turb}} \approx \rho \langle \sigma \rangle ^{2}
\end{equation}

\noindent
where $\langle \sigma \rangle$ is the average linewidth and $\rho$ is the gas volume density inside the region. This volume density can be approximated as $\rho$ $\approx$ $\frac{1}{2H}$ $\Sigma_{\mathrm{H_2}}$, with H being the scale height in the line of sight. Thus

\begin{equation}
    P_{\mathrm{turb}} \approx \frac{1}{2H} \Sigma_{\mathrm{H_2}} \langle \sigma \rangle^{2}
\end{equation}

\noindent
This expression provides a measurement of the kinetic energy density of the molecular gas; which is typically expressed in terms of the Boltzmann constant P$_{\mathrm{turb}}$/ $\kappa_{B}$ in units of K~cm$^{-3}$ \citep[e.g.,][]{Sun2018}.

The ISM turbulent pressure is given by:

\begin{equation}\label{eq_P_turb}
    \frac{P_{\mathrm{turb}}}{\kappa_{B}} = 2451 ~\mathrm{K} ~\mathrm{cm}^{-3} \left(\frac{\Sigma_{H_2}}{\mathrm{M}_{\odot} \mathrm{pc}^{-2}} \right) \left(\frac{\langle \sigma \rangle}{\mathrm{km}~\mathrm{s}^{-1}} \right)^{2} \left(\frac{H}{\mathrm{pc}} \right)^{-1}
\end{equation}


\noindent
Note that Equation \ref{eq_P_turb} requires knowledge of the scale height, which in turn necessitates modeling the density distribution of the molecular gas. We do not attempt to undertake this modeling; instead, we simplify our calculations by assuming a beam filling factor\footnote{The beam filling factor is the ratio of the solid angle subtended by the source to the solid angle of the synthesized beam.} of 1 and adopting a constant scale height of 100~pc throughout our analysis \citep[e.g.,][]{Leroy_2008, Wilson_2019}.

In the top panels (a-b) of Figure \ref{fig:figure_P_E}, we present maps of P$_{\mathrm{turb}}$ across each galaxy. Their probability density distribution is shown in panel (e) of the same Figure. NGC 5258 has a median value of (1.20 $\pm$ 0.20) $\times$ 10$^6$ K~cm$^{-3}$; while the median for NGC 5257 is (6.11 $\pm$ 0.11) $\times$ 10$^5$ K~cm$^{-3}$. We found that the maximum values of P$_{\mathrm{turb}}$, (1.65 $\pm$ 0.10) $\times$ 10$^7$ K~cm$^{-3}$ for NGC 5258 and  (2.34 $\pm$ 0.12) $\times$ 10$^7$ K~cm$^{-3}$ for NGC 5257 correspond to the highest value of $\Sigma_{\mathrm{SFR}}$ for both galaxies. In $\S$ \ref{P_SFE}, we will discuss these results in relation to the distribution of SFR.

\subsubsection{Free-Fall Time and SFE per Free-Fall Time}\label{e_ff}

In addition to the depletion time presented in $\S$ \ref{section_SFE}, another relevant timescale to quantify is the free-fall time (t$_{\mathrm{ff}}$), as it measures the time required for a cloud of gas to collapse under its own gravity in the absence of any opposing forces \citep[e.g.,][]{Krumholz_2005}. The general solution to the equations of motion for this idealized setup is: 

\begin{equation}\label{eq_t_ff_original}
    t_{\mathrm{ff}} = \sqrt{\frac{3~\pi}{32~G~\rho}}
\end{equation}

\noindent
where $\rho$ represents the volume density of the molecular gas. Following \cite{Wilson_2019}, we computed t$_{\mathrm{ff}}$ as follows:

\begin{equation}\label{t_ff}
    t_{\mathrm{ff}} = \frac{\sqrt{3}}{4G}\frac{\langle \sigma \rangle}{\Sigma_{\mathrm{H_2}}}
\end{equation}

We now use these results to quantify the SFE of each region over a period of one free-fall time. The SFE per free-fall time, denoted as $\epsilon_{\mathrm{ff}}$, is defined as:
\begin{equation}\label{e_ff}
    \epsilon_{\mathrm{ff}} = t_{\mathrm{ff}} ~SFE = t_{\mathrm{ff}} ~\frac{\Sigma_{\mathrm{SFR}}}{\Sigma_{\mathrm{H_2}}} = \frac{\sqrt{3}}{4G}~\frac{\langle \sigma \rangle ~\Sigma_{\mathrm{SFR}}}{\Sigma_{\mathrm{H_2}}^2} 
\end{equation}

In the middle panels (c-d) of Figure \ref{fig:figure_P_E}, we present maps of $\epsilon_{\mathrm{ff}}$, along with their probability density distribution shown in panel (f) of the same Figure. The median value of $\epsilon_{\mathrm{ff}}$ for NGC 5258 is 25\%, whereas NGC 5257 exhibits a higher median of 42\%. The highest values, exceeding 100\%, are predominantly located in the periphery of the bright star-forming complexes. Notably, \cite{He_2019} also reported exceptionally high $\epsilon_{\mathrm{ff}}$ values, exceeding 100\%, in some extranuclear regions of NGC 5257. In \S4, we will discuss this findings and offer an interpretation of these elevated $\epsilon_{\mathrm{ff}}$ values.

\begin{figure}
    \centering
    \includegraphics[width=0.9\columnwidth]{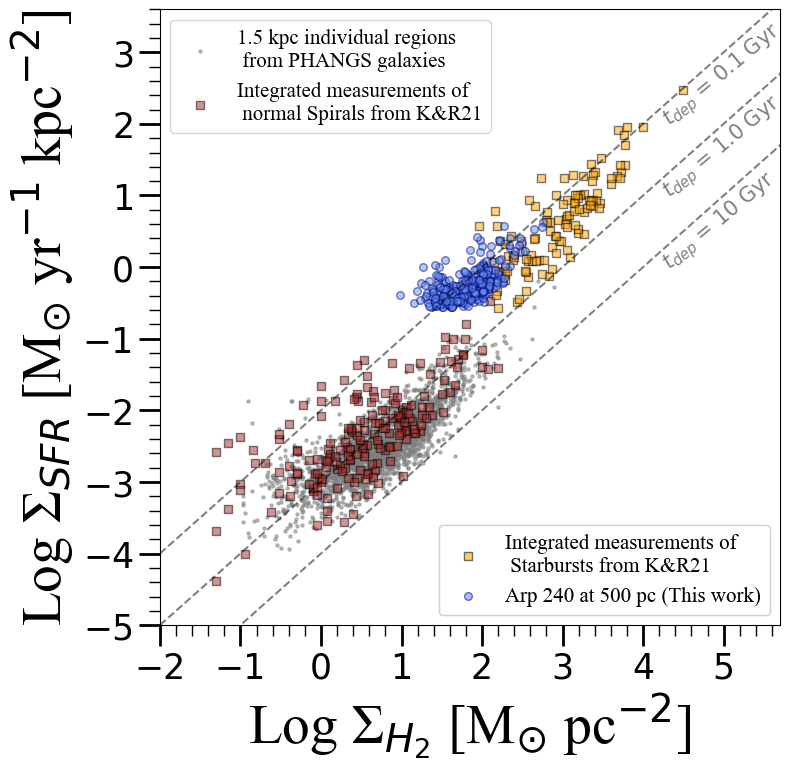}
    \caption{\small The mK-S law: Comparison between regions in Arp 240 and measurements from the literature at different scales. Blue circles represent regions within Arp 240 at a resolution of 500 pc. Brown and yellow squares denote integrated measurements of normal spirals and starbursts, respectively, from \citet{Kennicutt2021} (K$\&$R21). Gray dots correspond to individual 1.5-kpc-sized regions from normal spiral galaxies in the PHANGS sample, detailed in \citet{Sun_2023}. Error bars are omitted for clarity.}
    \label{fig:K-S_all}
\end{figure}

\begin{figure*}
    \centering
    \begin{tabular}{ccc}
        \includegraphics[width=0.32\textwidth]{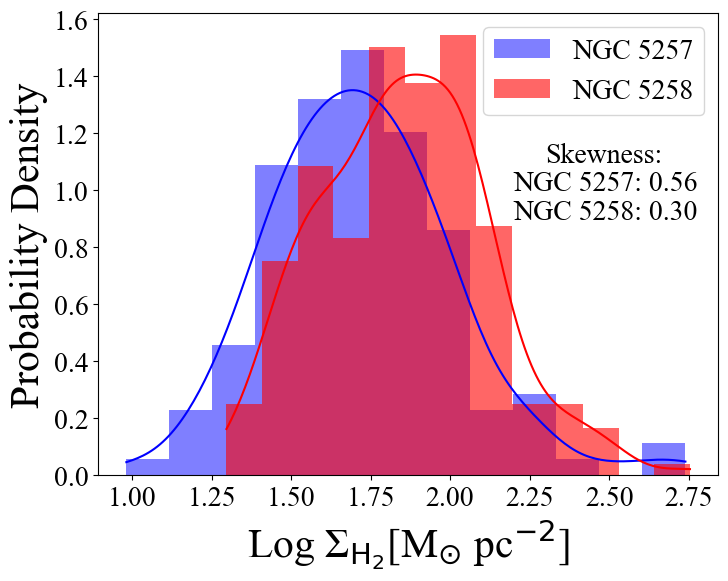} &
        \includegraphics[width=0.32\textwidth]{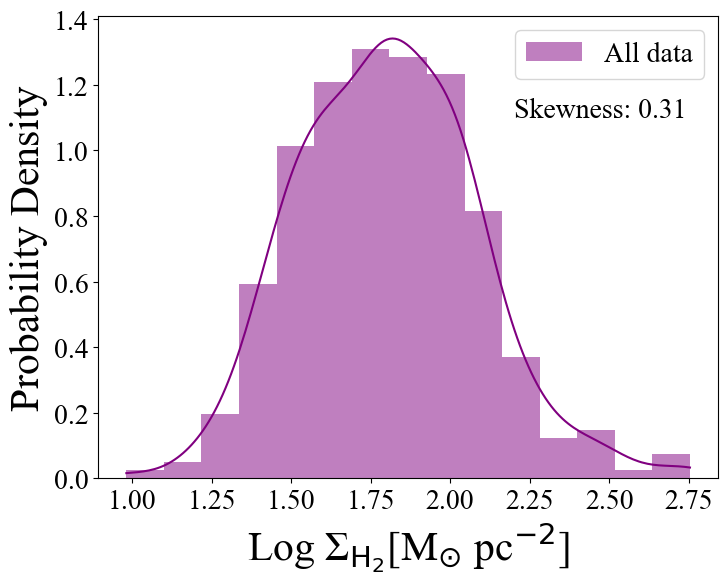} &
        \includegraphics[width=0.32\textwidth]{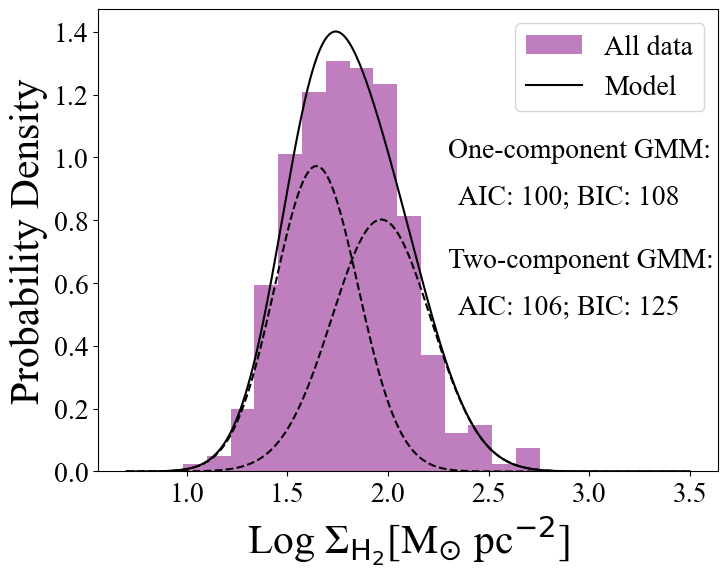} \\
    \end{tabular}
    \\[2ex]

    \begin{tabular}{ccc}
        \includegraphics[width=0.32\textwidth]{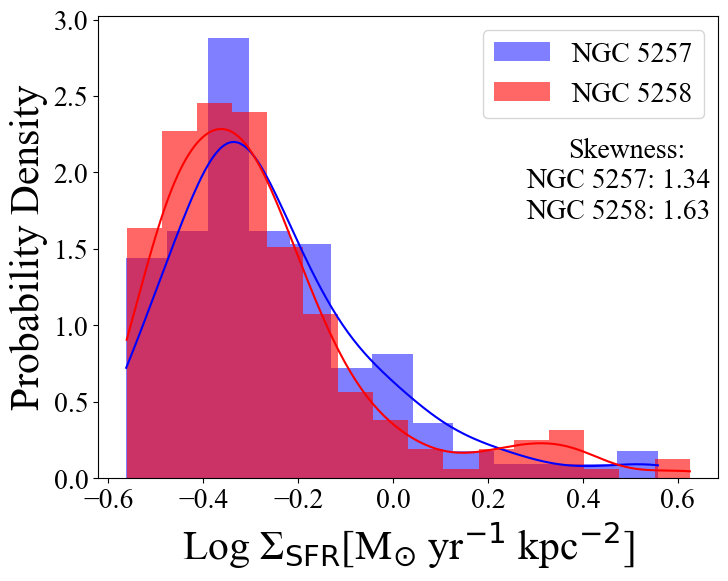} &
        \includegraphics[width=0.32\textwidth]{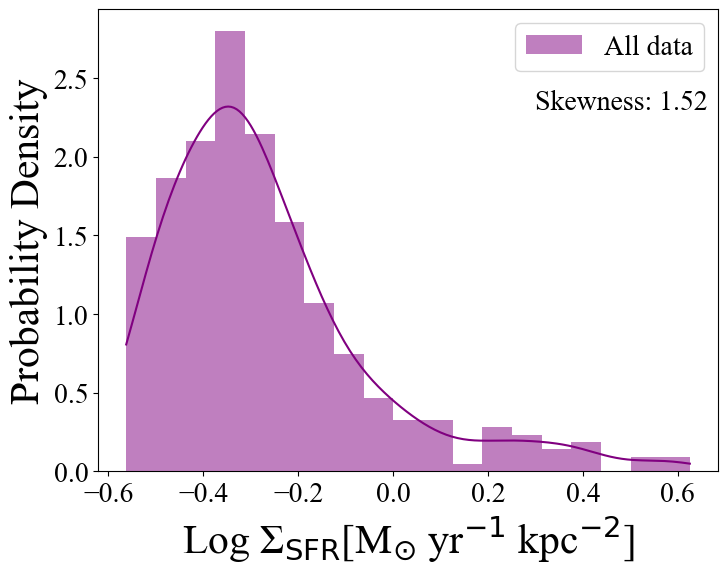} &
        \includegraphics[width=0.32\textwidth]{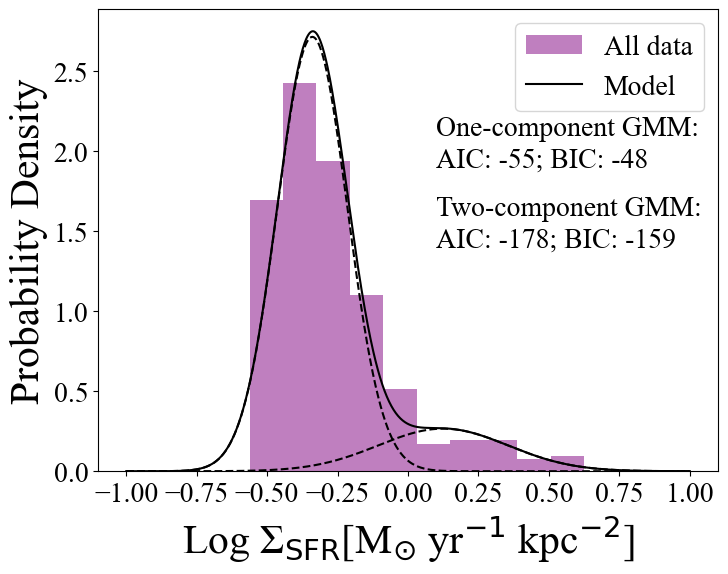} \\
    \end{tabular}

    \caption{\small Probability Density Distributions of Log~$\Sigma_{\mathrm{H_2}}$ (upper panels) and Log~$\Sigma_{\mathrm{SFR}}$ (lower panels). The left panels display the skewness of the distributions for each galaxy, while the middle panels show the skewness for the entire system. The right panels display the AIC and BIC indices for models with one and two Gaussian components.}
    \label{fig:skew_hist}
\end{figure*}

\subsection{The Molecular Kennicutt-Schmidt Law}\label{section_KS}
\subsubsection{Slope Determination}
In Figure \ref{fig:K-S_all}, we show the mK-S law, expressed as Log~($\Sigma_{\mathrm{SFR}}$)~$\propto$~N~Log($\Sigma_{\mathrm{H_2}}$), applied to our study of 500-pc-sized regions in Arp 240. To contextualize our measurements, we juxtapose them with observations from various scales reported in the literature. We incorporate integrated measurements of both normal spirals and starbursts from \citet{Kennicutt2021}\footnote{This dataset includes Kennicutt's original data from \citep{Kennicutt98}.}(brown and yellow squares). Additionally, we include 1.5-kpc-sized regions from a sample of 80 normal spiral galaxies from PHANGS \citep{Sun_2023}, represented by gray dots. In each of the aforementioned studies, the slope of the mK-S law generally converges on a value of N $\approx$ 1.

\begin{figure*}
    \centering

    \begin{tabular}{cc}
        \includegraphics[width=0.42\textwidth]{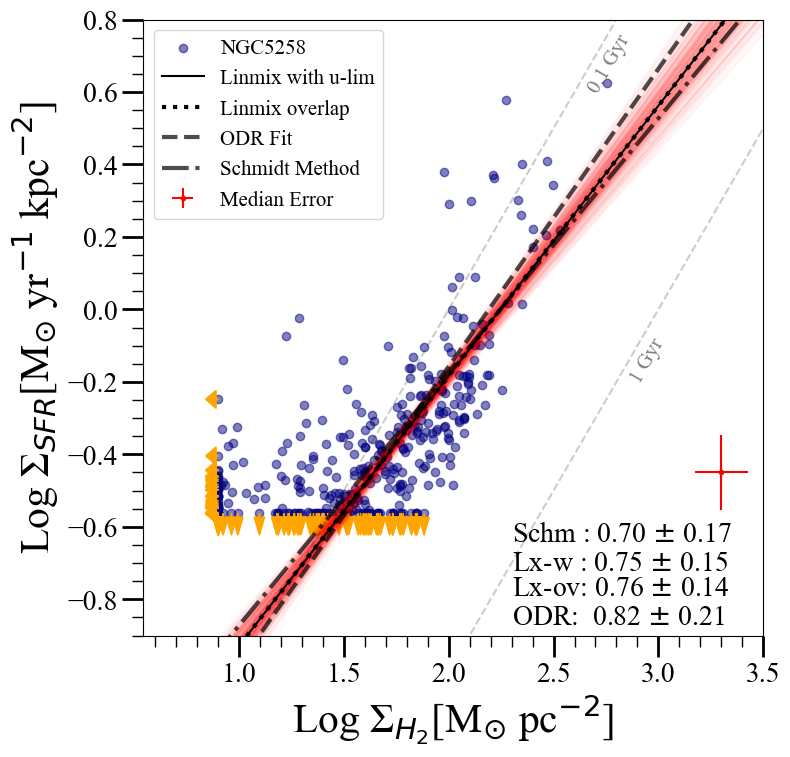} &
        \includegraphics[width=0.42\textwidth]{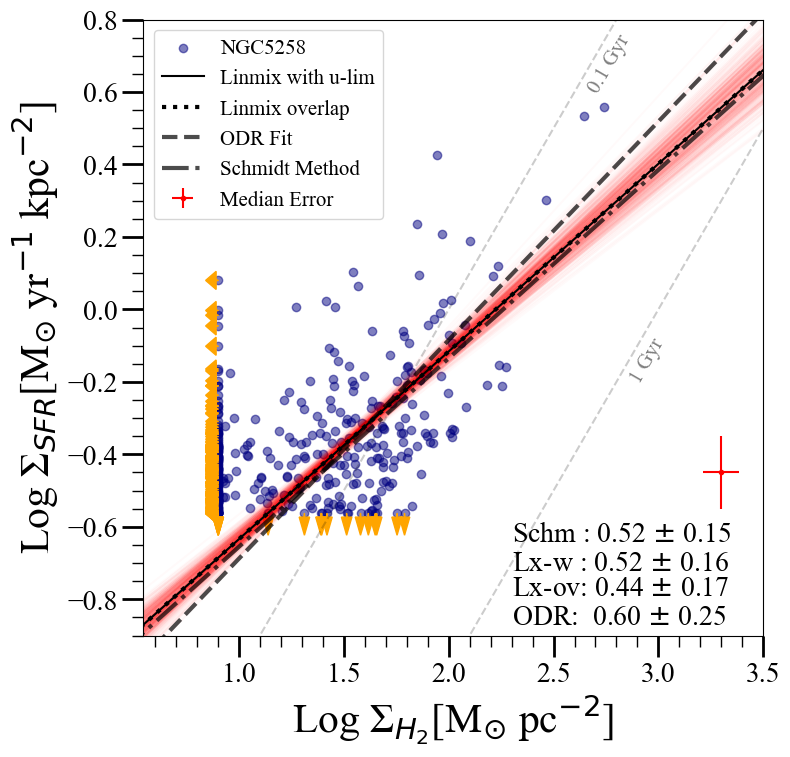} \\
    \end{tabular}
    \\[2ex]

    \begin{tabular}{c}
        \includegraphics[width=0.42\textwidth]{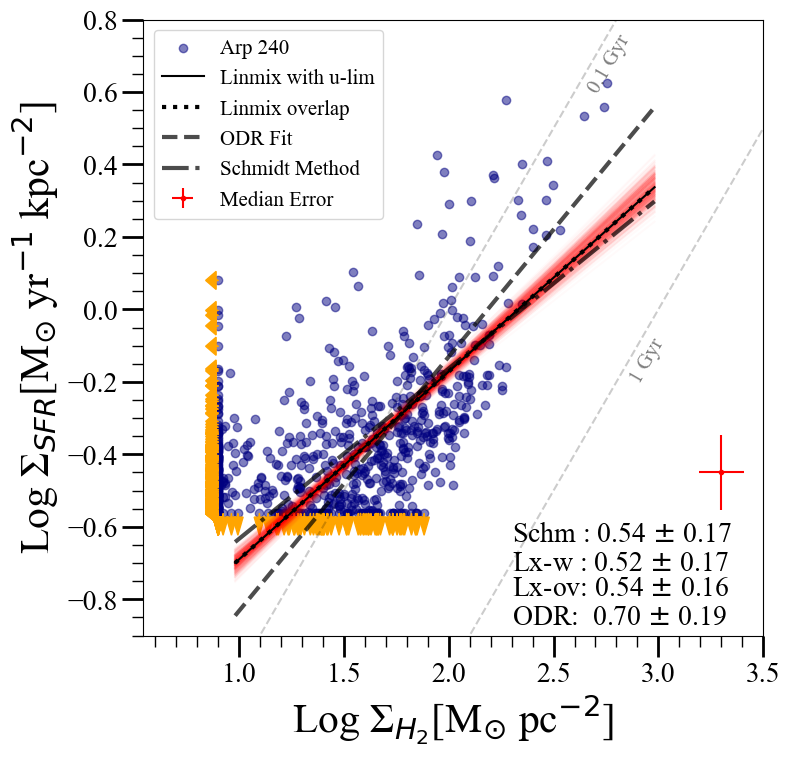} \\
    \end{tabular}

    \caption{\small The slope of the mK-S law in Arp 240. Top panels (left: NGC 5258, right: NGC 5257) and bottom panel (entire system): The dashed black line represents the fit with ODR, while the dotted black line indicates the median slope from \texttt{Linmix} using only the overlapped regions. The solid black line shows the median value from \texttt{Linmix} using upper limits in $\Sigma_{\mathrm{SFR}}$. Red thin lines represent the range of values from 10,000 MCMC runs performed by \texttt{Linmix}. The dashed-dotted black line indicates the slope using Schmitt’s method with \texttt{ASURV}. Orange arrows represent upper limits. In all cases, the slopes are found to be sub-linear, regardless of the fitting method used.}
    \label{fig:K-S_fit}
\end{figure*}

 To determine the slope of Arp 240, we first assessed the probability density distributions of both Log~$\Sigma_{\mathrm{H_2}}$ and Log~$\Sigma_{\mathrm{SFR}}$, evaluating the skewness of these quantities for each galaxy separately as well as for their collective ensemble. Our approach intends to identify potential deviations from unimodality, the presence of outliers, sub-populations, or other non-Gaussian distribution characteristics. Figure \ref{fig:skew_hist} shows the results of our analysis. For Log~$\Sigma_{\mathrm{H_2}}$,  the distributions exhibit approximate symmetry in NGC 5258 with a skewness value of 0.30, and a slight asymmetry in NGC 5257 with a skewness of 0.56, as illustrated in the top left panel. The ensemble trend, shown in the top middle panel, however, demonstrates overall symmetry with a skewness of 0.31. In contrast, the distributions of  Log~$\Sigma_{\mathrm{SFR}}$, shown in the bottom left panel, are markedly asymmetric, showing higher skewness values of 1.34 for NGC 5257 and 1.63 for NGC 5258, with the ensemble trend following suit at 1.52, as illustrated in the bottom middle panel of Figure \ref{fig:skew_hist}.

Moreover, we applied the Akaike Information Criterion (AIC; \citealt{Akaike_74}) and the Bayesian Information Criterion (BIC; \citealt{Schwarz_78}) to evaluate the fitting of Gaussian Mixture Models (GMMs) to our data. AIC and BIC are statistical metrics used to compare the goodness of fit of different models while penalizing for increased complexity, thus helping in determining the optimal number of components in a GMM. Lower values of AIC and BIC indicate a better model fit, with BIC giving more weight to model simplicity \citep[e.g.,][]{AIC_BIC}. This approach allowed us to ascertain whether a unimodal or bimodal distribution more accurately describes the observed data. We used the $\tt{GaussianMixture}$ function from $\tt{scikit-learn}$ in Python \citep{scikit-learn} to compute these values. In the right panels of Figure \ref{fig:skew_hist}, we display the AIC and BIC indices for Log~$\Sigma_{\mathrm{H_2}}$ and Log~$\Sigma_{\mathrm{SFR}}$, using models with one and two Gaussian components. While both indicators favor a single component for Log~$\Sigma_{\mathrm{H_2}}$, our analysis yields that Log~$\Sigma_{\mathrm{SFR}}$ is best modeled with two components. In principle, these results may suggest a bimodality only in SFR. We will address this result in $\S$ \ref{discussion}. 

We used three different fitting algorithms for determining the slope, allowing a comparative analysis of the relation $\Sigma_{\mathrm{SFR}}$-~$\Sigma_{\mathrm{H_2}}$ in Arp 240. These methods include Orthogonal Distance Regression (ODR; \citealt{ODR}), which minimizes the orthogonal distances from the data points to the regression line. We also used  $\tt{Linmix}$\footnote{The Linmix algorithm, as applied in this study for slope determination, follows a methodology similar to that employed by \citealt{Kennicutt2021} and \citealt{Sun_2023} in their analysis of the K-S law.} \citep{Kelly_2007}, an approach that accounts for measurement errors in both variables along with their intrinsic scatter and incorporates a Bayesian framework for linear regression. We applied $\tt{Linmix}$ to fit the slope of the regions where both radio continuum and CO($2\rightarrow1$) overlap. Additionally, we used $\tt{Linmix}$ to fit the slope of all regions, including single detections, and placed upper limits in the SFR tracer.

The third method uses survival analysis incorporated in the code $\tt{ASURV}$ Rev 1.2 (\citealt{ASURV_1990}; \citealt{Feigelson1992}), which implements the methods presented in \cite{ASURV_1986} for bivariate problems. For a comprehensive review of survival methods in astronomy, refer to \cite{Schmitt_1985} and references therein. This technique accounts for upper limits in both the independent and dependent variables, making it appropriate for handling non-detections in both tracers.

The top two panels of Figure \ref{fig:K-S_fit} display our results for NGC 5258 and NGC 5257. The bottom panel showcases the combined slope for the entire system. In these panels, the dashed thick line indicates the fit obtained using ODR. For $\tt{Linmix}$, the software executes 10$^4$ models using its built-in Markov Chain Monte Carlo (MCMC) routine. This modeling process accounts for the potential range of each variable, considering their respective errors and intrinsic scatter. The red thin lines in Figure \ref{fig:K-S_fit} illustrate the results of the $\tt{Linmix}$ models applied to all data, including single detections, with upper limits in the SFR tracer (y-axis) censored. The thick solid black line represents the median slope of these $\tt{Linmix}$ models. Additionally, the dotted black lines show the median slopes derived from $\tt{Linmix}$ but calculated exclusively for the overlapped regions. The slope derived using $\tt{ASURV}$ is depicted by a dashed-dotted black line.

In all cases, the slope of the mK-S law at 500 pc scales is sub-linear for this system as well as for its individual galaxies, regardless of the fitting technique employed. Table \ref{table:total_slopes} summarizes the results of the slope determination. Notably, the slopes differ between the two galaxies, with NGC 5258 consistently exhibiting a steeper slope across all methods compared to NGC 5257. These differences will be discussed in detail in $\S$ \ref{discussion} and $\S$ \ref{is_there_1_slope}.

\begin{table*}
    \centering
    \begin{tabular}{lcccc}
    \hline
    \hline
    Name & ODR & Linmix Overlap & Linmix with Upper Limits & Schmidt Method \\
    \hline
    NGC 5258 & $0.82 \pm 0.21$ & $0.76 \pm 0.14$ & $0.75 \pm 0.15$ & $0.70 \pm 0.17$ \\
    NGC 5257 & $0.60 \pm 0.25$ & $0.44 \pm 0.17$ & $0.52 \pm 0.16$ & $0.52 \pm 0.15$ \\
    Total    & $0.70 \pm 0.19$ & $0.54 \pm 0.16$ & $0.52 \pm 0.17$ & $0.54 \pm 0.17$ \\
    \hline
    \end{tabular}
    \caption{Table of mK-S Slope Values. The second column presents the fitting results obtained using ODR. The third and fourth columns display results from Linmix, first considering only detections, and then including non-detections, respectively. The fifth column details slope determinations conducted using the Schmidt Method implemented with ASURV.}
    \label{table:total_slopes}
\end{table*}

\begin{figure*}
    \centering

    \textbf{NGC 5258} \\[1ex]
    \begin{tabular}{ccc}
        \includegraphics[width=0.32\textwidth]{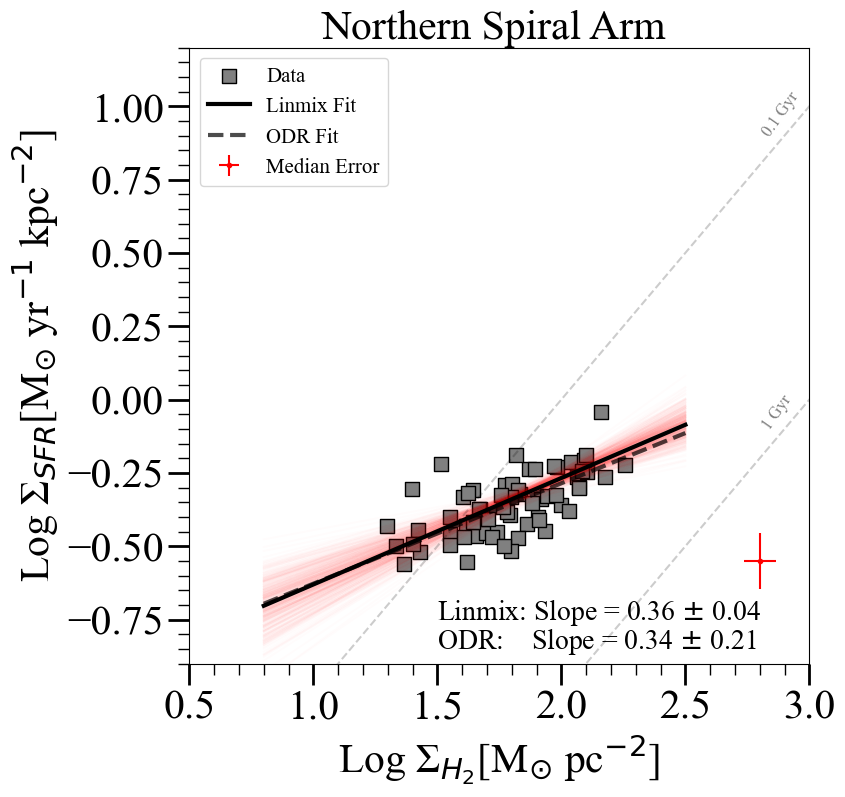} &
        \includegraphics[width=0.32\textwidth]{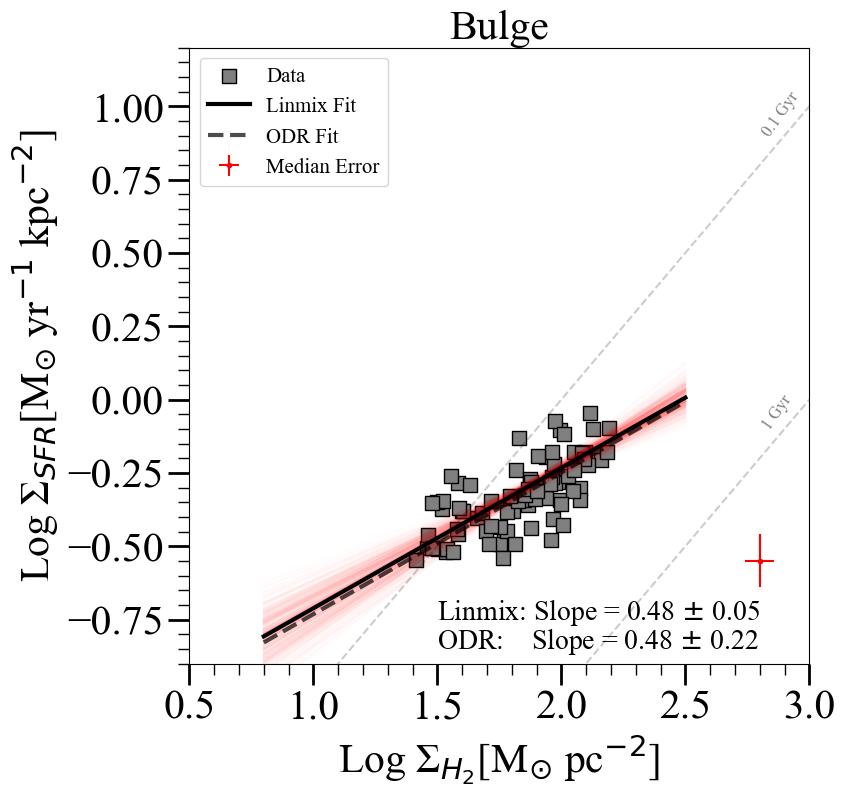} &
        \includegraphics[width=0.32\textwidth]{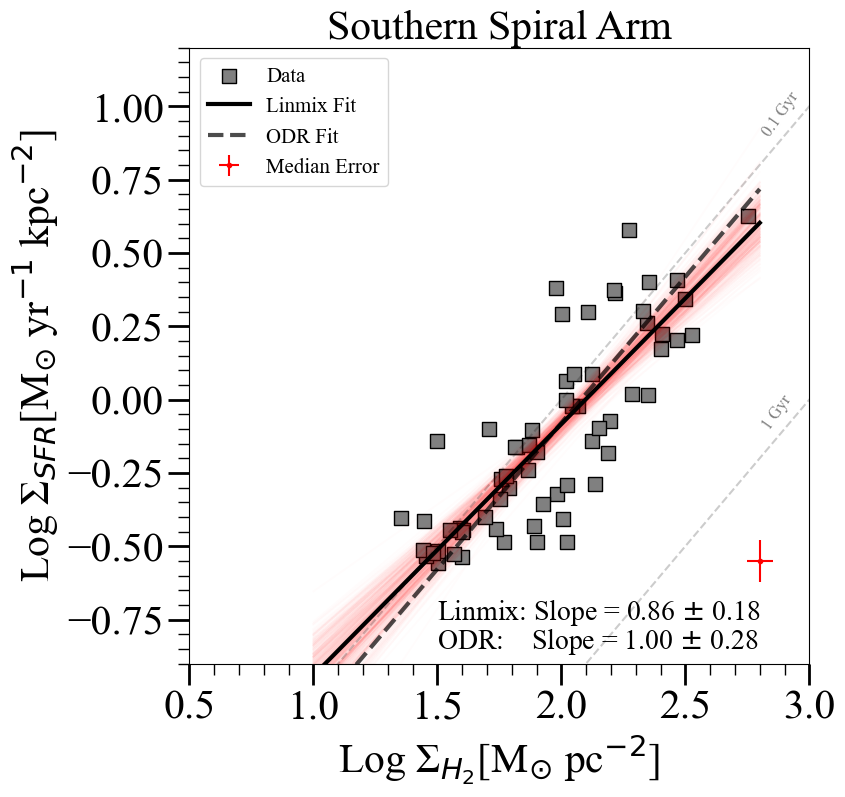} \\
    \end{tabular}
    \\[2ex]

    \textbf{NGC 5257} \\[1ex]
    \begin{tabular}{ccc}
        \includegraphics[width=0.32\textwidth]{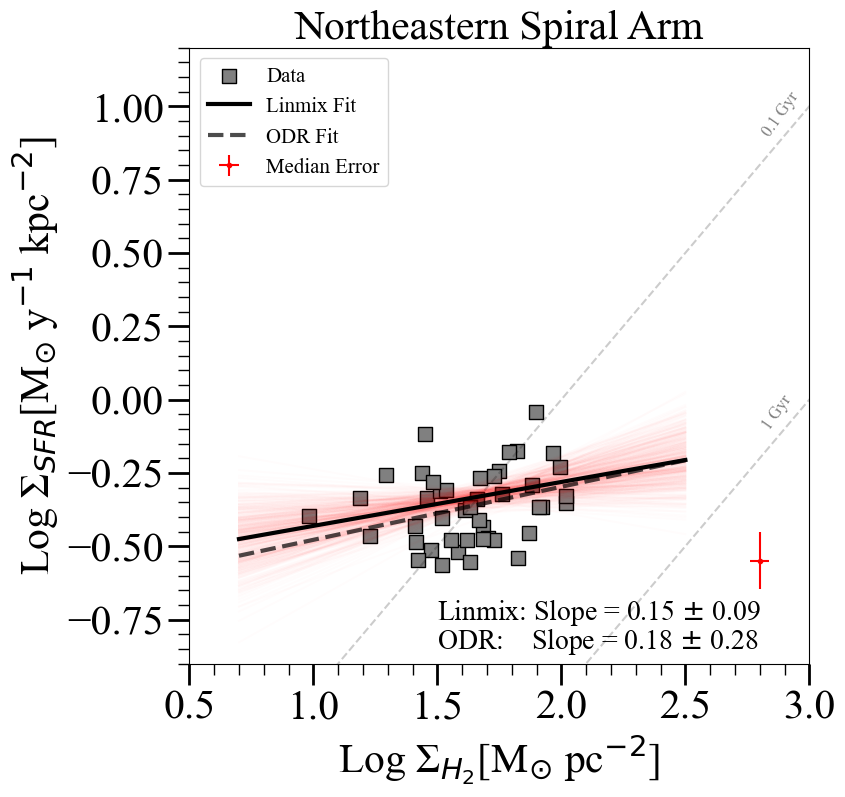} &
        \includegraphics[width=0.32\textwidth]{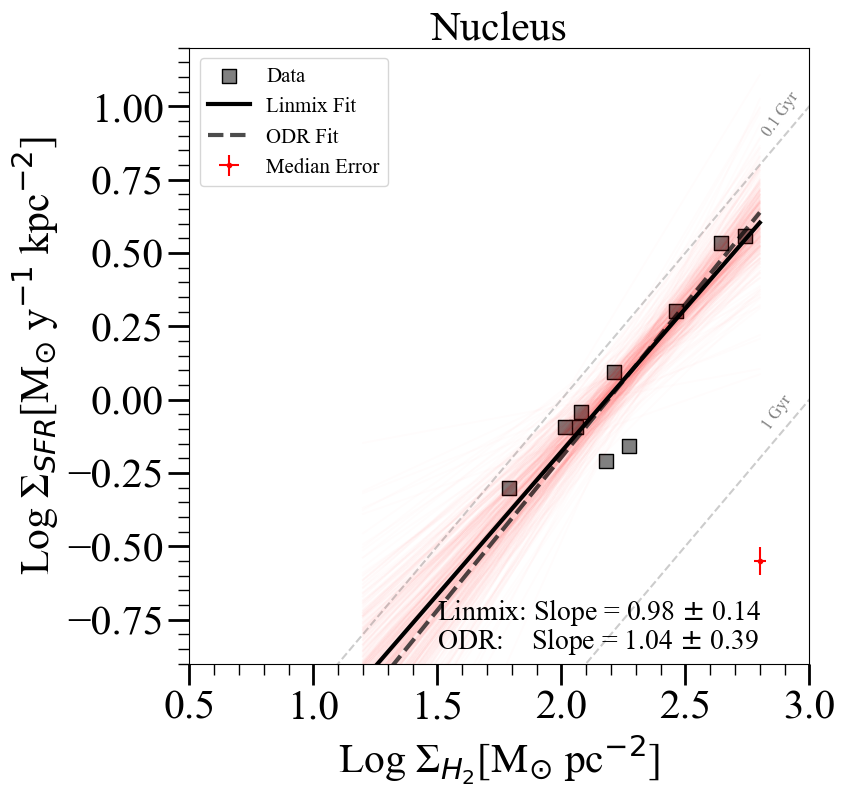} &
        \includegraphics[width=0.32\textwidth]{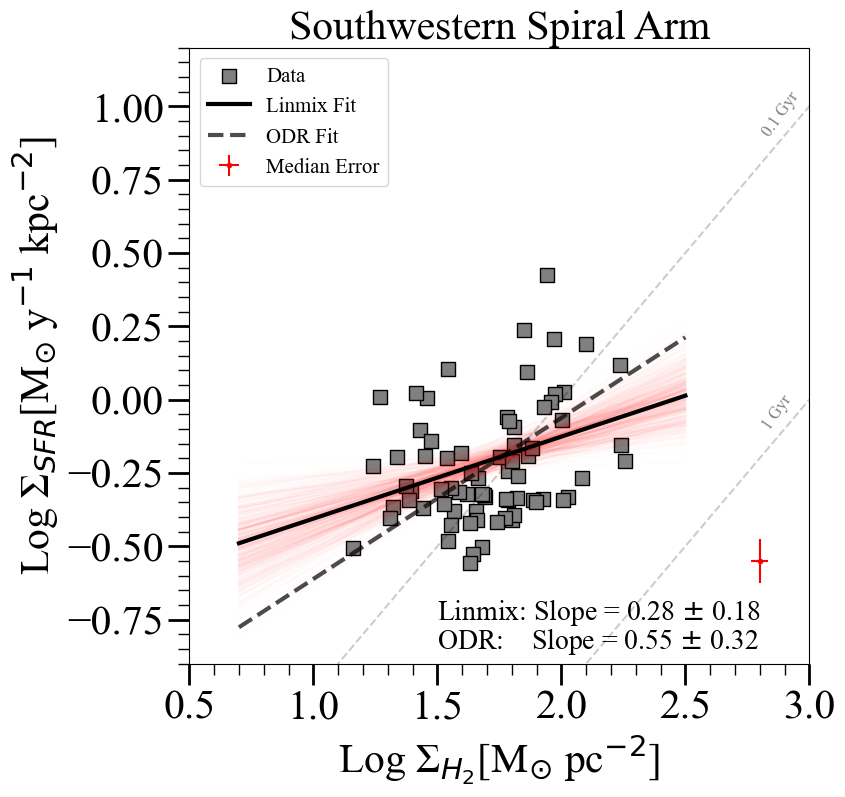} \\
    \end{tabular}

    \caption{\small Decomposition of the mK-S per morphology. Top panels from left to right display the northern arm, the bulge and the southern arm of NGC 5258. The bottom panels from left to right show the northeastern arm, the nucleus, and the southwestern arm of NGC 5257. Fits to the slope are shown: The thick dashed black line represents the slope derived using ODR, while the solid black line is the median slope obtained with \texttt{Linmix}. The red thin lines illustrate the range of potential values modeled by \texttt{Linmix} using 10$^4$ MCMC executions. This decomposition highlights that within Arp 240, the high surface brightness components exhibit steeper slopes compared to the low surface brightness components.}
    \label{fig:Per_region}
\end{figure*}

\subsubsection{mK-S Law as a Function of Location}\label{singles}
To further investigate the local effects on the mK-S law, we decomposed the data presented in Figure \ref{fig:K-S_fit} into their respective morphological components, adhering to the scheme introduced in $\S$ \ref{sfr}. This approach allows us to disentangle the contributions of each morphological component in shaping the mK-S law for the entire system at 500 pc scales. For this part of the analysis, we utilized only hexagons where both radio continuum and CO ($2\rightarrow1$) overlap. The slopes were fitted using $\tt{Linmix}$ to account for measurement uncertainties, and ODR was used for cross-validation to ensure the robustness of our results.

Figure \ref{fig:Per_region} illustrates how the components of each galaxy have very distinct slopes. The brightest components in each galaxy, the southern spiral arm in NGC 5258 and the nucleus in NGC 5257, have the steepest slopes, while the low surface brightness components have shallow slopes. Table \ref{table:slopes_per_compnent} summarizes the slope value for each component.

In the top panel of Figure \ref{fig:Per_region}, we present the slopes for the components of NGC 5258. It is noticeable that the slope increases for brighter components, with the brightest one in the southern spiral arm approximating 1. Similar to the overall results for this galaxy, both ODR and $\tt{Linmix}$ methods yield consistent results within the error margins. The greatest discrepancy between these methods is observed in the component with the highest scatter (0.19 dex), in contrast to the lower scatter values of 0.08 and 0.09 dex in the other components.

The bottom panels of Figure \ref{fig:Per_region} showcase similar trends for NGC 5257, where brighter components correspond to steeper slopes. Notably, the northeastern spiral arm, one of the low brightness components of this system, has the flattest slope at 0.15~$\pm$0.09~(ODR) and 0.18~$\pm$~0.28~($\tt{Linmix}$) suggesting a weak or negligible $\Sigma_{\mathrm{SFR}}$-~$\Sigma_{\mathrm{H_2}}$ correlation. The scatter significantly influences the concordance between the ODR and $\tt{Linmix}$ fits. For the northeastern spiral arm and the nucleus, where the scatter is 0.12 dex, the results are consistent. In contrast, the southwestern spiral arm, with a higher scatter of 0.19 dex, shows less consistent results between the two methods. 

The southwestern spiral arm of NGC 5257 on the bottom right panel of Figure \ref{fig:Per_region}, despite containing some of the brightest regions in both $\Sigma_{\mathrm{H_2}}$ and $\Sigma_{\mathrm{SFR}}$, exhibits a notably shallow slope at 0.28~$\pm$~0.18~($\tt{Linmix}$) and 0.55~$\pm$~0.32~(ODR).  It is noteworthy that the brightest $\Sigma_{\mathrm{SFR}}$ values do not align with the highest in $\Sigma_{\mathrm{H_2}}$, contrary to what would typically be expected from the correlation. In fact, several regions with high $\Sigma_{\mathrm{H_2}}$ (as indicated on the horizontal axis) correspond to lower  $\Sigma_{\mathrm{SFR}}$ values (on the vertical axis). 

\begin{table}
\centering
\setlength{\tabcolsep}{3.5pt} 
\begin{tabular*}{\columnwidth}{@{\extracolsep{\fill}} c c c }
\hline
Component & ODR & Linmix \\ 
\hline
\multicolumn{3}{c}{NGC 5258} \\
Northern Arm & 0.34 $\pm$ 0.21 & 0.36 $\pm$ 0.04 \\ 
Bulge & 0.48 $\pm$ 0.22 & 0.48 $\pm$ 0.04 \\ 
Southern Arm & 1.00 $\pm$ 0.28 & 0.86 $\pm$ 0.18 \\ 
\hline
\multicolumn{3}{c}{NGC 5257} \\
Northeastern Arm & 0.18 $\pm$ 0.28 & 0.15 $\pm$ 0.09 \\ 
Nucleus & 1.04 $\pm$ 0.39 & 0.98 $\pm$ 0.14 \\ 
Southwestern Arm & 0.55 $\pm$ 0.32 & 0.28 $\pm$ 0.18 \\ 
\hline
\end{tabular*}
\caption{Table of mK-S law slopes for various morphological components. The second and third columns are the slope values derived from the ODR and Linmix fitting algorithms, respectively.}
\label{table:slopes_per_compnent}
\end{table}

\begin{table*}[ht]
\centering
\begin{tabular}{@{}llcccccccccc@{}} 
\toprule
Galaxy & Component & \multicolumn{2}{c}{$\Sigma_{\mathrm{H_2}}$-$\langle \sigma \rangle$} & \multicolumn{2}{c}{$\Sigma_{\mathrm{SFR}}$-$\langle \sigma \rangle$} & \multicolumn{2}{c}{SFE-$\langle \sigma \rangle$} & \multicolumn{2}{c}{$\Sigma_{\mathrm{SFR}}$-P$_{\mathrm{turb}}$} & \multicolumn{2}{c}{SFE-P$_{\mathrm{turb}}$} \\
\cmidrule(lr){3-4} \cmidrule(lr){5-6} \cmidrule(lr){7-8} \cmidrule(lr){9-10} \cmidrule(lr){11-12}
       &           & $\rho$ & P-value & $\rho$ & P-value & $\rho$ & P-value & $\rho$ & P-value & $\rho$ & P-value \\
\midrule
\textbf{NGC 5258} & Galaxy           & 0.32 & $<$ 0.01 & 0.25 & $<$ 0.01 & -0.10 & 0.16    & 0.52         & $<$ 0.01 & -0.29 & $<$ 0.01 \\
                  & Southern arm     & 0.38 & $<$ 0.01 & 0.33 & 0.01     & -0.05 & 0.62    &\textbf{0.72} & $<$ 0.01 & -0.18 &0.15 \\
                  & Bulge            & 0.39 & $<$ 0.01 & 0.22 & 0.04     & -0.25 & 0.02    &0.38          & $<$ 0.01 & -0.41 & $<$ 0.01 \\
                  & Northern arm     & 0.02 & 0.77     & -0.02& 0.57     & 0.01  & 0.61    & 0.33         & 0.01     & -0.43 & $<$ 0.01 \\\hline
\textbf{NGC 5257} & Galaxy           & 0.49 & $<$ 0.01 & 0.27 & $<$ 0.01 & -0.27 & $<$ 0.01& 0.38         & $<$ 0.01 & -0.51 & $<$ 0.01 \\
                  & Northeastern arm & 0.23 & 0.13     & 0.16 &    0.29  & -0.10 & 0.51    & 0.23         & 0.12     & 0.46  & $<$ 0.01 \\
                  & Center           & \textbf{0.67} & 0.03     & 0.31 &     0.38 & -0.43 & 0.21    & \textbf{0.60}& 0.07     & -0.25  & 0.47 \\
                  & Southwestern arm & 0.50 & $<$ 0.01 & 0.14 &     0.25 & -0.35 &$<$ 0.01 & 0.26         & 0.03     & -0.52  & $<$ 0.01 \\
\bottomrule
\end{tabular}
\caption{Summary of Spearman's $\rho$ values and p-values at the 50th percentile for each galaxy and for each morphological component within the galaxies. This table shows the correlation strength and statistical significance of relationships between key mK-S law quantities—$\Sigma_{\mathrm{H_2}}$, $\Sigma_{\mathrm{SFR}}$, and SFE—with linewidth $\langle \sigma \rangle$ and turbulent pressure P$_{\mathrm{turb}}$. The complete table, including the 16th, 50th, and 84th percentiles, is available in Tables \ref{tab:correlation_system}, \ref{tab:correlation_ngc5258} and \ref{tab:correlation_ngc5257} in Apendix \ref{Apendix_tables}. Bolded values indicate the strongest correlations.}
\label{stats_components}
\end{table*}

\subsubsection{Star Formation and Molecular Gas Kinematics}\label{SFR_n_kinematics}

In addition to examining the $\Sigma_{\mathrm{SFR}}$-~$\Sigma_{\mathrm{H_2}}$ relationship on its own, we also explore the connection between the mK-S law and the kinematic properties of the molecular gas at the 500 pc scales under investigation. As previously noted, our analysis of the kinematics is based on the assumption that the linewidth is broadened by turbulence. This assumption will be further discussed in \S \ref{vel_mkS} and \S \ref{P_SFE}.

Figures \ref{fig:linewidth} and \ref{fig:figure_P_E} indicate that the Probability Density Distributions of these kinematic quantities do not follow a normal distribution. Given this non-normality, we utilize the Spearman's $\rho$ statistic to quantify the correlations. Spearman's rank correlation coefficient is a non-parametric measure, making it suitable for data that do not adhere to normal distribution assumptions and it is robust against outliers \citep[e.g.,][]{Spearman_stat}. 

\begin{figure}
    \centering
    \includegraphics[width=0.49\textwidth]{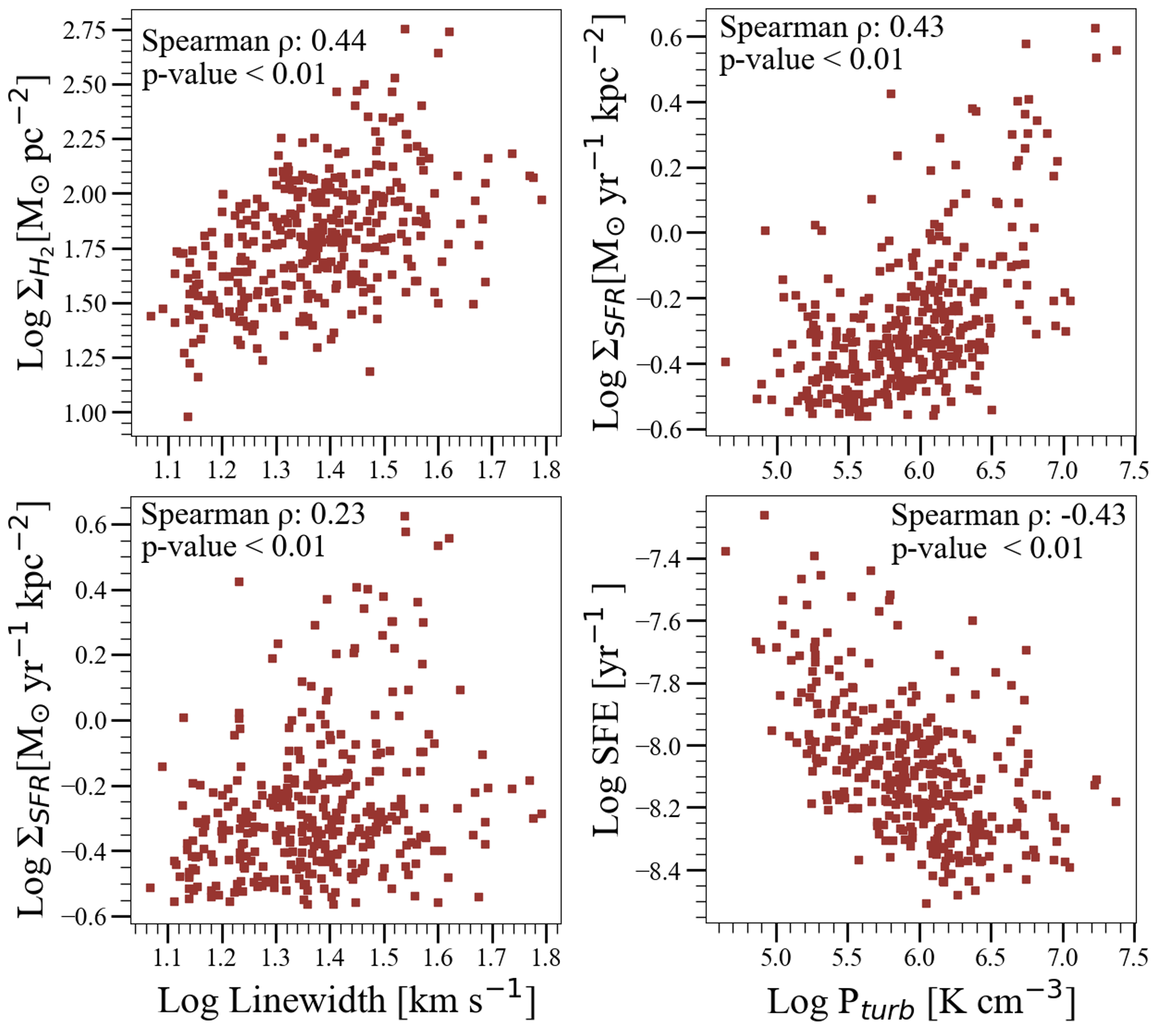}
    \caption{\small Spearman correlation coefficients for $\Sigma_{\mathrm{SFR}}$, $\Sigma_{\mathrm{H_2}}$, SFE, and the kinematic properties of the molecular gas in Arp 240. The left panels show the strength of the correlation between $\Sigma_{\mathrm{SFR}}$, $\Sigma_{\mathrm{H_2}}$, and linewidth. The right panels display $\Sigma_{\mathrm{SFR}}$ and SFE versus P$_{\mathrm{turb}}$. The $\rho$-values shown in each panel correspond to the 50th percentile calculated for the entire system. A detailed breakdown per galaxy and per component at the 50th percentile is provided in Table \ref{stats_components}.}
    \label{fig:SFR_kin}
\end{figure}

Figure \ref{fig:SFR_kin} displays the key quantities in the mK-S law—$\Sigma_{\mathrm{H_2}}$, $\Sigma_{\mathrm{SFR}}$, and SFE—alongside their corresponding relationships with linewidth in the left column and turbulent pressure in the right column for the entire system. For a detailed breakdown of these statistics across the galaxies and their morphological components, see Table \ref{stats_components}, which presents the results of the Spearman's $\rho$ and p-values at the 50th percentile separately for each galaxy and component. Additionally, the complete table, including the 16th, 50th, and 84th percentiles, is available in Tables \ref{tab:correlation_system}, \ref{tab:correlation_ngc5258} and \ref{tab:correlation_ngc5257} in Apendix \ref{Apendix_tables}. We consider a correlation to be statistically significant if $\rho$ $>$ 0.30 and its corresponding p-value $<$~0.03. Correlations with $\rho$ values between 0.30 and 0.40 are classified as weak, those between 0.40 and 0.60 as moderate, and those greater than 0.60 as strong.

The top left panel of Figure \ref{fig:SFR_kin} demonstrates that, at 500 pc scales in Arp 240, $\Sigma_{\mathrm{H_2}}$ correlates moderately with linewidth, with a $\rho$-value of 0.44. In contrast, the bottom-left panel shows no significant correlation between $\Sigma_{\mathrm{SFR}}$ and linewidth, with a $\rho$-value of 0.23.

The right column of Figure \ref{fig:SFR_kin} shows that $\Sigma_{\mathrm{SFR}}$ has a moderate positive correlation with turbulent pressure, with a $\rho$-value of 0.43, indicating that SFR increases as turbulent pressure rises. Conversely, SFE negatively correlates with turbulent pressure, as indicated by a $\rho$-value of -0.43. In all cases, the $p$-values for these relationships across the entire system are much less than 0.01. 

We will discuss these results and analyze their implications in terms of the morphological environments and their relation with the mK-S law in $\S$ \ref{vel_mkS} and $\S$ \ref{P_SFE}.

\section{Discussion}\label{discussion}

Both the resolution and the sensitivity of the observations influence the determination of the mK-S law's slope, as they dictate the quantity and distribution of emitting regions. The sensitivity of the VLA at 3GHz enables the detection of SF across the entire extent of the spiral arms in both galaxies, extending beyond the prominently bright clumps and the nuclei. Notably, the $\Sigma_{\mathrm{SFR}}$ in these low surface brightness components matches that of starbursts, with similar depletion times, as illustrated in Figure \ref{fig:K-S_all}. 

Moreover, approximately 60$\%$ of all hexagons detected in the radio continuum and $\sim$ 90$\%$ of those detected in CO ($2\rightarrow1$) overlap, suggesting that both tracers are significantly co-spatial in both the spiral arms and central components. This result is further supported by our synthetic maps in Figures \ref{fig:maps}, \ref{fig:linewidth} and \ref{fig:figure_P_E}, where the morphology of each galaxy can be readily discerned. Nonetheless, we identified regions within each tracer, even above 5-$\sigma$, without a corresponding counterpart in the other. 

These single detections could be due to sensitivity limitations but may also hint at a possible decoupling between the tracers of SFR and molecular gas in some areas. Whether from sensitivity issues or intrinsic factors, they may affect our slope determination. For instance, NGC 5257, which has the highest rate of single detections and only $\sim$~36$\%$ overlap (compared to NGC 5258's $\sim$~72$\%$ overlap), also has the shallowest slope (detailed results can be found in Table \ref{table:total_slopes}). Remarkably, by employing different techniques to treat upper limits, the slope of the mK-S law for the entire system is only adjusted from 0.52 $\pm$ 0.17 in the overlap regions to 0.54 $\pm$ 0.16 ($\tt{Linmix}$) and 0.54 $\pm$ 0.17 ($\tt{ASURV}$), thereby not substantially altering our overall result of a sub-linear slope. While sensitivity issues could play a role, the possibility of intrinsic physical decoupling driven by differing timescales, as suggested by other high-resolution studies \citep[e.g.,][]{Schinnerer2019}, remains plausible. This will be discussed further in \S \ref{decouple_cause}.

\begin{figure*}
    \centering
    \begin{tabular}{cc}
        \includegraphics[width=0.45\textwidth]{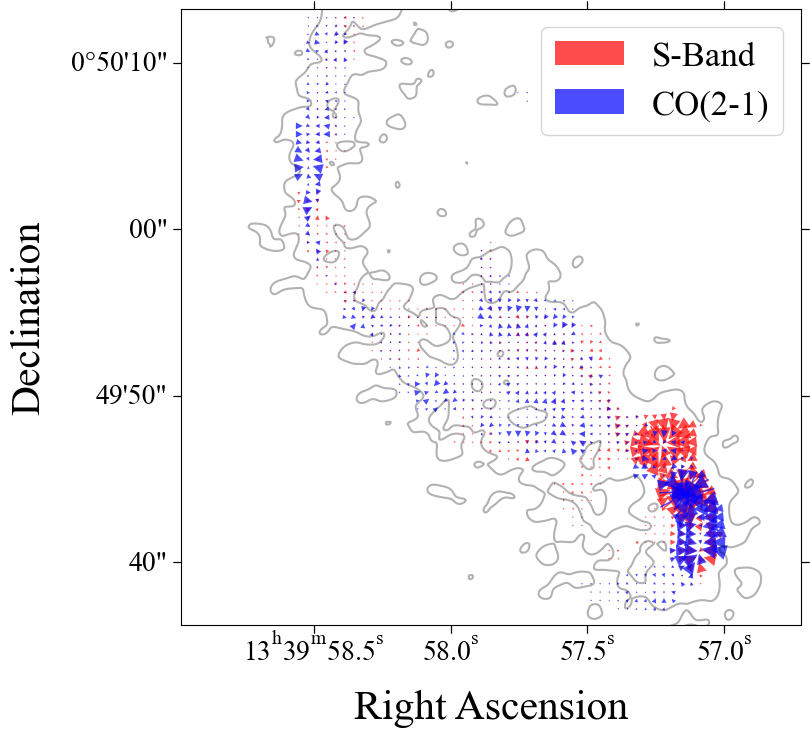} &
        \includegraphics[width=0.45\textwidth]{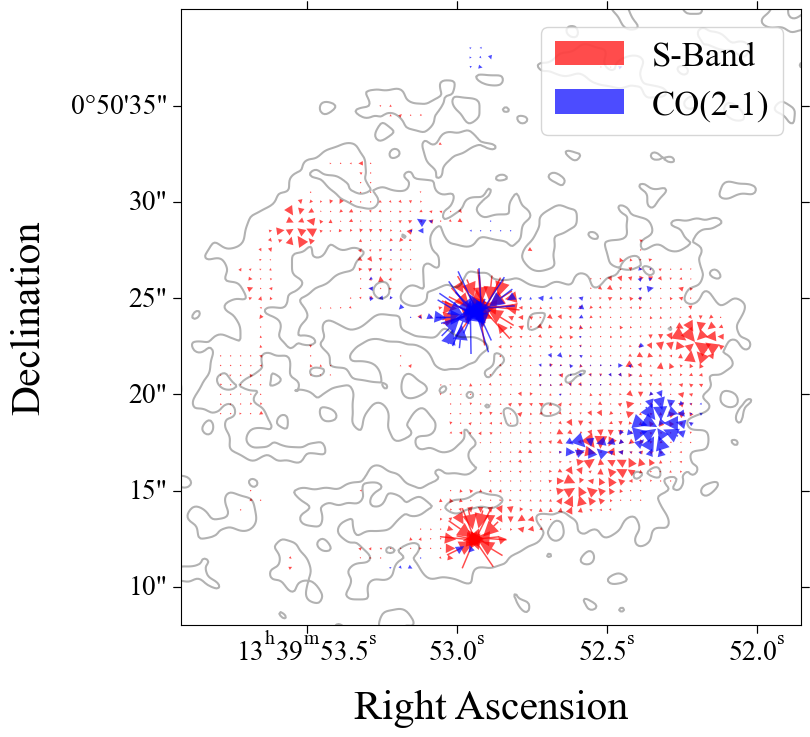} \\
    \end{tabular}

    \caption{\small Intensity gradient maps for NGC 5258 (left) and NGC 5257 (right). Gray contours correspond to the radio continuum at 3 GHz (S-Band) observed at a 3-$\sigma$ level. Red and blue arrows indicate the gradient vectors of emissions for the S-Band and CO ($2\rightarrow1$), respectively. The magnitude of each arrow is normalized to the maximum gradient vector observed for each tracer, with arrows pointing towards the positions of local emission peaks. These gradient maps highlight regions where the tracers exhibit coupled local maxima, specifically the nucleus of NGC 5257 and the southern spiral arm of NGC 5258, as well as areas of decoupling, such as the southwestern spiral arm of NGC 5257.}
    \label{fig:contours}
\end{figure*}

Indeed, our observations reveal that variations in the intensity of the SFR and molecular gas tracers are not entirely coupled. This is illustrated in Figure \ref{fig:contours}, where we explicitly plot the intensity gradients of both tracers. In Figure \ref{fig:contours}, blue and red arrows represent CO ($2\rightarrow1$) and radio continuum, respectively. The magnitude of each arrow has been normalized to the maximum gradient vector for each tracer, with arrows pointing to the positions of local peaks. Figure \ref{fig:contours}  clearly demonstrates that although emission is spatially coupled in the overlapped regions—as both tracers are detected above their respective thresholds—their spatial variations are decoupled in some areas. Note the southwestern spiral arm of NGC 5257, where this effect is more evident. By plotting these gradients separately rather than their ratio (as in the SFE map), we can more effectively identify the distinct distributions of molecular gas and SF across the observed area.

In a perfectly coupled scenario, faint regions of molecular gas would correspond with low SFR within scatter, placing data points in the bottom left of the mK-S relation. However, if decoupled, faint molecular gas emission in regions of high SFR shifts data points from the bottom left to the upper left, contributing to a shallower slope. Conversely, low SFR detections in regions of high molecular gas emission shift data points from the upper right to the lower right, also leading to a shallower slope. 

The sensitivity and spatial resolution of our observations enable us to effectively sample each morphological component in both galaxies. This allows us to identify and analyze the effects of intensity gradient decoupling in the mK-S law of each individual component, shedding light on how the system's mK-S slope is assembled and adding context to the results in Figure \ref{fig:Per_region}. For instance, in the right panel of Figure \ref{fig:contours}, the southwestern spiral arm of NGC 5257 displays a more pronounced decoupling. This decoupling is reflected in the lower right panel of Figure \ref{fig:Per_region}, which shows the mK-S law for the same region, revealing a notably shallow slope. In contrast, regions where the gradients align—particularly the center of NGC 5257 and, to a lesser extent, the southern spiral arm of NGC 5258—exhibit steeper slopes in their respective mK-S plots in Figure \ref{fig:Per_region}. Thus, intensity gradients provide critical insight into the underlying causes driving the varied slopes observed in Figure \ref{fig:Per_region}.

It is worth noting that this decoupling offers one possible interpretation for the high $\epsilon_{\mathrm{ff}}$ values shown in Figure \ref{e_ff}, which were also reported by \cite{He_2019}. Some regions in the periphery of star-forming complexes have faint molecular gas but significant radio continuum, which elevates the $\epsilon_{\mathrm{ff}}$ to values greater than 100\%. This highlights how decoupling between tracers can artificially inflate $\epsilon_{\mathrm{ff}}$, creating values that surpass physically plausible thresholds. Indeed, Figure \ref{fig:Per_region} shows that the southwestern spiral arm of NGC 5257, where the decoupling is more pronounced, contains more regions above the 0.1 Gyr t$_\mathrm{d}$ line, compared to any other component in the system. A similar trend is observed, to a lesser extent, in the northeastern component. Notably, 20\% of the regions in the southwestern component of NGC 5257 exceed $\epsilon_{\mathrm{ff}}$ = 100\%, underscoring the extent of this effect. Note that this component also exhibits the strongest signs of disruption driven by the merger.

\subsection{What Causes the Decoupling of Molecular Gas and SFR Tracers?}\label{decouple_cause}
Observations of spatial decoupling between SF, traced by H$\alpha$, and the amount of gas as traced by CO, have been reported in nearby spiral galaxies from the PHANGS sample \citep[e.g.,][]{Schinnerer2019}. They found that while at low resolution ($\geq$ 1 kpc) both H$\alpha$ and CO overlap almost entirely, at higher resolution (e.g. from 250 to 140 pc), the two tracers decoupled, revealing regions where only molecular gas is detected but no SF, and vice versa. 

Additionally, theoretical work suggests that molecular gas may be abundant in some regions during the earliest stages of the SF cycle, making it easier to detect compared to incipient SF. In contrast, regions at later stages might have depleted their star-forming molecular gas reservoir, leaving only the SF to be detected \citep[e.g.,][]{Kruijssen_2018}. High-resolution measurements might be able to capture regions at different stages in the SF cycle by distinguishing between diffuse, non-star forming gas from its denser actively star-forming phase. 

Despite this, PHANGS galaxies showed minimal spatial decoupling between H$\alpha$ and CO at 500 pc scales—the resolution we are investigating in Arp 240. However, the primary distinction between our findings and those of \cite{Schinnerer2019} lies not only in the scales at which spatial decoupling is observed but, more critically, in our identification of decoupling within the peaks of the intensity gradients in some regions, even where both tracers overlap. It is worth noting that galaxies in PHANGS are mostly isolated, nearby spirals, where gas dynamics and distribution are primarily influenced by galactic structure and SF. In contrast, for Arp 240, gravitational perturbations and tidal forces between the galaxies may further redistribute gas \citep[e.g.,][]{Barnes_1996}, potentially enhancing decoupling at scales as large as 500~pc. The exploration of connections between smaller scales (e.g., GMC scales of approximately 100 pc) and the larger scale addressed in this study, along with a more direct comparison with the PHANGS sample, will be the subject of a follow-up paper by Saravia et al. (in preparation), with higher resolution analysis.

In addition to the misalignment of tracer maxima observed in certain regions, the bright, compact complexes and the more extended diffuse component exhibit distinct behaviors in the slope of the mK-S law, as discussed in $\S$~\ref{singles}. Furthermore, the different timescales of the tracers themselves may also play a role, as each tracer relies on distinct physical processes with different characteristic timescales. This contributes to the observed complexity of the SF process. For instance, H$\alpha$ traces SF with a survival timescale of approximately 5 Myr, while the total radio continuum emission survives for significantly longer periods, ranging from 2 to 100 Myr, with 40 Myr being the expected value for the non-thermal component, dominant at 3 GHz, to drop by 50$\%$  \citep{Murphy_2012}. Meanwhile, the median free-fall time of the molecular gas in Arp 240 is approximately 12 Myr (see Appendix B), in agreement with numerical simulations suggesting that GMC lifetimes are on the order of a few Myr \citep[e.g.,][]{Grudic_2018, Benincasa_2019}. 

Although both the thermal and non-thermal components of the total radio continuum trace recent SF, the extended timescale of the non-thermal component may allow for a decoupling of gas and SFR tracers in some regions, disrupting the 1-to-1 correlation expected from the mK-S law. Alternatively, in low surface brightness regions, SF may be sparse and thus diluted within the larger apertures used in our analysis, which can also weaken the correlation by underestimating one or both of the tracers. 

Under either scenario, our current observations lack the angular resolution and SED sampling needed to fully disentangle the radio continuum emission and probe the distinct characteristics of these two surface brightness regimes. To address this, we plan to revisit this with observations at 100~pc resolution, increased sensitivity, and wider radio frequency coverage to separate the thermal and non-thermal components (Saravia et al., in preparation).

\subsection{Velocity Dispersion at 500~pc Scales and $\alpha_{\mathrm{CO}}$}\label{vel_mkS}
Studies of star-forming regions within the PHANGS sample, at 120-150~pc scales, have shown that the velocity dispersion of molecular gas significantly influences the CO emissivity, and consequently, the value of $\alpha_{\mathrm{CO}}$ \citep[e.g.,][]{Teng_2023}. These authors found an inverse proportionality between $\alpha_{\mathrm{CO}}$ and the velocity dispersion. A greater velocity dispersion results in lower optical depths, hence leading to an overestimation of $\alpha_{\mathrm{CO}}$. These results align with previous findings of $\alpha_{\mathrm{CO}}$ decreasing with increasing $\Sigma_{\mathrm{H_2}}$ in GMC, as the velocity dispersion of GMCs scales with mass \citep[e.g.,][]{Bolatto_2013}. Such variations in $\alpha_{\mathrm{CO}}$ could significantly alter the slope of the mK-S relationship. 

In our observations, although there is a statistically significant correlation between linewidth and $\Sigma_{\mathrm{H_2}}$, as shown in the top left panel of Figure \ref{fig:SFR_kin}, this correlation is moderate ($\rho$ = 0.44) compared to the stronger correlation ($\rho$ = 0.57 and  $\rho$ = 0.79)\footnote{The two Spearman’s $\rho$-values referenced here represent different methodologies used by \cite{Sun_2022}. The first value ($\rho$ = 0.57) is derived from an object-by-object approach, while the second value ($\rho$ = 0.79) is from a pixel-by-pixel approach. The latter is more comparable to our analysis.} observed at GMC scales (120-150 pc) in the PHANGS sample \citep{Sun2018, Sun2020, Sun_2022}. Similar weak to moderate correlation strengths are observed in each galaxy separately, with NGC 5258 having a $\rho$-value of 0.32 and NGC 5257 showing a moderate correlation with $\rho$-value of 0.49, see Table \ref{stats_components} for reference. 

The weaker correlation that we observed in Arp 240 could be attributed to several factors. One such factor is the spatial resolution of our observations. At 500~pc scales, the presence of intercloud gas not associated with SF could contaminate our linewidth determinations. Spatially unresolved clouds within our resolution element may exhibit differing velocity dispersion, despite our efforts to disentangle them. Additionally, our assumption of uniform line-of-sight depth could introduce another source of variation, as it may not always be realistic. Pronounced velocity gradients in certain areas, especially toward the centers of the galaxies, can lead to beam smearing from unresolved velocity components, complicating the interpretation of the line profiles. This effect might explain the largest velocity dispersion values, predominantly measured in the central inner parts, as illustrated in Figure \ref{fig:linewidth}. These variations can lead to intrinsic differences in $\alpha_{\mathrm{CO}}$ (at $\sim$100~pc scales) across clouds that we would not be able to detect with our current observations. Hence, the linewidth we measure is an average quantity encompassing these diverse contributions, potentially obscuring true local variations in $\alpha_{\mathrm{CO}}$. In light of these complexities, we have opted to apply a uniform value of $\alpha_{\mathrm{CO}}$ in this study, as detailed in $\S$ \ref{h2}, and defer the analysis of a variable $\alpha_{\mathrm{CO}}$ to a forthcoming paper focused on Arp 240 at approximately 100 pc resolution (Saravia et al. in preparation).

\subsection{SFR and SFE vs. Kinematics}\label{P_SFE}
We observe no significant correlation between $\Sigma_{\mathrm{SFR}}$ and linewidth at 500~pc~scales, as illustrated in the bottom left panel of Figure \ref{fig:SFR_kin}. This result contrasts starkly with findings from the PHANGS sample of nearby galaxies at 120-150~pc scales, where they find $\rho$-values of 0.57 and 0.74 \citep{Sun_2022}. However, it aligns with results from numerical simulations of spiral galaxies at 750~pc~ scales, calibrated to the same timescale as our SFR tracer \citep{Orr_FIRE_2020}, using FIRE-2 \citep{FIRE_2}.

In contrast, we observe a moderate correlation between $\Sigma_{\mathrm{SFR}}$ and P$-{\mathrm{turb}}$, with $\rho$ = 0.43 for the entire system. Individually, $\rho$~=~0.52 for NGC 5258 and $\rho$~=~0.38 for NGC 5257, as shown in Table \ref{stats_components}. Notably, this correlation is stronger in brightest regions, such as the southern spiral arm of NGC 5258 ($\rho$ = 0.72) and the center of NGC 5257 ($\rho$ = 0.60). The high density of recent SF in the brightest regions of Arp 240 correlates more strongly with turbulent pressure due to immediate and continuous feedback injection. In contrast, the diffuse component of the SFR tracer exhibits weaker correlations, reflecting the cumulative impact of feedback over time. For example, if the survival timescale of the SFR tracer is comparable to the dissipation time of turbulence—about 80 Myr in Milky Way-like galaxies \citep[e.g.,][]{Orr_FIRE_2020}—turbulence is likely to decay unless balanced by new feedback injection.

Additionally, Figure \ref{fig:SFR_kin} and Table \ref{stats_components} illustrate the regulatory effect of feedback. SFE inversely correlates with P$_{\mathrm{turb}}$. We observe a moderate anti-correlation for the entire system with $\rho$ = -0.43, and for individual galaxies, with $\rho$ = -0.29 for NGC 5258 and $\rho$ = -0.51 for NGC 5257. Remarkably, this trend is absent in the brightest regions, such as the southern spiral arm of NGC 5258 and the center of NGC 5257, where no significant correlation between SFE and P$_{\mathrm{turb}}$ is found, as shown in the last two columns of Table \ref{stats_components}. These regions also have the highest amounts of molecular gas, suggesting that the high SFR observed here is driven strongly by the abundance of molecular gas. This aligns with the mK-S law premise that the amount of gas drives the SFR, resulting in a slope of $\sim$1. In our observations, regions where the SFR is driven by the amount of gas indeed have slopes closest to 1, consistent with the expectation of the mK-S law.

\subsection{Is there one single slope of the mK-S law?}\label{is_there_1_slope}
The mK-S star formation law successfully describes the correlation between the SFR of galaxies and their molecular gas supply at large scales ($>$ tens of kpc), where ISM conditions are unresolved. Generally, a larger molecular gas reservoir increases the likelihood of forming stars. However, star-forming galaxies are dynamic systems, and complexities arise when more physics is involved at resolved scales. These complexities include feedback, chemistry, dynamical influences, environment-induced perturbations, and the various timescales at which these processes occur within a galaxy. Phenomena affecting the distribution and conditions of the molecular gas and its capacity to form stars are likely to be reflected in the mK-S relation if the observations can resolve them. 

Our case study of Arp 240 reveals distinct slopes in the mK-S law for each galaxy and their morphological components. We identified two modes of SF within Arp 240: brighter components exhibit steeper slopes, whereas low-surface-brightness regions show shallower slopes. Moreover, decoupling in local maxima in some regions disrupts the correlation, affecting the overall slope of the entire galaxy. Our results are similar to those in \cite{Sanchez-Garcia_2022}. In their study of 90 pc scale regions in local LIRGs, some galaxies exhibit a bimodality in the mK-S law while others do not. Indeed, the slopes varied from sub-linear (slope of $\sim$0.34) to highly super-linear (slope of $\sim$4.74). Interestingly, in Arp 240, the bimodality is observed even at 500~pc scales, whereas in \cite{Sanchez-Garcia_2022}, the bimodality disappears at lower resolution. Similarly, our calculations of individual slopes for galaxies in the PHANGS sample at 1.5~kpc scales, as studied by \cite{Sun_2023} , show values ranging from $\sim$0.3 to $\sim$1.9, although the ensemble slope converges to $\sim$1. 

These results suggest that measurements at small scales within any particular galaxy do not inherently produce a linear slope (or a single value) for the mK-S law, but rather a range of values dictated by the local ISM conditions. Exploring this parameter space could provide valuable physical information about the boundaries and limits of the different SF regimes and SFE observed in galaxies.

\section{Conclusions}\label{Summary}
We utilized 3 GHz (S-Band) radio continuum observations to determine $\Sigma_{\mathrm{SFR}}$ in 500 pc-sized regions of the early-stage merger Arp 240 (NGC 5257/8), applying a uniform hexagonal grid. Matching resolution ALMA CO~(2$\rightarrow$1) observations were used to derive $\Sigma_{\mathrm{H_2}}$, linewidths, and mean turbulent pressures for these regions. Our principal findings are as follows:

\begin{enumerate}
    \item Shallower mK-S Law Slopes at 500 pc Scale: The slope of the mK-S law observed within Arp 240’s regions at a 500 pc scale is notably shallower than the classical slope of approximately 1 for integrated measurements. Specifically:
    \begin{itemize}
        \item NGC 5258: The median slope is 0.75 $\pm$ 0.15
        \item NGC 5257: The median slope is 0.52 $\pm$ 0.16
        \item Entire system: The slope converges to a median value of 0.52 $\pm$ 0.17 
    \end{itemize}
    
    \item Two Regimes or Modes of SF: Our analysis identifies two distinct SF regimes within Arp 240: bright, compact SF complexes and a diffuse, low-surface-brightness component. Despite differences in brightness and emission characteristics, all identified regions exhibit high $\Sigma_{\mathrm{SFR}}$ and $\Sigma_{\mathrm{H_2}}$ relative to nearby normal star-forming galaxies. Additionally, depletion times in these regions are shorter than those observed in typical star-forming galaxies.
    
    \item mK-S Slope by Morphological Components: When calculating the mK-S slope for specific morphological components, such as spiral arms and the nucleus, the observed slopes in both galaxies range from 0.15 to 0.98. This range encompasses both the shallower slopes associated with low surface brightness components in the radio continuum and the steeper slopes found in the brighter components. 
    \item Intensity Gradient Misalignment and mK-S Slope: Misalignments of local peaks in the intensity gradients of radio continuum and CO (2$\rightarrow$1) emissions, in some regions within both galaxies, weaken the correlation between $\Sigma_{\mathrm{SFR}}$ and $\Sigma_{\mathrm{H_2}}$, resulting in a shallower mK-S slope. This suggest a decoupling between the SF and molecular gas tracers observed at 500~pc scales, in contrast to other studies where the decoupling is only observed at scales of $\sim$100~pc. 
    \item Kinematics and the mK-S Law: We observed a weak correlation between linewidth and $\Sigma_{\mathrm{H_2}}$, and no significant correlation between $\Sigma_{\mathrm{SFR}}$ and linewidth. However, moderate to strong correlations were found between $\Sigma_{\mathrm{SFR}}$ and $P_{\mathrm{turb}}$ in different morphological components. The strongest correlations were associated with the brightest regions in radio continuum (newer SF), indicative of immediate and continuous feedback injection. In contrast, moderate correlations were found in low surface brightness regions, suggesting a cumulative effect of feedback over longer times.

    \item Regulatory Effect of Feedback: Our analysis shows that SFE inversely correlates with $P_{\mathrm{turb}}$. We observe a moderate anti-correlation for the entire system ($\rho$ = -0.43), with $\rho$ = -0.29 for NGC 5258 and $\rho$ = -0.51 for NGC 5257. This trend is absent in the brightest regions in radio continuum (newer SF), which also have the highest amounts of molecular gas, suggesting that the high SFR in these areas is primarily driven by molecular gas abundance.
\end{enumerate}

\section{Acknowledgements}
This paper makes use of the following ALMA data: ADS/JAO.ALMA$\#$2015.1.00804.S. ALMA is a partnership of ESO (representing its member states), NSF (USA) and NINS (Japan), together with NRC (Canada), NSTC and ASIAA (Taiwan), and KASI (Republic of Korea), in cooperation with the Republic of Chile. The Joint ALMA Observatory is operated by ESO, AUI/NRAO and NAOJ.
The National Radio Astronomy Observatory is a facility of the National Science Foundation operated under cooperative agreement by Associated Universities, Inc.

A.S. would like to thank J. Sun and Y.H.E. Teng for their valuable insights and discussions on the PHANGS galaxy sample results; J. Hibbard for enriching discussions on galaxy mergers; E. Murphy for helpful discussions on radio continuum emission as an SFR tracer; and the VLA and ALMA help desk staff for their assistance with data products. Additionally, A.S. acknowledges the NRAO-NINE program for their support and encouragement; M.R. Meza for providing useful discussions and writing tips; and Y. Tu for insightful discussions on SF at smaller scales. Support for this work was provided by the NSF through the Grote Reber Fellowship Program administered by Associated Universities, Inc./National Radio Astronomy Observatory; ALMA Student Observing Support (SOS) program. A.S.E was supported by NASA through grants HST-GO10592.01-A, HST-GO11196.01-A and HST-GO13364 from the Space Telescope Science Institute, which is operated by the Association of Universities for Research in Astronomy, Inc., under NASA contract NAS5-26555. V.U acknowledges funding support from $\#$JWST-GO-01717.001-A, $\#$HST-AR-17065.005-A, $\#$HST-GO-17285.001-A, and NASA ADAP grants $\#$80NSSC20K0450 and $\#$80NSSC23K0750. C. R acknowledges support from Fondecyt Regular grant 1230345 and ANID BASAL project FB210003.

\appendix

\section{Spearman's Correlation Coeffients}\label{Apendix_tables}
\begin{table}[h]
    \centering
    \begin{tabular}{llccc}
        \toprule
        \textbf{Entire System} & & 16$\%$ & 50$\%$ & 84$\%$ \\
        \midrule
        $\Sigma_{\mathrm{H_2}}$-$\langle \sigma \rangle$ & $\rho$ & 0.42 & 0.44 & 0.45 \\
                      & $p$ & $<$0.01 & $<$0.01 & $<$0.01 \\
                      \hline
        $\Sigma_{\mathrm{SFR}}$-$\langle \sigma \rangle$ & $\rho$ & 0.20 & 0.23 & 0.27 \\
                      & $p$ & $<$0.01 & $<$0.01 & $<$0.01 \\
                      \hline
        SFE-$\langle \sigma \rangle$ & $\rho$ & -0.26 & -0.23 & -0.20 \\
                      & $p$ & $<$0.01 & $<$0.01 & 0.01 \\
                      \hline
        $\Sigma_{\mathrm{SFR}}$-P$_{turb}$ & $\rho$ & 0.40 & 0.43 & 0.45 \\
                  & $p$ & $<$0.01 & $<$0.01 & $<$0.01 \\
                      \hline
        SFE-P$_{turb}$ & $\rho$ & -0.46 & -0.43 & -0.40 \\
                       & $p$ & $<$0.01 & $<$0.01 & $<$0.01 \\
        \bottomrule
    \end{tabular}
    \caption{Spearman's $\rho$ and $p$ values for quantities in the mK-S law versus linewidth and turbulent pressure for the entire system, encompassing both galaxies. The 16th, 50th, and 84th percentiles are presented.}

    \label{tab:correlation_system}
\end{table}

\begin{table*}[h]
    \centering
    \small
    \begin{tabular}{llcccccccccc}
        \hline
        \hline
        \textbf{NGC 5258} & & \multicolumn{2}{c}{$\Sigma_{\mathrm{H_2}}$-$\langle \sigma \rangle$} & \multicolumn{2}{c}{$\Sigma_{\mathrm{SFR}}$-$\langle \sigma \rangle$} & \multicolumn{2}{c}{SFE-$\langle \sigma \rangle$} & \multicolumn{2}{c}{$\Sigma_{\mathrm{SFR}}$-P$_{\mathrm{turb}}$} & \multicolumn{2}{c}{SFE-P$_{\mathrm{turb}}$} \\
        \cline{3-4} \cline{5-6} \cline{7-8} \cline{9-10} \cline{11-12}
        & & $\rho$ & $p$ & $\rho$ & $p$ & $\rho$ & $p$ & $\rho$ & $p$ & $\rho$ & $p$ \\
        \hline
        Galaxy & & 0.31 & $<$0.01 & 0.20 & $<$0.01 & -0.14 & 0.04 & 0.49 & $<$0.01 & -0.35 & $<$0.01 \\
        & & 0.32 & $<$0.01 & 0.25 & $<$0.01 & -0.10 & 0.16 & 0.52 & $<$0.01 & -0.29 & $<$0.01 \\
        & & 0.35 & $<$0.01 & 0.28 & $<$0.01 & -0.05 & 0.50 & 0.56 & $<$0.01 & -0.25 & $<$0.01 \\
        \hline
        Southern Arm & & 0.35 & $<$0.01 & 0.29 & $<$0.01 & -0.14 & 0.26 & 0.69 & $<$0.01 & -0.27 & 0.03 \\
        & & 0.38 & $<$0.01 & 0.33 & 0.01 & -0.05 & 0.62 & 0.72 & $<$0.01 & -0.18 & 0.15 \\
        & & 0.40 & $<$0.01 & 0.37 & 0.02 & 0.03 & 0.87 & 0.74 & $<$0.01 & -0.11 & 0.37 \\
        \hline
        Bulge & & 0.35 & $<$0.01 & 0.14 & 0.01 & -0.32 & $<$0.01 & 0.32 & $<$0.01 & -0.50 & $<$0.01 \\
        & & 0.39 & $<$0.01 & 0.22 & 0.04 & -0.25 & 0.02 & 0.38 & $<$0.01 & -0.41 & $<$0.01 \\
        & & 0.42 & $<$0.01 & 0.26 & 0.19 & -0.18 & 0.10 & 0.45 & $<$0.01 & -0.33 & $<$0.01 \\
        \hline
        Northern Arm & & -0.02 & 0.55 & -0.11 & 0.30 & -0.09 & 0.33 & 0.21 & $<$0.01 & -0.51 & $<$0.01 \\
        & & 0.02 & 0.77 & 0.02 & 0.57 & -0.01 & 0.62 & 0.33 & 0.01 & -0.43 & $<$0.01 \\
        & & 0.07 & 0.94 & 0.08 & 0.82 & 0.08 & 0.86 & 0.42 & 0.09 & -0.34 & $<$0.01 \\
        \hline
    \end{tabular}
    \caption{Spearman's $\rho$ and $p$ values for various regions and quantities in NGC 5258. The 16th, 50th, and 84th percentiles are listed from top to bottom within each section for $\Sigma_{\mathrm{H_2}}$-$\langle \sigma \rangle$, $\Sigma_{\mathrm{SFR}}$-$\langle \sigma \rangle$, SFE-$\langle \sigma \rangle$, $\Sigma_{\mathrm{SFR}}$-P$_{\mathrm{turb}}$, and SFE-P$_{\mathrm{turb}}$.}
    \label{tab:correlation_ngc5258}
\end{table*}

\begin{table*}[h]
    \centering
    \small
    \begin{tabular}{llcccccccccc}
        \hline
        \hline
        \textbf{NGC 5257} & & \multicolumn{2}{c}{$\Sigma_{\mathrm{H_2}}$-$\langle \sigma \rangle$} & 
        \multicolumn{2}{c}{$\Sigma_{\mathrm{SFR}}$-$\langle \sigma \rangle$} & 
        \multicolumn{2}{c}{SFE-$\langle \sigma \rangle$} & 
        \multicolumn{2}{c}{$\Sigma_{\mathrm{SFR}}$-P$_{\mathrm{turb}}$} & 
        \multicolumn{2}{c}{SFE-P$_{\mathrm{turb}}$} \\
        \cline{3-4} \cline{5-6} \cline{7-8} \cline{9-10} \cline{11-12}
        & & $\rho$ & $p$ & $\rho$ & $p$ & $\rho$ & $p$ & $\rho$ & $p$ & $\rho$ & $p$ \\
        \hline
        Galaxy & & 0.46 & $<$0.01 & 0.23 & $<$0.01 & -0.31 & $<$0.01 & 0.34 & $<$0.01 & -0.55 & $<$0.01 \\
        & & 0.49 & $<$0.01 & 0.27 & $<$0.01 & -0.27 & $<$0.01 & 0.38 & $<$0.01 & -0.51 & $<$0.01 \\
        & & 0.50 & $<$0.01 & 0.31 & 0.01 & -0.24 & $<$0.01 & 0.42 & $<$0.01 & -0.47 & $<$0.01 \\
        \hline
        Southwestern Arm & & 0.48 & $<$0.01 & 0.08 & 0.11 & -0.39 & $<$0.01 & 0.21 & 0.01 & -0.57 & $<$0.01 \\
        & & 0.50 & $<$0.01 & 0.14 & 0.25 & -0.35 & $<$0.01 & 0.26 & 0.03 & -0.52 & $<$0.01 \\
        & & 0.52 & $<$0.01 & 0.19 & 0.48 & -0.30 & 0.01 & 0.31 & 0.08 & -0.46 & $<$0.01 \\
        \hline
        Center & & 0.60 & 0.08 & 0.22 & 0.26 & -0.61 & 0.06 & 0.54 & 0.03 & -0.47 & 0.17 \\
        & & 0.67 & 0.16 & 0.31 & 0.38 & -0.43 & 0.21 & 0.60 & 0.07 & -0.25 & 0.47 \\
        & & 0.71 & 0.37 & 0.40 & 0.53 & -0.15 & 0.68 & 0.67 & 0.11 & -0.04 & 0.78 \\
        \hline
        Northeastern Arm & & 0.18 & 0.07 & 0.08 & 0.05 & -0.20 & 0.20 & 0.12 & 0.02 & -0.53 & $<$0.01 \\
        & & 0.23 & 0.13 & 0.16 & 0.29 & -0.10 & 0.51 & 0.23 & 0.13 & -0.46 & $<$0.01 \\
        & & 0.27 & 0.24 & 0.29 & 0.62 & -0.01 & 0.83 & 0.34 & 0.43 & -0.38 & 0.01 \\
        \hline
    \end{tabular}
    \caption{Spearman's $\rho$ and $p$ values for various regions and quantities in NGC 5257. The 16th, 50th, and 84th percentiles are listed from top to bottom within each section for $\Sigma_{\mathrm{H_2}}$-$\langle \sigma \rangle$, $\Sigma_{\mathrm{SFR}}$-$\langle \sigma \rangle$, SFE-$\langle \sigma \rangle$, $\Sigma_{\mathrm{SFR}}$-P$_{\mathrm{turb}}$, and SFE-P$_{\mathrm{turb}}$.}
    \label{tab:correlation_ngc5257}
\end{table*}

\textcolor{white}{space}

\textcolor{white}{space}

\textcolor{white}{space}

\textcolor{white}{space}

\textcolor{white}{space}

\textcolor{white}{space}

\textcolor{white}{space}

\textcolor{white}{space}

\textcolor{white}{space}

\textcolor{white}{space}

\section{Suplementary Plots}\label{Apendix_plots}
\begin{figure}[ht]
    \centering

    \begin{tabular}{cc}
        \includegraphics[width=0.35\textwidth]{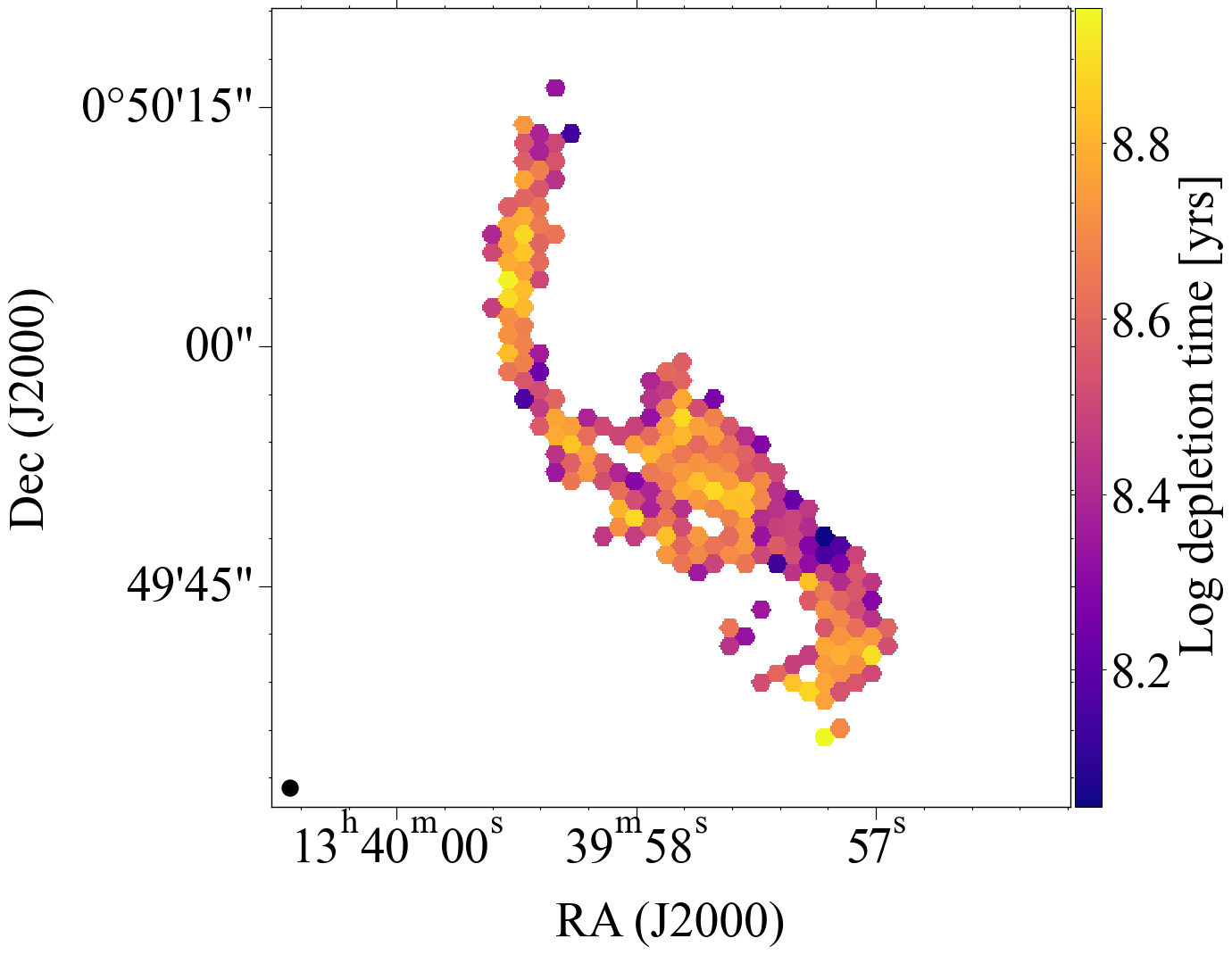} &
        \includegraphics[width=0.35\textwidth]{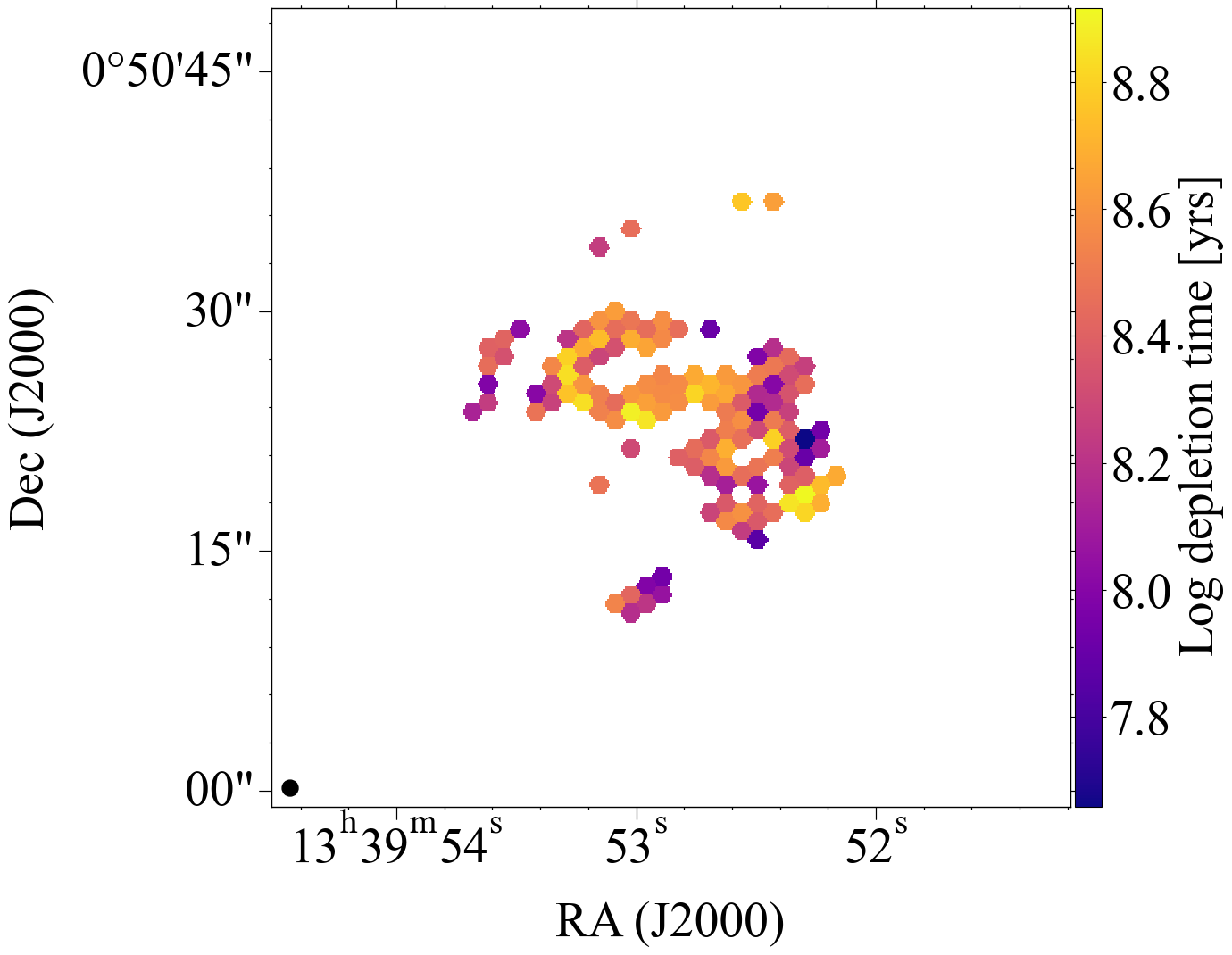} \\
        \textbf{(a) t$_{\mathrm{dep}}$ NGC 5258} & \textbf{(b) t$_{\mathrm{dep}}$ NGC 5257} \\
    \end{tabular}
    \\[2ex]

    \begin{tabular}{cc}
        \includegraphics[width=0.35\textwidth]{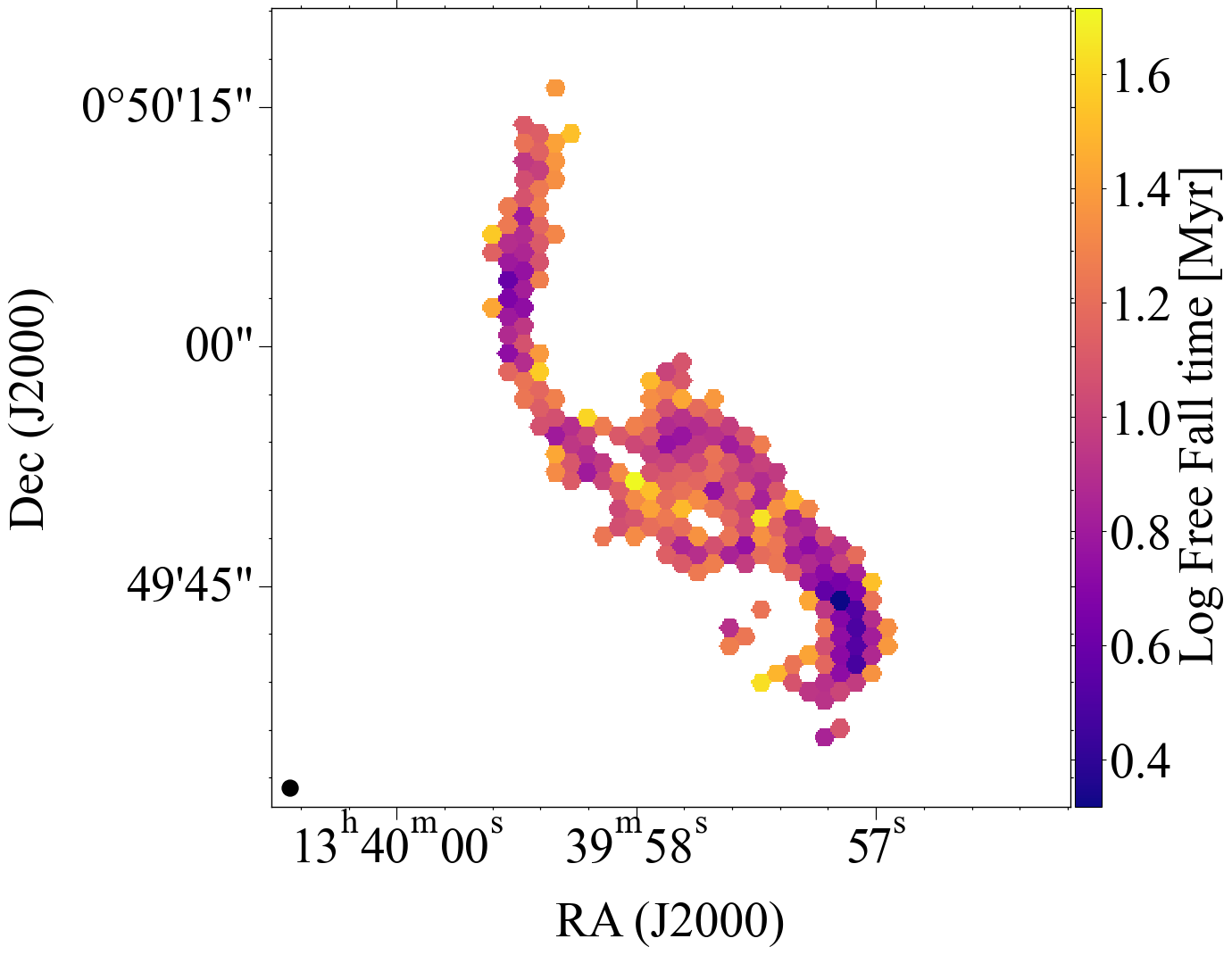} &
        \includegraphics[width=0.35\textwidth]{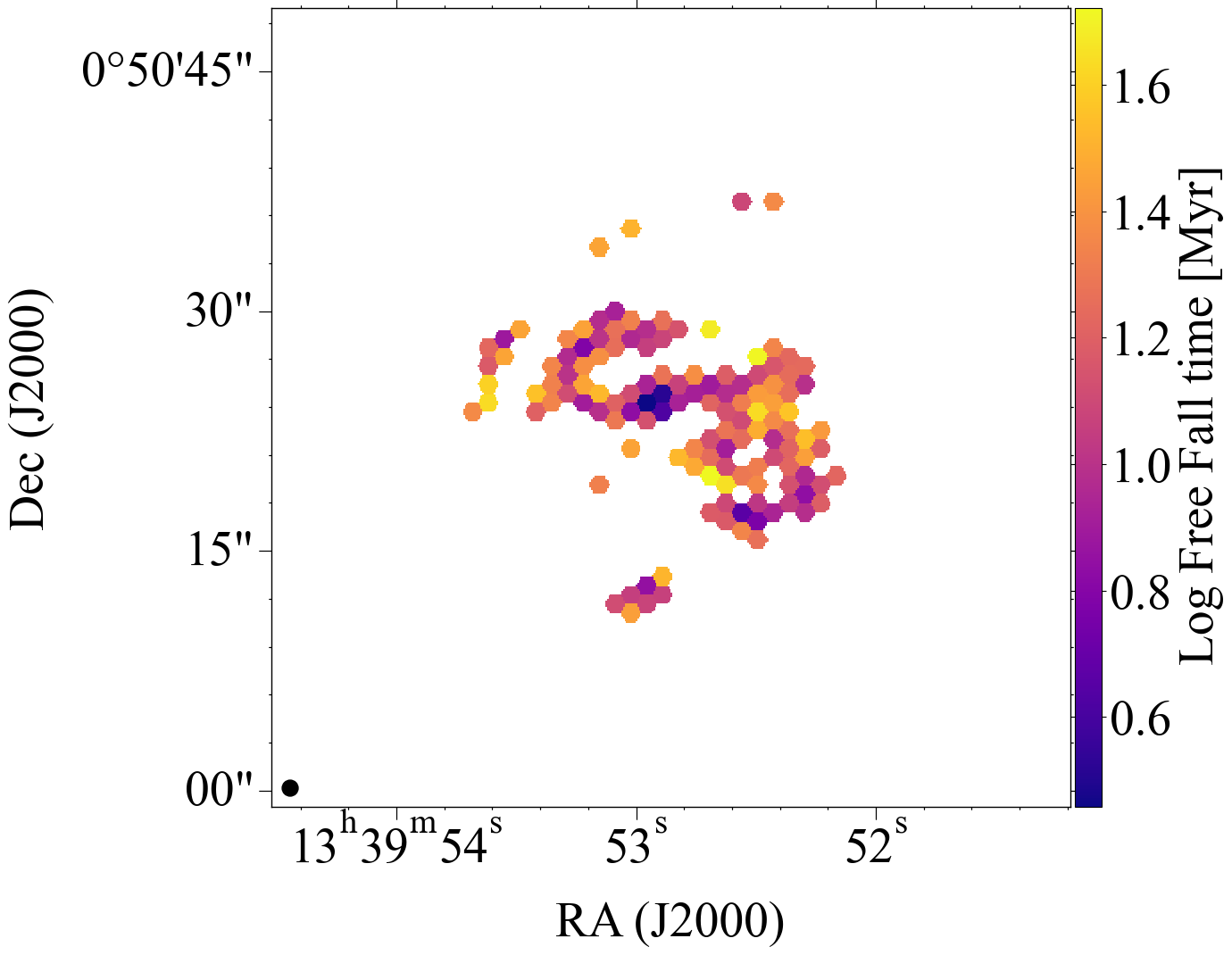} \\
        \textbf{(c) t$_{\mathrm{ff}}$ NGC 5258} & \textbf{(d) t$_{\mathrm{ff}}$ NGC 5257} \\
    \end{tabular}

    \caption{\small Supplementary plots: Maps of depletion times (upper panels) and free-fall times (lower panels) for NGC 5258 (left panels) and NGC 5257 (right panels).}
    \label{fig:figure_td}
\end{figure}


\begin{thebibliography}{}
\expandafter\ifx\csname natexlab\endcsname\relax\def\natexlab#1{#1}\fi
\providecommand{\url}[1]{\href{#1}{#1}}
\providecommand{\dodoi}[1]{doi:~\href{http://doi.org/#1}{\nolinkurl{#1}}}
\providecommand{\doeprint}[1]{\href{http://ascl.net/#1}{\nolinkurl{http://ascl.net/#1}}}
\providecommand{\doarXiv}[1]{\href{https://arxiv.org/abs/#1}{\nolinkurl{https://arxiv.org/abs/#1}}}

\bibitem[{Akaike(1974)}]{Akaike_74}
Akaike, H. 1974, IEEE Transactions on Automatic Control, 19, 716,
  \dodoi{10.1109/TAC.1974.1100705}

\bibitem[{{Armus} {et~al.}(2009){Armus}, {Mazzarella}, {Evans}, {Surace},
  {Sanders}, {Iwasawa}, {Frayer}, {Howell}, {Chan}, {Petric}, {Vavilkin},
  {Kim}, {Haan}, {Inami}, {Murphy}, {Appleton}, {Barnes}, {Bothun}, {Bridge},
  {Charmandaris}, {Jensen}, {Kewley}, {Lord}, {Madore}, {Marshall},
  {Melbourne}, {Rich}, {Satyapal}, {Schulz}, {Spoon}, {Sturm}, {U}, {Veilleux},
  \& {Xu}}]{GOALS_2009}
{Armus}, L., {Mazzarella}, J.~M., {Evans}, A.~S., {et~al.} 2009, \pasp, 121,
  559, \dodoi{10.1086/600092}

\bibitem[{Barcos-Muñoz {et~al.}(2017)Barcos-Muñoz, Leroy, Evans, Condon,
  Privon, Thompson, Armus, Díaz-Santos, Mazzarella, Meier, Momjian, Murphy,
  Ott, Sanders, Schinnerer, Stierwalt, Surace, \& Walter}]{Barcos-Muñoz_2017}
Barcos-Muñoz, L., Leroy, A.~K., Evans, A.~S., {et~al.} 2017, The Astrophysical
  Journal, 843, 117, \dodoi{10.3847/1538-4357/aa789a}

\bibitem[{{Barnes} \& {Hernquist}(1996)}]{Barnes_1996}
{Barnes}, J.~E., \& {Hernquist}, L. 1996, \apj, 471, 115,
  \dodoi{10.1086/177957}

\bibitem[{Benincasa {et~al.}(2020)Benincasa, Loebman, Wetzel, Hopkins, Murray,
  Bellardini, Faucher-Giguère, Guszejnov, \& Orr}]{Benincasa_2019}
Benincasa, S.~M., Loebman, S.~R., Wetzel, A., {et~al.} 2020, Monthly Notices of
  the Royal Astronomical Society, 497, 3993, \dodoi{10.1093/mnras/staa2116}

\bibitem[{{Bigiel} {et~al.}(2008){Bigiel}, {Leroy}, {Walter}, {Brinks}, {de
  Blok}, {Madore}, \& {Thornley}}]{Bigiel_2008}
{Bigiel}, F., {Leroy}, A., {Walter}, F., {et~al.} 2008, \aj, 136, 2846,
  \dodoi{10.1088/0004-6256/136/6/2846}

\bibitem[{Boggs {et~al.}(1987)Boggs, Byrd, \& Schnabel}]{ODR}
Boggs, P.~T., Byrd, R.~H., \& Schnabel, R.~B. 1987, SIAM Journal on Scientific
  and Statistical Computing, 8, 1052, \dodoi{10.1137/0908085}

\bibitem[{Bolatto {et~al.}(2013)Bolatto, Wolfire, \& Leroy}]{Bolatto_2013}
Bolatto, A.~D., Wolfire, M., \& Leroy, A.~K. 2013, Annual Review of Astronomy
  and Astrophysics, 51, 207, \dodoi{10.1146/annurev-astro-082812-140944}

\bibitem[{{Condon} \& {Yin}(1990)}]{Condon1990}
{Condon}, J.~J., \& {Yin}, Q.~F. 1990, \apj, 357, 97, \dodoi{10.1086/168894}

\bibitem[{{Dekel} {et~al.}(2009){Dekel}, {Birnboim}, {Engel}, {Freundlich},
  {Goerdt}, {Mumcuoglu}, {Neistein}, {Pichon}, {Teyssier}, \&
  {Zinger}}]{Dekel_2009}
{Dekel}, A., {Birnboim}, Y., {Engel}, G., {et~al.} 2009, \nat, 457, 451,
  \dodoi{10.1038/nature07648}

\bibitem[{Evans {et~al.}(2022)Evans, Frayer, Charmandaris, Armus, Inami,
  Surace, Linden, Soifer, Diaz-Santos, Larson, Rich, Song, Barcos-Munoz,
  Mazzarella, Privon, U, Medling, Böker, Aalto, Iwasawa, Howell, van~der Werf,
  Appleton, Bohn, Brown, Hayward, Hoshioka, Kemper, Lai, Law, Malkan, Marshall,
  Murphy, Sanders, \& Stierwalt}]{Evans_2022}
Evans, A.~S., Frayer, D.~T., Charmandaris, V., {et~al.} 2022, The Astrophysical
  Journal Letters, 940, L8, \dodoi{10.3847/2041-8213/ac9971}

\bibitem[{Feigelson(1992)}]{Feigelson1992}
Feigelson, E.~D. 1992, Censoring in Astronomical Data Due to Nondetections (New
  York, NY: Springer New York), 221--237, \dodoi{10.1007/978-1-4613-9290-3_24}

\bibitem[{{Grudi{\'c}} {et~al.}(2018){Grudi{\'c}}, {Hopkins},
  {Faucher-Gigu{\`e}re}, {Quataert}, {Murray}, \& {Kere{\v{s}}}}]{Grudic_2018}
{Grudi{\'c}}, M.~Y., {Hopkins}, P.~F., {Faucher-Gigu{\`e}re}, C.-A., {et~al.}
  2018, \mnras, 475, 3511, \dodoi{10.1093/mnras/sty035}

\bibitem[{{Haan} {et~al.}(2011){Haan}, {Surace}, {Armus}, {Evans}, {Howell},
  {Mazzarella}, {Kim}, {Vavilkin}, {Inami}, {Sanders}, {Petric}, {Bridge},
  {Melbourne}, {Charmandaris}, {Diaz-Santos}, {Murphy}, {U}, {Stierwalt}, \&
  {Marshall}}]{Haan_2011}
{Haan}, S., {Surace}, J.~A., {Armus}, L., {et~al.} 2011, \aj, 141, 100,
  \dodoi{10.1088/0004-6256/141/3/100}

\bibitem[{{Hauke} \& {Kossowski}(2011)}]{Spearman_stat}
{Hauke}, J., \& {Kossowski}, T. 2011, Quaestiones Geographicae, 30, 87,
  \dodoi{10.2478/v10117-011-0021-1}

\bibitem[{He {et~al.}(2020)He, Wilson, Sliwa, Iono, \& Saito}]{He_2019}
He, H., Wilson, C.~D., Sliwa, K., Iono, D., \& Saito, T. 2020, Monthly Notices
  of the Royal Astronomical Society, 496, 5243, \dodoi{10.1093/mnras/staa1826}

\bibitem[{{Helou} {et~al.}(1985){Helou}, {Soifer}, \& {Rowan-Robinson}}]{Helou}
{Helou}, G., {Soifer}, B.~T., \& {Rowan-Robinson}, M. 1985, \apjl, 298, L7,
  \dodoi{10.1086/184556}

\bibitem[{{H{\"o}gbom}(1974)}]{Hogbom_1974}
{H{\"o}gbom}, J.~A. 1974, \aaps, 15, 417

\bibitem[{{Hopkins} {et~al.}(2008){Hopkins}, {Hernquist}, {Cox}, \&
  {Kere{\v{s}}}}]{Hopkins_2008}
{Hopkins}, P.~F., {Hernquist}, L., {Cox}, T.~J., \& {Kere{\v{s}}}, D. 2008,
  \apjs, 175, 356, \dodoi{10.1086/524362}

\bibitem[{Hopkins {et~al.}(2018)Hopkins, Wetzel, Kereš, Faucher-Giguère,
  Quataert, Boylan-Kolchin, Murray, Hayward, Garrison-Kimmel, Hummels,
  Feldmann, Torrey, Ma, Anglés-Alcázar, Su, Orr, Schmitz, Escala, Sanderson,
  Grudić, Hafen, Kim, Fitts, Bullock, Wheeler, Chan, Elbert, \&
  Narayanan}]{FIRE_2}
Hopkins, P.~F., Wetzel, A., Kereš, D., {et~al.} 2018, Monthly Notices of the
  Royal Astronomical Society, 480, 800, \dodoi{10.1093/mnras/sty1690}

\bibitem[{Iono {et~al.}(2005)Iono, Yun, \& Ho}]{Iono_2005}
Iono, D., Yun, M.~S., \& Ho, P. T.~P. 2005, The Astrophysical Journal
  Supplement Series, 158, 1, \dodoi{10.1086/429093}

\bibitem[{{Isobe} {et~al.}(1990){Isobe}, {Feigelson}, {Akritas}, \&
  {Babu}}]{ASURV_1990}
{Isobe}, T., {Feigelson}, E.~D., {Akritas}, M.~G., \& {Babu}, G.~J. 1990, \apj,
  364, 104, \dodoi{10.1086/169390}

\bibitem[{{Isobe} {et~al.}(1986){Isobe}, {Feigelson}, \& {Nelson}}]{ASURV_1986}
{Isobe}, T., {Feigelson}, E.~D., \& {Nelson}, P.~I. 1986, \apj, 306, 490,
  \dodoi{10.1086/164359}

\bibitem[{Kelly(2007)}]{Kelly_2007}
Kelly, B.~C. 2007, The Astrophysical Journal, 665, 1489, \dodoi{10.1086/519947}

\bibitem[{{Kennicutt}(1998)}]{Kennicutt98}
{Kennicutt}, Robert~C., J. 1998, \apj, 498, 541, \dodoi{10.1086/305588}

\bibitem[{{Kennicutt} \& {De Los Reyes}(2021)}]{Kennicutt2021}
{Kennicutt}, Robert~C., J., \& {De Los Reyes}, M. A.~C. 2021, \apj, 908, 61,
  \dodoi{10.3847/1538-4357/abd3a2}

\bibitem[{{Kennicutt} \& {Evans}(2012)}]{Kennicutt_n_Evans_2012}
{Kennicutt}, R.~C., \& {Evans}, N.~J. 2012, \araa, 50, 531,
  \dodoi{10.1146/annurev-astro-081811-125610}

\bibitem[{Kepley {et~al.}(2020)Kepley, Tsutsumi, Brogan, Indebetouw, Yoon,
  Mason, \& Meyer}]{Kepley_2020}
Kepley, A.~A., Tsutsumi, T., Brogan, C.~L., {et~al.} 2020, Publications of the
  Astronomical Society of the Pacific, 132, 024505,
  \dodoi{10.1088/1538-3873/ab5e14}

\bibitem[{{Kormendy} \& {Ho}(2013)}]{Kormendy_2013}
{Kormendy}, J., \& {Ho}, L.~C. 2013, \araa, 51, 511,
  \dodoi{10.1146/annurev-astro-082708-101811}

\bibitem[{Kreckel {et~al.}(2018)Kreckel, Faesi, Kruijssen, Schruba, Groves,
  Leroy, Bigiel, Blanc, Chevance, Herrera, Hughes, McElroy, Pety, Querejeta,
  Rosolowsky, Schinnerer, Sun, Usero, \& Utomo}]{Kreckel_2018}
Kreckel, K., Faesi, C., Kruijssen, J. M.~D., {et~al.} 2018, The Astrophysical
  Journal Letters, 863, L21, \dodoi{10.3847/2041-8213/aad77d}

\bibitem[{Kroupa(2001)}]{Kroupa_2001}
Kroupa, P. 2001, Monthly Notices of the Royal Astronomical Society, 322, 231,
  \dodoi{10.1046/j.1365-8711.2001.04022.x}

\bibitem[{Kruijssen {et~al.}(2018)Kruijssen, Schruba, Hygate, Hu, Haydon, \&
  Longmore}]{Kruijssen_2018}
Kruijssen, J. M.~D., Schruba, A., Hygate, A. P.~S., {et~al.} 2018, Monthly
  Notices of the Royal Astronomical Society, 479, 1866,
  \dodoi{10.1093/mnras/sty1128}

\bibitem[{Krumholz \& McKee(2005)}]{Krumholz_2005}
Krumholz, M.~R., \& McKee, C.~F. 2005, The Astrophysical Journal, 630, 250,
  \dodoi{10.1086/431734}

\bibitem[{{Krumholz} {et~al.}(2009){Krumholz}, {McKee}, \&
  {Tumlinson}}]{Krumholz_2009}
{Krumholz}, M.~R., {McKee}, C.~F., \& {Tumlinson}, J. 2009, \apj, 693, 216,
  \dodoi{10.1088/0004-637X/693/1/216}

\bibitem[{Leitherer {et~al.}(1999)Leitherer, Schaerer, Goldader, Delgado,
  Robert, Kune, de~Mello, Devost, \& Heckman}]{Leitherer_1999}
Leitherer, C., Schaerer, D., Goldader, J.~D., {et~al.} 1999, The Astrophysical
  Journal Supplement Series, 123, 3, \dodoi{10.1086/313233}

\bibitem[{Leroy {et~al.}(2008)Leroy, Walter, Brinks, Bigiel, de~Blok, Madore,
  \& Thornley}]{Leroy_2008}
Leroy, A.~K., Walter, F., Brinks, E., {et~al.} 2008, The Astronomical Journal,
  136, 2782, \dodoi{10.1088/0004-6256/136/6/2782}

\bibitem[{{Leroy} {et~al.}(2013){Leroy}, {Walter}, {Sandstrom}, {Schruba},
  {Munoz-Mateos}, {Bigiel}, {Bolatto}, {Brinks}, {de Blok}, {Meidt}, {Rix},
  {Rosolowsky}, {Schinnerer}, {Schuster}, \& {Usero}}]{Leroy_2013}
{Leroy}, A.~K., {Walter}, F., {Sandstrom}, K., {et~al.} 2013, \aj, 146, 19,
  \dodoi{10.1088/0004-6256/146/2/19}

\bibitem[{{Leroy} {et~al.}(2021){Leroy}, {Schinnerer}, {Hughes}, {Rosolowsky},
  {Pety}, {Schruba}, {Usero}, {Blanc}, {Chevance}, {Emsellem}, {Faesi},
  {Herrera}, {Liu}, {Meidt}, {Querejeta}, {Saito}, {Sandstrom}, {Sun},
  {Williams}, {Anand}, {Barnes}, {Behrens}, {Belfiore}, {Benincasa},
  {Be{\v{s}}li{\'c}}, {Bigiel}, {Bolatto}, {den Brok}, {Cao}, {Chandar},
  {Chastenet}, {Chiang}, {Congiu}, {Dale}, {Deger}, {Eibensteiner}, {Egorov},
  {Garc{\'\i}a-Rodr{\'\i}guez}, {Glover}, {Grasha}, {Henshaw}, {Ho}, {Kepley},
  {Kim}, {Klessen}, {Kreckel}, {Koch}, {Kruijssen}, {Larson}, {Lee}, {Lopez},
  {Machado}, {Mayker}, {McElroy}, {Murphy}, {Ostriker}, {Pan}, {Pessa},
  {Puschnig}, {Razza}, {S{\'a}nchez-Bl{\'a}zquez}, {Santoro}, {Sardone},
  {Scheuermann}, {Sliwa}, {Sormani}, {Stuber}, {Thilker}, {Turner}, {Utomo},
  {Watkins}, \& {Whitmore}}]{Phangs2021}
{Leroy}, A.~K., {Schinnerer}, E., {Hughes}, A., {et~al.} 2021, \apjs, 257, 43,
  \dodoi{10.3847/1538-4365/ac17f3}

\bibitem[{Liddle(2007)}]{AIC_BIC}
Liddle, A.~R. 2007, Monthly Notices of the Royal Astronomical Society: Letters,
  377, L74, \dodoi{10.1111/j.1745-3933.2007.00306.x}

\bibitem[{Linden {et~al.}(2019)Linden, Song, Evans, Murphy, Armus,
  Barcos-Muñoz, Larson, Díaz-Santos, Privon, Howell, Surace, Charmandaris,
  Vivian, Medling, Chu, \& Momjian}]{Linden_2019}
Linden, S.~T., Song, Y., Evans, A.~S., {et~al.} 2019, The Astrophysical
  Journal, 881, 70, \dodoi{10.3847/1538-4357/ab2872}

\bibitem[{{McMullin} {et~al.}(2007){McMullin}, {Waters}, {Schiebel}, {Young},
  \& {Golap}}]{CASA}
{McMullin}, J.~P., {Waters}, B., {Schiebel}, D., {Young}, W., \& {Golap}, K.
  2007, in Astronomical Society of the Pacific Conference Series, Vol. 376,
  Astronomical Data Analysis Software and Systems XVI, ed. R.~A. {Shaw},
  F.~{Hill}, \& D.~J. {Bell}, 127

\bibitem[{{Murphy} {et~al.}(2011){Murphy}, {Condon}, {Schinnerer}, {Kennicutt},
  {Calzetti}, {Armus}, {Helou}, {Turner}, {Aniano}, {Beir{\~a}o}, {Bolatto},
  {Brandl}, {Croxall}, {Dale}, {Donovan Meyer}, {Draine}, {Engelbracht},
  {Hunt}, {Hao}, {Koda}, {Roussel}, {Skibba}, \& {Smith}}]{Murphy_2011}
{Murphy}, E.~J., {Condon}, J.~J., {Schinnerer}, E., {et~al.} 2011, \apj, 737,
  67, \dodoi{10.1088/0004-637X/737/2/67}

\bibitem[{Murphy {et~al.}(2012)Murphy, Bremseth, Mason, Condon, Schinnerer,
  Aniano, Armus, Helou, Turner, \& Jarrett}]{Murphy_2012}
Murphy, E.~J., Bremseth, J., Mason, B.~S., {et~al.} 2012, The Astrophysical
  Journal, 761, 97, \dodoi{10.1088/0004-637X/761/2/97}

\bibitem[{{Murray} {et~al.}(2010){Murray}, {Quataert}, \&
  {Thompson}}]{Murray_2010}
{Murray}, N., {Quataert}, E., \& {Thompson}, T.~A. 2010, \apj, 709, 191,
  \dodoi{10.1088/0004-637X/709/1/191}

\bibitem[{{Niklas} {et~al.}(1997){Niklas}, {Klein}, \&
  {Wielebinski}}]{Niklas_97}
{Niklas}, S., {Klein}, U., \& {Wielebinski}, R. 1997, \aap, 322, 19

\bibitem[{Orr {et~al.}(2020)Orr, Hayward, Medling, Gurvich, Hopkins, Murray,
  Pineda, Faucher-Giguère, Kereš, Wetzel, \& Su}]{Orr_FIRE_2020}
Orr, M.~E., Hayward, C.~C., Medling, A.~M., {et~al.} 2020, Monthly Notices of
  the Royal Astronomical Society, 496, 1620, \dodoi{10.1093/mnras/staa1619}

\bibitem[{Pedregosa {et~al.}(2011)Pedregosa, Varoquaux, Gramfort, Michel,
  Thirion, Grisel, Blondel, Prettenhofer, Weiss, Dubourg, Vanderplas, Passos,
  Cournapeau, Brucher, Perrot, \& Duchesnay}]{scikit-learn}
Pedregosa, F., Varoquaux, G., Gramfort, A., {et~al.} 2011, Journal of Machine
  Learning Research, 12, 2825

\bibitem[{{Pessa, I.} {et~al.}(2021){Pessa, I.}, {Schinnerer, E.}, {Belfiore,
  F.}, {Emsellem, E.}, {Leroy, A. K.}, {Schruba, A.}, {Kruijssen, J. M. D.},
  {Pan, H.-A.}, {Blanc, G. A.}, {Sanchez-Blazquez, P.}, {Bigiel, F.},
  {Chevance, M.}, {Congiu, E.}, {Dale, D.}, {Faesi, C. M.}, {Glover, S. C. O.},
  {Grasha, K.}, {Groves, B.}, {Ho, I.}, {Jiménez-Donaire, M.}, {Klessen, R.},
  {Kreckel, K.}, {Koch, E. W.}, {Liu, D.}, {Meidt, S.}, {Pety, J.}, {Querejeta,
  M.}, {Rosolowsky, E.}, {Saito, T.}, {Santoro, F.}, {Sun, J.}, {Usero, A.},
  {Watkins, E. J.}, \& {Williams, T. G.}}]{Pessa_2021}
{Pessa, I.}, {Schinnerer, E.}, {Belfiore, F.}, {et~al.} 2021, A\&A, 650, A134,
  \dodoi{10.1051/0004-6361/202140733}

\bibitem[{Privon {et~al.}(2013)Privon, Barnes, Evans, Hibbard, Yun, Mazzarella,
  Armus, \& Surace}]{Privon_2013}
Privon, G.~C., Barnes, J.~E., Evans, A.~S., {et~al.} 2013, The Astrophysical
  Journal, 771, 120, \dodoi{10.1088/0004-637X/771/2/120}

\bibitem[{{Querejeta, M.} {et~al.}(2021){Querejeta, M.}, {Schinnerer, E.},
  {Meidt, S.}, {Sun, J.}, {Leroy, A. K.}, {Emsellem, E.}, {Klessen, R. S.},
  {Muñoz-Mateos, J. C.}, {Salo, H.}, {Laurikainen, E.}, {Bešlić, I.},
  {Blanc, G. A.}, {Chevance, M.}, {Dale, D. A.}, {Eibensteiner, C.}, {Faesi,
  C.}, {García-Rodríguez, A.}, {Glover, S. C. O.}, {Grasha, K.}, {Henshaw,
  J.}, {Herrera, C.}, {Hughes, A.}, {Kreckel, K.}, {Kruijssen, J. M. D.}, {Liu,
  D.}, {Murphy, E. J.}, {Pan, H.-A.}, {Pety, J.}, {Razza, A.}, {Rosolowsky,
  E.}, {Saito, T.}, {Schruba, A.}, {Usero, A.}, {Watkins, E. J.}, \& {Williams,
  T. G.}}]{Querejeta_2021}
{Querejeta, M.}, {Schinnerer, E.}, {Meidt, S.}, {et~al.} 2021, A\&A, 656, A133,
  \dodoi{10.1051/0004-6361/202140695}

\bibitem[{{Rau} \& {Cornwell}(2011)}]{msmf_deconvolver}
{Rau}, U., \& {Cornwell}, T.~J. 2011, \aap, 532, A71,
  \dodoi{10.1051/0004-6361/201117104}

\bibitem[{Rich {et~al.}(2023)Rich, Aalto, Evans, Charmandaris, Privon, Lai,
  Inami, Linden, Armus, Diaz-Santos, Appleton, Barcos-Muñoz, Böker, Larson,
  Law, Malkan, Medling, Song, U, van~der Werf, Bohn, Brown, Finnerty, Hayward,
  Howell, Iwasawa, Kemper, Marshall, Mazzarella, McKinney, Muller-Sanchez,
  Murphy, Sanders, Soifer, Stierwalt, \& Surace}]{Rich_2023}
Rich, J., Aalto, S., Evans, A.~S., {et~al.} 2023, The Astrophysical Journal
  Letters, 944, L50, \dodoi{10.3847/2041-8213/acb2b8}

\bibitem[{{S{\'a}nchez-Garc{\'\i}a} {et~al.}(2022){S{\'a}nchez-Garc{\'\i}a},
  {Pereira-Santaella}, {Garc{\'\i}a-Burillo}, {Colina}, {Alonso-Herrero},
  {Villar-Mart{\'\i}n}, {Saito}, {D{\'\i}az-Santos}, {Piqueras L{\'o}pez},
  {Arribas}, {Bellocchi}, {Cazzoli}, \& {Labiano}}]{Sanchez-Garcia_2022}
{S{\'a}nchez-Garc{\'\i}a}, M., {Pereira-Santaella}, M., {Garc{\'\i}a-Burillo},
  S., {et~al.} 2022, \aap, 659, A102, \dodoi{10.1051/0004-6361/202141963}

\bibitem[{Sanders \& Mirabel(1996)}]{Sanders1996}
Sanders, D.~B., \& Mirabel, I.~F. 1996, Annual Review of Astronomy and
  Astrophysics, 34, 749, \dodoi{10.1146/annurev.astro.34.1.749}

\bibitem[{{Schinnerer} {et~al.}(2019){Schinnerer}, {Hughes}, {Leroy}, {Groves},
  {Blanc}, {Kreckel}, {Bigiel}, {Chevance}, {Dale}, {Emsellem}, {Faesi},
  {Glover}, {Grasha}, {Henshaw}, {Hygate}, {Kruijssen}, {Meidt}, {Pety},
  {Querejeta}, {Rosolowsky}, {Saito}, {Schruba}, {Sun}, \&
  {Utomo}}]{Schinnerer2019}
{Schinnerer}, E., {Hughes}, A., {Leroy}, A., {et~al.} 2019, \apj, 887, 49,
  \dodoi{10.3847/1538-4357/ab50c2}

\bibitem[{{Schmidt}(1959)}]{Schmidt_1959}
{Schmidt}, M. 1959, \apj, 129, 243, \dodoi{10.1086/146614}

\bibitem[{{Schmitt}(1985)}]{Schmitt_1985}
{Schmitt}, J.~H.~M.~M. 1985, \apj, 293, 178, \dodoi{10.1086/163224}

\bibitem[{Schwarz(1978)}]{Schwarz_78}
Schwarz, G. 1978, The Annals of Statistics, 6, 461 ,
  \dodoi{10.1214/aos/1176344136}

\bibitem[{Semenov {et~al.}(2019)Semenov, Kravtsov, \& Gnedin}]{Semenov_2019}
Semenov, V.~A., Kravtsov, A.~V., \& Gnedin, N.~Y. 2019, The Astrophysical
  Journal, 870, 79, \dodoi{10.3847/1538-4357/aaf163}

\bibitem[{Shi {et~al.}(2018)Shi, Yan, Armus, Gu, Helou, Qiu, Gwyn, Stierwalt,
  Fang, Chen, Zhou, Wu, Zheng, Zhang, Gao, \& Wang}]{Shi_2018}
Shi, Y., Yan, L., Armus, L., {et~al.} 2018, The Astrophysical Journal, 853,
  149, \dodoi{10.3847/1538-4357/aaa3e6}

\bibitem[{Song {et~al.}(2021)Song, Linden, Evans, Barcos-Muñoz, Privon, Yoon,
  Murphy, Larson, Díaz-Santos, Armus, Mazzarella, Howell, Inami, Torres-Albà,
  U, Charmandaris, McKinney, Kunneriath, \& Momjian}]{Song_2021}
Song, Y., Linden, S.~T., Evans, A.~S., {et~al.} 2021, The Astrophysical
  Journal, 916, 73, \dodoi{10.3847/1538-4357/ac05c2}

\bibitem[{Song {et~al.}(2022)Song, Linden, Evans, Barcos-Muñoz, Murphy,
  Momjian, Díaz-Santos, Larson, Privon, Huang, Armus, Mazzarella, U, Inami,
  Charmandaris, Ricci, Emig, McKinney, Yoon, Kunneriath, Lai, Rodas-Quito,
  Saravia, Gao, Meynardie, \& Sanders}]{Song_2022}
---. 2022, The Astrophysical Journal, 940, 52, \dodoi{10.3847/1538-4357/ac923b}

\bibitem[{Stierwalt {et~al.}(2013)Stierwalt, Armus, Surace, Inami, Petric,
  Diaz-Santos, Haan, Charmandaris, Howell, Kim, Marshall, Mazzarella, Spoon,
  Veilleux, Evans, Sanders, Appleton, Bothun, Bridge, Chan, Frayer, Iwasawa,
  Kewley, Lord, Madore, Melbourne, Murphy, Rich, Schulz, Sturm, U, Vavilkin, \&
  Xu}]{Stierwalt_2013}
Stierwalt, S., Armus, L., Surace, J.~A., {et~al.} 2013, The Astrophysical
  Journal Supplement Series, 206, 1, \dodoi{10.1088/0067-0049/206/1/1}

\bibitem[{{Sun} {et~al.}(2018){Sun}, {Leroy}, {Schruba}, {Rosolowsky},
  {Hughes}, {Kruijssen}, {Meidt}, {Schinnerer}, {Blanc}, {Bigiel}, {Bolatto},
  {Chevance}, {Groves}, {Herrera}, {Hygate}, {Pety}, {Querejeta}, {Usero}, \&
  {Utomo}}]{Sun2018}
{Sun}, J., {Leroy}, A.~K., {Schruba}, A., {et~al.} 2018, \apj, 860, 172,
  \dodoi{10.3847/1538-4357/aac326}

\bibitem[{{Sun} {et~al.}(2020){Sun}, {Leroy}, {Schinnerer}, {Hughes},
  {Rosolowsky}, {Querejeta}, {Schruba}, {Liu}, {Saito}, {Herrera}, {Faesi},
  {Usero}, {Pety}, {Kruijssen}, {Ostriker}, {Bigiel}, {Blanc}, {Bolatto},
  {Boquien}, {Chevance}, {Dale}, {Deger}, {Emsellem}, {Glover}, {Grasha},
  {Groves}, {Henshaw}, {Jimenez-Donaire}, {Kim}, {Klessen}, {Kreckel}, {Lee},
  {Meidt}, {Sandstrom}, {Sardone}, {Utomo}, \& {Williams}}]{Sun2020}
{Sun}, J., {Leroy}, A.~K., {Schinnerer}, E., {et~al.} 2020, \apjl, 901, L8,
  \dodoi{10.3847/2041-8213/abb3be}

\bibitem[{Sun {et~al.}(2022)Sun, Leroy, Rosolowsky, Hughes, Schinnerer,
  Schruba, Koch, Blanc, Chiang, Groves, Liu, Meidt, Pan, Pety, Querejeta,
  Saito, Sandstrom, Sardone, Usero, Utomo, Williams, Barnes, Benincasa, Bigiel,
  Bolatto, Boquien, Chevance, Dale, Deger, Emsellem, Glover, Grasha, Henshaw,
  Klessen, Kreckel, Kruijssen, Ostriker, \& Thilker}]{Sun_2022}
Sun, J., Leroy, A.~K., Rosolowsky, E., {et~al.} 2022, The Astronomical Journal,
  164, 43, \dodoi{10.3847/1538-3881/ac74bd}

\bibitem[{Sun {et~al.}(2023)Sun, Leroy, Ostriker, Meidt, Rosolowsky,
  Schinnerer, Wilson, Utomo, Belfiore, Blanc, Emsellem, Faesi, Groves, Hughes,
  Koch, Kreckel, Liu, Pan, Pety, Querejeta, Razza, Saito, Sardone, Usero,
  Williams, Bigiel, Bolatto, Chevance, Dale, Gensior, Glover, Grasha, Henshaw,
  Jiménez-Donaire, Klessen, Kruijssen, Murphy, Neumann, Teng, \&
  Thilker}]{Sun_2023}
Sun, J., Leroy, A.~K., Ostriker, E.~C., {et~al.} 2023, The Astrophysical
  Journal Letters, 945, L19, \dodoi{10.3847/2041-8213/acbd9c}

\bibitem[{Team {et~al.}(2022)Team, Bean, Bhatnagar, Castro, Meyer, Emonts,
  Garcia, Garwood, Golap, Villalba, Harris, Hayashi, Hoskins, Hsieh,
  Jagannathan, Kawasaki, Keimpema, Kettenis, Lopez, Marvil, Masters, McNichols,
  Mehringer, Miel, Moellenbrock, Montesino, Nakazato, Ott, Petry, Pokorny,
  Raba, Rau, Schiebel, Schweighart, Sekhar, Shimada, Small, Steeb, Sugimoto,
  Suoranta, Tsutsumi, van Bemmel, Verkouter, Wells, Xiong, Szomoru, Griffith,
  Glendenning, \& Kern}]{CASA_2022}
Team, T.~C., Bean, B., Bhatnagar, S., {et~al.} 2022, Publications of the
  Astronomical Society of the Pacific, 134, 114501,
  \dodoi{10.1088/1538-3873/ac9642}

\bibitem[{Teng {et~al.}(2023)Teng, Sandstrom, Sun, Gong, Bolatto, Chiang,
  Leroy, Usero, Glover, Klessen, Liu, Querejeta, Schinnerer, Bigiel, Cao,
  Chevance, Eibensteiner, Grasha, Israel, Murphy, Neumann, Pan, Pinna, Sormani,
  Smith, Walter, \& Williams}]{Teng_2023}
Teng, Y.-H., Sandstrom, K.~M., Sun, J., {et~al.} 2023, The Astrophysical
  Journal, 950, 119, \dodoi{10.3847/1538-4357/accb86}

\bibitem[{Virtanen {et~al.}(2020)Virtanen, Gommers, Oliphant, Haberland, Reddy,
  Cournapeau, Burovski, Peterson, Weckesser, Bright, {van der Walt}, Brett,
  Wilson, Millman, Mayorov, Nelson, Jones, Kern, Larson, Carey, Polat, Feng,
  Moore, {VanderPlas}, Laxalde, Perktold, Cimrman, Henriksen, Quintero, Harris,
  Archibald, Ribeiro, Pedregosa, {van Mulbregt}, \& {SciPy 1.0
  Contributors}}]{SciPy}
Virtanen, P., Gommers, R., Oliphant, T.~E., {et~al.} 2020, Nature Methods, 17,
  261, \dodoi{10.1038/s41592-019-0686-2}

\bibitem[{Wilson {et~al.}(2019)Wilson, Elmegreen, Bemis, \&
  Brunetti}]{Wilson_2019}
Wilson, C.~D., Elmegreen, B.~G., Bemis, A., \& Brunetti, N. 2019, The
  Astrophysical Journal, 882, 5, \dodoi{10.3847/1538-4357/ab31f3}

\bibitem[{Wong \& Blitz(2002)}]{Wong_2002}
Wong, T., \& Blitz, L. 2002, The Astrophysical Journal, 569, 157,
  \dodoi{10.1086/339287}

\bibitem[{Yun {et~al.}(2001)Yun, Reddy, \& Condon}]{Yun_2001}
Yun, M.~S., Reddy, N.~A., \& Condon, J.~J. 2001, The Astrophysical Journal,
  554, 803, \dodoi{10.1086/323145}

\end{thebibliography}
\end{document}